\documentclass[11pt,a4paper]{article}
\usepackage[top=3cm,bottom=3cm,left=3cm,right=3cm]{geometry}

\usepackage{amsmath,amssymb}
\usepackage{graphicx}
\usepackage{array}
\usepackage{booktabs}
\usepackage{microtype}
\usepackage[usenames,dvipsnames]{xcolor}
\usepackage{multirow}
\usepackage{cite}
\usepackage{tikz}
\usepackage{layouts}

\usepackage[framemethod=TikZ]{mdframed}
\definecolor{boxcolor}{RGB}{230,230,230}
\newenvironment{highlightbox}[1][]{%
\mdfsetup{%
frametitle={%
\tikz[baseline=(current bounding box.east),outer sep=0pt]
\node[anchor=east,rectangle,fill=boxcolor]
{\strut #1};}}%
\mdfsetup{innertopmargin=10pt,linecolor=boxcolor,%
linewidth=2pt,topline=true,%
frametitleaboveskip=\dimexpr-\ht\strutbox\relax
}
\begin{mdframed}[]\relax%
}{\end{mdframed}}

\definecolor{cLink}{RGB}{0, 64, 255}
\definecolor{cURL}{RGB}{0, 64, 255}
\definecolor{cCite}{RGB}{0, 64, 255}
\usepackage[colorlinks=true, linkcolor=cLink, urlcolor=cURL, citecolor=cCite]{hyperref}
\hypersetup{linktocpage}
\numberwithin{equation}{section}
\numberwithin{table}{section}

\newcommand{\bbZ}{\mathbb{Z}}

\begin{document}

\baselineskip=14pt
\parskip 5pt plus 1pt

\vspace*{-1.5cm}
\begin{flushright}    
  {\small MITP/21-034
  }
\end{flushright}

\vspace{2cm}
\begin{center}        
  {\LARGE Modular Curves and the Refined Distance Conjecture}
\end{center}

\vspace{0.5cm}
\begin{center}        
{\large  Daniel Kl\"awer}
\end{center}

\vspace{0.15cm}
\begin{center}        
\emph{PRISMA$^{+}$ Cluster of Excellence and Mainz Institute for Theoretical Physics, \\
Johannes Gutenberg-Universit\"at, 55099 Mainz, Germany\\[0.5cm]
II. Institut für Theoretische Physik, Universität Hamburg,\\
Luruper Chaussee 149, 22607 Hamburg, Germany}
\\[0.15cm]
 
\end{center}

\vspace{2cm}


\begin{abstract}
\noindent   
    We test the refined distance conjecture in the vector multiplet moduli space of 4D $\mathcal{N}=2$ compactifications of the type IIA string that admit a dual heterotic description. In the weakly coupled regime of the heterotic string, the moduli space geometry is governed by the perturbative heterotic dualities, which allows for exact computations. This is reflected in the type IIA frame through the existence of a K3 fibration. We identify the degree $d=2N$ of the K3 fiber as a parameter that could potentially lead to large distances, which is substantiated by studying several explicit models. The moduli space geometry degenerates into the modular curve for the congruence subgroup $\Gamma_0(N)^+$. In order to probe the large $N$ regime, we initiate the study of Calabi-Yau threefolds fibered by general degree $d>8$ K3 surfaces by suggesting a construction as complete intersections in Grassmann bundles.
\end{abstract}

\thispagestyle{empty}
\clearpage

\tableofcontents

\setcounter{page}{1}


\newpage

\section{Introduction}
\label{sec:intro}

The swampland program of classifying consistent quantum gravity vacua in the form of low energy effective field theory data, as first envisioned in~\cite{Vafa:2005ui,Ooguri:2006in}, has by now grown into a large sub-field of contemporary string theory research\footnote{Several reviews exist, presenting a snapshot of this growing field at the respective time of their publication~\cite{Brennan:2017rbf,Palti:2019pca,vanBeest:2021lhn,Grana:2021zvf}.}. It has been revealed that the subset of effective field theories coupled to Einstein gravity that arise in string compactifications satisfies a web of conjectural properties, the swampland conjectures.

As of now, what are the most fundamental constraints on effective field theories that imply consistent coupling to quantum gravity is still under debate. Abstract principles like the absence of global symmetries~\cite{Banks:1988yz,Kallosh:1995hi,Banks:2010zn,Harlow:2018tng,Harlow:2020bee,Chen:2020ojn,Heidenreich:2020pkc} and topological operators~\cite{Rudelius:2020orz,Heidenreich:2021tna}, completeness of the spectrum~\cite{Polchinski:2003bq}, or the cobordism conjecture~\cite{McNamara:2019rup} seem very promising and have been used to argue for string universality in supersymmetric theories with a large number of uncompactified spacetime dimensions~\cite{Cvetic:2020kuw,Montero:2020icj,Dierigl:2020lai,Hamada:2021bbz}.

Presumably closer to our usual physical intuition but still central in the swampland program are the weak gravity~\cite{Arkani-Hamed:2006emk} and distance~\cite{Ooguri:2006in} conjectures. The weak gravity conjecture (WGC) places a constraint on the charge-to-mass ratio of particles in a gravitationally coupled gauge theory. Although it was pointed out already in the original paper that the WGC constraint is trivially satisfied by a wide margin in the standard model of particle physics, the statement and its generalizations are sufficiently strong and interesting that it has since been indirectly used to put constraints on BSM physics and to explain hierarchies~\cite{Ibanez:2017kvh,Ibanez:2017oqr,Gonzalo:2018tpb,Hamada:2017yji,Reece:2018zvv,Montero:2019ekk,Montero:2021otb}. The distance conjecture (DC) states that infinite distance limits $\Delta\phi\to\infty$ in the moduli space of massless scalar fields are accompanied by an infinite tower of exponentially light states $m\sim e^{-\alpha\Delta\phi}$~\footnote{Distances and masses are measured in Planck units.}. This is interesting because a suitably modified statement, adapted to the non-vanishing mass of light scalar fields in phenomenological models, could be used to rule out such theories that feature vastly trans-Planckian scalar field variations without explicitly taking into account the effect of the light tower.

The weak gravity and distance conjectures are at least partly motivated from the properties of BPS objects in supersymmetric string compactifications. Although tempting, it may be dangerous to extrapolate such strong statements to phenomenologically interesting settings with $\mathcal{N}=1$ or $\mathcal{N}=0$ supersymmetry. In doing so, one has to be particularly careful about quantum corrections that become relevant when we reduce the number of supercharges. The process of ``sharpening'' swampland conjectures by probing them in string compactifications of increasing complexity has already lead to advances in our understanding of the web of swampland conjectures. For example, it was realized that in order for the WGC to be consistent under compactification, there should be a whole tower or even sub-lattice of states satisfying the WGC inequality on the charge-to-mass ratio~\cite{Heidenreich:2015nta,Heidenreich:2016aqi,Andriolo:2018lvp}. This tower is in many cases identified with the DC tower, providing a partial unification of these two a priori unrelated conjectures. Subsequently, it was argued that the DC tower is always a tower of string excitations or a Kaluza-Klein tower, signalling the transition to a different duality frame or a decompactification~\cite{Lee:2018urn,Lee:2018spm,Lee:2019tst,Lee:2019xtm,Lee:2019wij,Klaewer:2020lfg}. Scrutinizing swampland conjectures in this way, it is likely that we will also have to weaken their statements in order to increase their scope of application. 

The main phenomenological loopholes of the distance conjecture are:
\begin{itemize}
    \item The exponent $\alpha$ could be tiny, so we have $\exp(-\alpha\Delta\phi)\simeq 1-\alpha\Delta\phi$ even for large $\Delta\phi$.
    \item The conjecture applies only to geodesic motion of moduli fields, but in absence of SUSY and global symmetries we expect all scalars to acquire a mass. Physical trajectories are then also influenced by the scalar potential.
    \item It is an asymptotic statement, so it may not have any relevance in the deep interior of field space.
\end{itemize}

All of these points have received some attention in the literature. In particular, the first two loopholes have been studied extensively. The ``decay constant'' $\alpha$ is known to be of order one in many controlled string constructions and consequently lower bounds have been proposed~\cite{Klaewer:2016kiy,Lanza:2020qmt,Lanza:2021qsu,Grimm:2018ohb,Blumenhagen:2018nts,Erkinger:2019umg,Joshi:2019nzi,Bedroya:2019snp,EnriquezRojo:2020hzi,Andriot:2020lea,Baume:2020dqd,Perlmutter:2020buo}. The way the conjecture should apply to scalar fields with a potential is not completely clear, but recent work suggests bounds on the amount of non-geodesicity that a potential is allowed to introduce in order to be compatible with the DC~\cite{Calderon-Infante:2020dhm}.

\begin{figure}
    \centering
    \begin{tikzpicture}
        \node at (0,0) {\includegraphics[width=13cm]{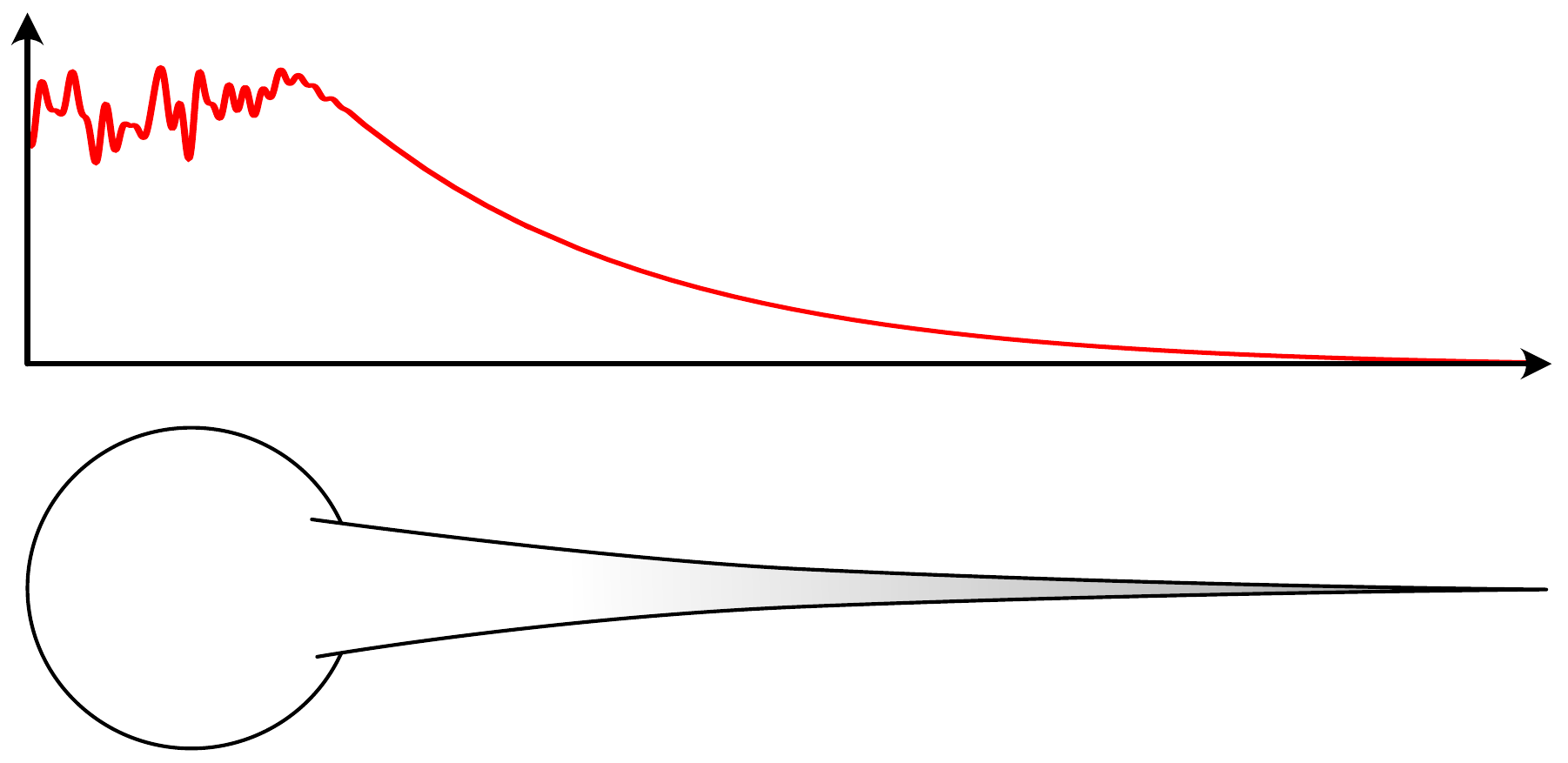}};
        \node at (-4.85,-1.7) {$\mathcal{M}_N$};
        \node at (6.7,0) {$\Delta\phi$};
        \node at (-6,3.3) {$M_{\textup{tower}}$};
    \end{tikzpicture}
    \caption{Sketch of the behavior of the DC tower mass scale over moduli space. In the bulk $\mathcal{M}_N$ of the moduli space the distance conjecture does not impose any constraint on $M_\textup{tower}(\phi)$, while in the throat region we approach $M_\textup{tower}\sim \exp(-\alpha\Delta\phi)$.}
    \label{fig:RDCModuliSpace}
\end{figure}
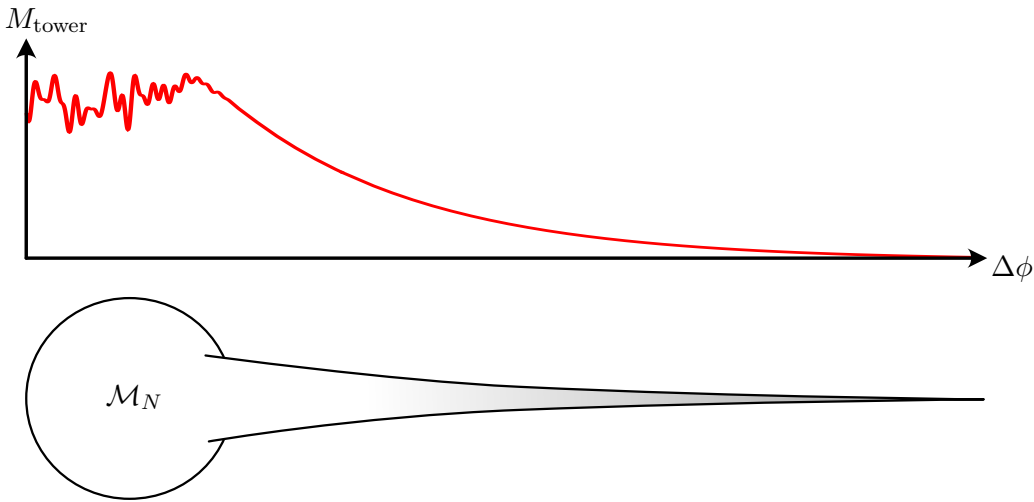

In this paper we will focus on the third point. Without any further qualification, we can picture a situation where we have a sequence of disjoint theories labeled by some integer $N$, with moduli spaces $\overline{\mathcal{M}}_N$, which decompose into bulk moduli spaces $\mathcal{M}_N$ and some infinite distance throats, see Figure~\ref{fig:RDCModuliSpace}. The distance conjecture does not place any constraints on the behavior of the DC tower in each individual bulk $\mathcal{M}_N$. It is then conceivable that we have such a sequence of theories with the property that $\Delta\phi_N\overset{N\to\infty}{\longrightarrow}\infty$, so we can generate large distances without having to deal with an infinite tower of states. This would render the distance conjecture essentially powerless. Whether such behavior could occur in actual string compactifications is another question. First of all, finiteness of the string landscape would imply some upper bound $N_\textup{max}$ on the sequence of compactifications. From our experience in studying Calabi-Yau compactifications, $N_\textup{max}$ may in principle be some astronomically large number~\cite{Demirtas:2020dbm}, which could allow for a numerically large $\Delta\phi_{N_\textup{max}}$. Secondly, we may expect that distinct $\mathcal{M}_N$ are actually connected through (possibly non-perturbative) transitions, modifying the naive diameter of the bulk.

The refined distance conjecture (RDC) postulates that this kind of behavior should not occur in a theory of quantum gravity:
\begin{highlightbox}[Refined Distance Conjecture]
    Consider two points $P,Q$ in moduli space. If $\textup{dist}(P,Q)\geq c$, where $c$ is a universal $\mathcal{O}(1)$ number, there is an infinite tower of states with mass scale
    \begin{equation}
        M_\textup{tower}(Q)<M_\textup{tower}(P)e^{-\alpha\,\textup{dist}(P,Q)}\;.
    \end{equation}
\end{highlightbox}
This conjecture was first motivated by observations in simple models of axion monodromy \cite{Baume:2016psm,Blumenhagen:2017cxt}, where the scalars in question are not true moduli. Axions are ordinarily not expected to control the mass scale of some infinite tower of states. Nevertheless, it was found that displacing them from a minimum of the scalar potential there was a backreaction on the saxions in the $\mathcal{N}=1$ chiral multiplet leading to a tower coming down. It was found that the critical distance after which this happened was always $\mathcal{O}(1)$ and could not be tuned by adjusting fluxes. Further motivation for the conjecture came from studying moduli in black hole backgrounds~\cite{Klaewer:2016kiy}.

\begin{figure}
    \centering
    \begin{tikzpicture}
        \node at (0,0) {\includegraphics[height=7cm]{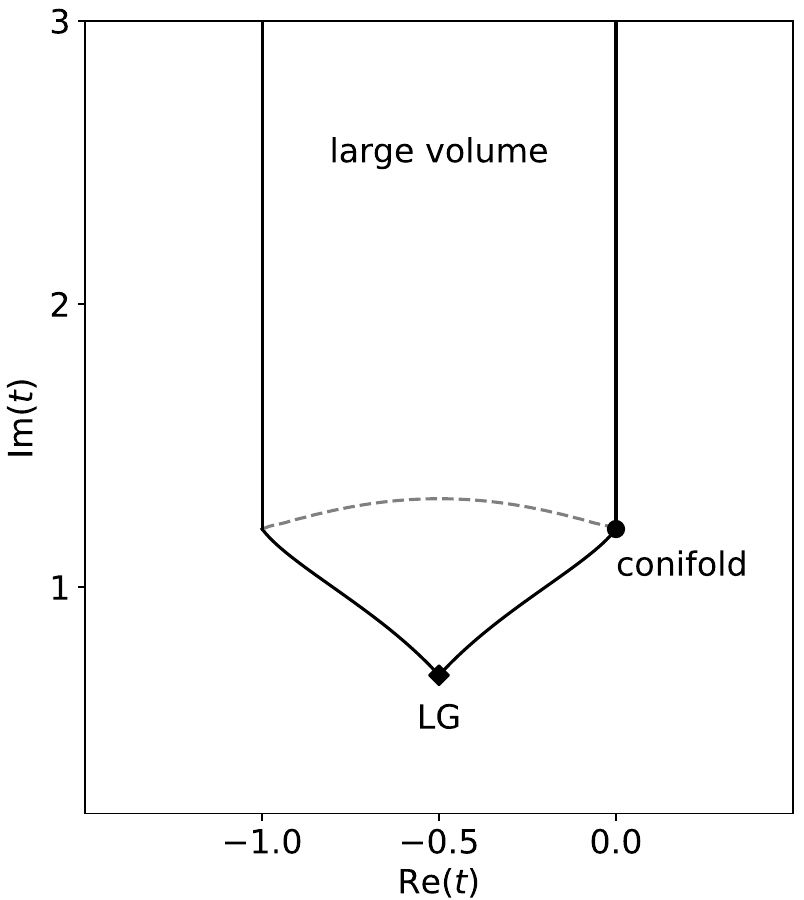}};
        \draw[->,red] (0.295,-1.7) -> (0.295,2.1);
    \end{tikzpicture}
    \hspace{1cm}
    \begin{tikzpicture}
        \node at (0,0) {\includegraphics[height=7cm]{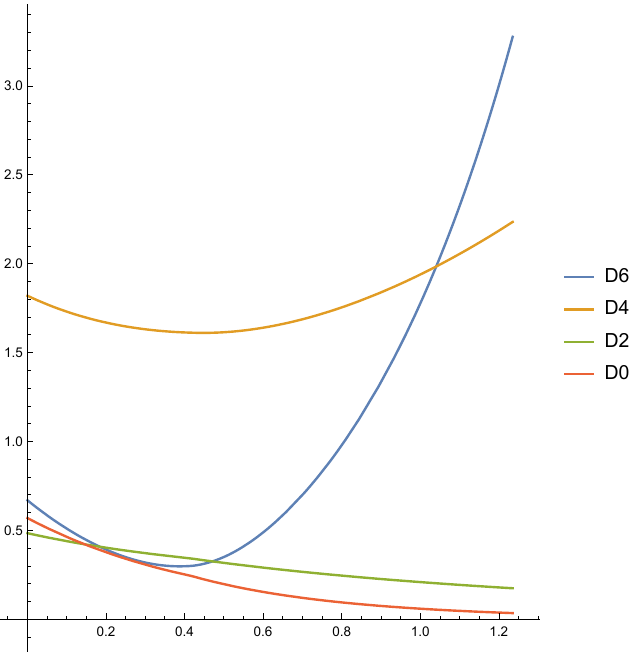}};
        \node at (2.8,-3.2) {$\Delta\phi$};
        \node at (-2.1,3.25) {$M_\textup{BPS}/M_\textup{pl}$};
    \end{tikzpicture}
    \caption{Left: A geodesic (red) in the moduli space of the quintic~\cite{Blumenhagen:2018nts,Klawer:2019czy} starting at the mirror of the Gepner/LG point and approaching large volume. Right: Value of the central charge for states with $D0/D2/D4/D6$ charge along this trajectory. The exponential behavior $\sim\exp(-\Delta\phi)$ is reached for $\Delta\phi\approx 1$. Qualitatively similar behavior was observed for the mass scale of the KK tower in~\cite{Blumenhagen:2018nts}. The existence or stability of a BPS state with these charges along the trajectory is not implied.}
    \label{fig:QuinticModuliSpace}
\end{figure}
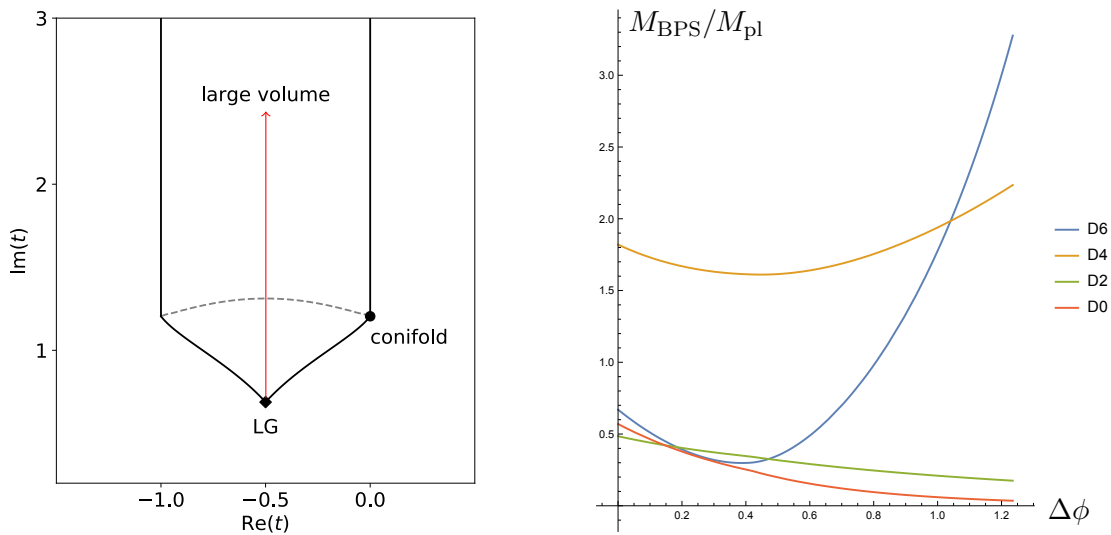

The RDC has been tested in various well-controlled settings with eight supercharges~\cite{Blumenhagen:2018nts,Joshi:2019nzi,Erkinger:2019umg,Brodie:2021ain}. We will focus here on the 4D $\mathcal{N}=2$ vector multiplet moduli space, since in this case the geometry in the interior of moduli space can often be exactly determined using dualities. We will mostly consider type IIA string theory compactified on a Calabi-Yau threefold $X$. In this case the vector multiplet moduli space is spanned by $h^{11}(X)$ Kähler moduli and the metric on moduli space receives an infinite series of world-sheet instanton corrections. Under mirror symmetry, this is mapped to type IIB string theory on the mirror Calabi-Yau $\check{X}$, where now the vector multiplet moduli space is classically exact and equivalent to the $h^{21}(\check{X})$-dimensional moduli space of complex structures of $\check{X}$. This was used in~\cite{Blumenhagen:2018nts} to check the RDC in examples with $h^{11}(X)\in \{1,2,101\}$, with no indication of a failure of the conjecture. An illustrative example is given by type IIA on the quintic hypersurface, see Figure~\ref{fig:QuinticModuliSpace}.
We will revisit and partially extend the results of~\cite{Blumenhagen:2018nts} in light of the duality of type IIA with the heterotic string.

The existence of a heterotic dual imposes the structure of a K3 fibration onto $X$ and in particular implies $h^{11}\geq 2$. We will restrict our attention to smooth, projective Calabi-Yau threefolds over $\mathbb{P}^1$ with smooth generic K3 fiber. Furthermore, we require $h^{11}(X)=2$. These assumptions are very restrictive and represent a small corner in the $\mathcal{N}=2$ landscape, but we will see that they nonetheless provide an interesting challenge for the RDC. Just as described above, will be able to realize a series of moduli spaces $\mathcal{M}_N$ with parametrically increasing diameter of a certain hybrid phase. To be more precise, the volume of the $\mathbb{P}^1$ base is dual to the heterotic dilaton $S_\textup{het}$. In the limit of large base, or equivalently weak heterotic string coupling, the moduli space degenerates into the modular curve $\mathbb{H}/\Gamma_0(N)^+$ and we can compute geodesics and distances exactly using the geometry of the upper half plane\footnote{We would like to point out that attempts to obtain large distances in scalar field space by breaking $SL(2,\mathbb{Z})$ to congruence subroups were also made in~\cite{Hebecker:2017lxm}, although in the $\mathcal{N}=1$ setting.}. The group $\Gamma_0(N)^+$ is composed out of the congruence subgroup $\Gamma_0(N)$ together with the Fricke involution $T\to -1/(NT)$ and can be thought of as the heterotic T-duality group acting on the torus factor of a very special $\textup{K3}\times T^2$ compactification\footnote{These groups (for prime $N$) were famously studied in the context of monstrous moonshine~\cite{Conway:1979qga}.}. In the type IIA frame, the parameter $2N$ is the degree of the K3 fiber, which is polarized by an ample divisor with $L^2=2N$. We find empirically that the length of finite distance geodesics that we can realize in the moduli space $\mathcal{M}_N$ grows like $\sim\log(N)$. Hence, we would require exponentially large $N$ to generate tension with the RDC. In order to construct models with $N>4$, we have to venture outside the realm of toric constructions. We will construct several examples as complete intersections in Grassmann bundles\footnote{The problem is somewhat analogous to constructing genus-zero fibered Calabi-Yau threefolds with an $N$-section, which also requires Grassmannian/Pfaffian constructions beyond $N=4$. Explicit such Calabi-Yau threefolds with $5$-sections are constructed in~\cite{Knapp:2021vkm}. See also~\cite{Kimura:2016crs,Kimura:2019bzv} for related constructions involving Grassmannians.}.

The paper is structured as follows. In section~\ref{sec:RDC_LowDegree} we will revisit the results of~\cite{Blumenhagen:2018nts} for the particular Calabi-Yau threefolds $\mathbb{P}^4_{11226}[12]$ and $\mathbb{P}^4_{11222}[8]$ in light of heterotic / IIA duality. We furthermore discuss several other well-known and explicit examples of toric Calabi-Yau threefold fibered by degree $2N\leq 8$ K3 surfaces in light of the refined distance conjecture. Section~\ref{sec:PolarizedK3Fibrations} deals with the generalization of the setup to arbitrary degree $d=2N$ and presents explicit constructions of $N>4$ as complete intersections in Grassmann bundles over $\mathbb{P}^1$. We conclude by discussing our results and some future directions. Mathematical facts about congruence subgroups, fundamental domains and K3 fibrations can be found in appendices~\ref{app:CongruenceSubgroups},~\ref{app:FunDomains} and~\ref{app:K3SurfacesAndFibrations}. Appendix~\ref{app:HetDuals} contains some speculations about heterotic duals.

\section{Refined Distance Conjecture for Simple K3 Fibrations}
\label{sec:RDC_LowDegree}

Here we will discuss the refined distance conjecture in some explicit K3 fibered Calabi-Yau threefolds with $h^{11}=2$. The discussion will be limited to fibrations by K3 surfaces of degree $2N<8$, in which case the geometry can be described as a complete intersection in some toric ambient space. The main examples that we discuss and their modular properties have been first discussed in~\cite{Klemm:1995tj}.

\subsection{Fibration by \texorpdfstring{$\mathbb{P}^3_{1113}[6]$}{sextic K3} - \texorpdfstring{$SL(2,\bbZ)$}{SL(2)}}
\label{subsec:11226}

Let us start by discussing the canonical and best understood pair of dual type IIA and heterotic compactifications. The type IIA string is compactified on the Calabi-Yau threefold $X$ with Hodge numbers $(h^{11},h^{21})=(2,128)$ given as a toric resolution of the degree 12 hypersurface in weighted projective space $\mathbb{P}^4_{11226}[12]$~\cite{Candelas:1993dm}. This manifold admits a fibration by sextic K3 surfaces $\mathbb{P}^3_{1113}[6]$, which can be made manifest by considering the linear system $x^1=\lambda x^2$, where $x^1,x^2$ are the first two coordinates of the weighted projective space and $\lambda$ parameterizes the $\mathbb{P}^1$ base of the fibration.

The dual is constructed by considering the heterotic string on $\textup{K3}\times T^2$ with the $T^2$ factor at the point $T=U$, where the gauge group enhances to $G=E_8\times E_8\times SU(2)$. Then one considers an instanton backgrounds with instanton numbers $(10,10,4)$ in $G$, which freezes the moduli at $T=U$~\cite{Kachru:1995wm}. One is left with a model with only two vector multiplets, the heterotic torus volume $T$ and four-dimensional dilaton $S$, matching the type IIA spectrum. Non-perturbative checks of this duality were performed in~\cite{Kachru:1995fv}.

The vector multiplet moduli space geometry of type IIA on $\mathbb{P}^4_{11226}[12]$, which is accessible via mirror symmetry, was used to probe the RDC in~\cite{Blumenhagen:2018nts}. The mirror manifold can be constructed as~\cite{Candelas:1993dm}
\begin{equation}
    V\left(x_1^{12}+x_2^{12}+x_3^6+x_4^6+x_5^2-12\psi\,\prod_i x_i-2\phi\, x_1^6 x_2^6\,\right)\Big/\mathbb{Z}_6^2\times\mathbb{Z}_2
\end{equation}
in the same weighted projective space. Under the mirror map, the two complex deformations parameterized by $\psi$ and $\phi$ get mapped to the two Kähler moduli $t^i(\psi,\phi)$ of $\mathbb{P}^4_{11226}[12]$.
\begin{figure}
    \centering
    \begin{tikzpicture}[xscale=1.2,yscale=1.2]
        \node at (-3.3,6) {$\psi$};
        \node at (3,-0.3) {$\phi$};
        \node[align=center,below] at (0,0) {\scriptsize $\phi=1$};
        \node[align=left,below] at (-3,0) {$0$};
        
        \node[align=center,above] at (2.5,1.5) {\scriptsize $|864 \psi^6| \lessgtr |\phi \pm 1|$};
        \draw[dashed] (-3,4.5) -> (-3,5.5);
        \draw[-] (-3,0) -> (-3,4.5);
        
        \draw[->] (-3,5.5) -> (-3,6);
        \draw[dashed] (1.5,0) -> (2.5,0);
        \draw (-3,0) -> (1.5,0);
        
        \draw[->] (2.5,0) -> (3,0);
        \draw[dashed,blue] (0,0) -> (0,6);
        \coordinate (A) at (-3,3);
        \coordinate (B) at (3,1.5);
        \draw[dashed, blue]    (A) to[out=-35,in=180] (B);
        
        \draw[<->] (-2.7,0.3) to (-0.3,0.3);
        \draw[<->] (-2.7,0.4) to (-2.7,2.7);
        \node at (-1.4,0.6) {\footnotesize 0.21};
        \node at (-2.3,2) {\footnotesize 0.32};
        
        \draw[<->] (0.3,0.3) to (0.3,1.5);
        \draw[<->] (2.5,0.3) to (2.5,1.3);
        \draw[<-] (0.3,0.2) to (1,0.2);
        \draw[dashed] (1,0.2) to (2,0.2);
        \draw[->] (2,0.2) to (2.5,0.2);
        \node at (0.7,1) {\footnotesize 0.10};
        \node at (2.9,1) {\footnotesize 0.27};
        \node at (1.5,0.4) {\footnotesize $\infty$};
        
        \draw[<->] (-2.7,3.3) to (-0.3,3.3);
        \draw[<->] (-2.7,5.7) to (-0.3,5.7);
        \draw[<-] (-2.7,3.4) to (-2.7,4.5);
        \draw[dashed] (-2.7,4.5) to (-2.7,5.1);
        \draw[->] (-2.7,5.1) to (-2.7,5.6);
        \node at (-1.5,3.6) {\footnotesize 0.16};
        \node at (-1.5,5.4) {\footnotesize $0$};
        \node at (-2.4,4.5) {\footnotesize $\infty$};
        
        \node[align=center,below] at (1.5,4.5) {\footnotesize LV};
        \node[align=center,below] at (1.5,1.3) {$\mathbb P^1$};
        \node[align=center,below] at (-1.5,1.8) {\footnotesize Landau- \\ \footnotesize{Ginzburg}};
        \node[align=center,below] at (-1.5,4.5) {\footnotesize orbifold};
    \end{tikzpicture}
    \caption{Sketch of the (mirror) moduli space of $\mathbb{P}^4_{11226}[12]$ with indicated finite and infinite diameters of phases~\cite{Blumenhagen:2018nts}.}
    \label{fig:11226ModuliSpace}
\end{figure}
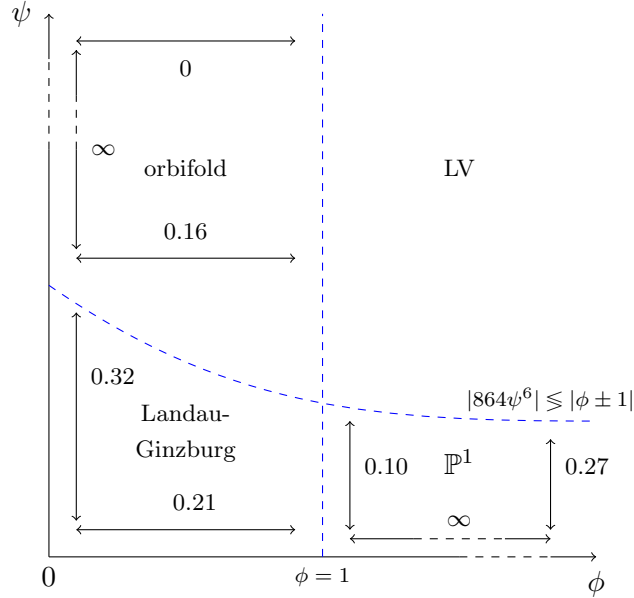

The moduli space in the $(\psi,\phi)$ coordinates is sketched in Figure~\ref{fig:11226ModuliSpace}. It is partioned into four different phases, where the periods assume qualitatively different expansions in terms of the coordinates. In the large volume phase (labeled LV), the moduli space geometry is appropriately described by the classical prepotential supplemented by perturbative and instanton corrections.
\begin{equation}
    \label{eq:GVPrepotential}
    \mathcal{F}=\frac16 k_{ijk}t^i t^j t^k+\frac12 a_{ij}t^i t^j+b_i t^i+\frac12 c+\sum_{\beta>0}\,n_{\beta}^0\,\textup{Li}_3\big(e^{2\pi i t^\beta}\big)\;.
\end{equation}
Here $k_{ijk}$ are the triple intersection numbers, $a_{ij}$ and $b_i\sim J_i\cdot c_2$ are related to the choice of an integral symplectic basis of periods, $c\sim\chi(X)$, and $n_\beta^0$ are the genus zero Gopakumar-Vafa invariants, see for example~\cite{Hosono:1994ax,Hori:2003ic}. The instanton expansion~\eqref{eq:GVPrepotential} breaks down in the other phases and one has to perform an analytic continuation, which is feasible since the periods for this Calabi-Yau have rather explicit expressions in terms of hypergeometric functions. The phases labeled ``orbifold'' and ``$\mathbb{P}^1$''\footnote{The terminology is explained in~\cite{Blumenhagen:2018nts}.} in~\ref{fig:11226ModuliSpace} are hybrid phases with one infinite direction and one finite direction. The ``Landau-Ginzburg'' phase represents the compact interior of the moduli space and is of finite sub-Planckian diameter, in agreement with the RDC.

To explain the behavior in the hybrid phases, let us focus on the $\mathbb{P}^1$-phase. Following~\cite{Klemm:1995tj}, we can define coordinates
\begin{equation}
    \label{eq:11226SWCoordinates}
    x=-\frac{1}{864}\frac{\phi}{\psi^{6}}\;,\qquad y=\frac{1}{\phi^2}\;.
\end{equation}
The periods and mirror map take the schematic form
\begin{equation}
    \label{eq:11226PeriodsMirrorMapP1Phase}
    \begin{gathered}
        \Pi\sim\sum_{a=0}^{1}(\log y)^a P_a\left(x^{-1/3},y\right)\;,\\
        \qquad t^1=Q_1\left(x^{-1/3},y\right)\;,\qquad t^2=\frac{1}{2\pi i}\log(y)+Q_2\left(x^{-1/3},y\right)\;,
    \end{gathered}
\end{equation}
where $P_a,Q_i$ are power series in the indicated variables. Using the explicit expressions for the periods, one can see that for small $\psi$ and large $\phi$ the metric on moduli space asymptotes to
\begin{equation}
    \label{eq:11226MetricP1Phase}
    g_{\alpha\bar{\beta}}=
    \begin{pmatrix}
        \frac{1}{4|\phi|^2(\log|\phi|)^2} & 0\\
        0 & \frac{27.23|\psi|^2}{|\phi|^{2/3}}
    \end{pmatrix}\;.
\end{equation}
We can see immediately that
\begin{equation}
    \label{eq:11226InfiniteDistance}
    \int_{|\phi_0|}^{\infty}d|\phi|\sqrt{g_{\phi\bar{\phi}}}=\infty\;,
\end{equation}
which means that the $\mathbb{P}^1$ hybrid phase has infinite diameter in the $\phi$ direction. On the contrary, the diameter of the $\mathbb{P}^1$ phase is bounded in the $\psi$ direction
\begin{equation}
    \label{eq:11226FiniteDistanceApprox}
    \Delta\phi_c\approx\int_{0}^{|\psi_\textup{max}|}d|\psi|\sqrt{g_{\psi\bar{\psi}}}\Bigg|_{|\phi|\to\infty}
    \approx\sqrt{27.23}\int_{0}^{|\psi_\textup{max}|}d|\psi|\frac{|\psi|}{|\phi|^{1/3}}\approx 0.27\;,
\end{equation}
where $|864\psi_\textup{max}^6|=|\phi|$. After displacing a short sub-Planckian distance, we reach the large volume phase and see a light tower of states. From equation~\eqref{eq:11226PeriodsMirrorMapP1Phase} we see that for finite $\phi$ the periods have a generic polynomial behavior with respect to $\psi$, so we do not expect a universally behaving DC tower either. Hence, the diameter of the finite directions of the hybrid phases poses an interesting challenge for the RDC.

The main drawback of the previous discussion is the perturbative approach to computing the periods, leading to the approximate result of $\Delta\phi_c\approx 0.27$ in equation~\eqref{eq:11226FiniteDistanceApprox} and similar for the other distances in Figure~\ref{fig:11226ModuliSpace}. Although we have an exact expression for the periods in terms of hypergeometric functions~\cite{Blumenhagen:2018nts}, actual computations on a computer require series expansions of these. Even for $\phi\to\infty$, the value of $0.27$ will receive corrections from expanding the integrand as a power series in $\psi$. It turns out that the convergence of the series expansion is rather slow. The goal of the remainder of this section is to explain the approximate result and give an exact expression.

The first step is to realize that the fact that the Calabi-Yau $X$ has a K3 fibration leads to constraints on the intersection numbers. In the limit $\phi\to\infty$ / $y\to 0$, we have $v^2\equiv\textup{Im} t^2\to \infty$. Here $v^2$ is the volume modulus of the $\mathbb{P}^1$ base of the fibration. At large $v^2$, the variation of the fibers over the base is approximately constant and the volume of X factorizes $\mathcal{V}(X)\sim v^2\cdot (v^1)^2$. Correspondingly, any instanton corrections to the prepotential are subleading and it asymptotes to
\begin{equation}
    \label{eq:K3FibrationAsymptoticModuliMetric}
    \mathcal{F}\sim t^2\cdot (t^1)^2\qquad\Rightarrow\qquad g_{\alpha\bar{\beta}}
    =\begin{pmatrix}
        \frac{1}{2 (\textup{Im}t^1)^2}& 0\\
        0& \frac{1}{4 (\textup{Im}t^2)^2}\\
    \end{pmatrix}\;,
\end{equation}
where we now display the metric components in type IIA variables. Restricting to any of the two Kähler moduli, we see that the local moduli space geometry is just the hyperbolic geometry of the upper half-plane $\mathcal{H}$. Since geodesics in the upper half-plane are just semi-circles with center on the real line, the problem of computing distances locally reduces to simple trigonometric integrals. The only complication that we have to take care of are global identifications, which include the B-field shifts $t^i\to t^i+1$.

In order to obtain the actual physical domain of the modulus $t^1$, it is instructive to consider the heterotic dual. It is expected that as $t^2\to\infty$, the moduli are identified as $t^2=S$ and $t^1=T$, such that the large base limit coincides with the heterotic weak copling limit. From this map and from the previous description of the heterotic compactification, it is clear that the B-field shifts acting on $t^1$ are completed into a full $SL(2,\mathbb{Z})$ group of identifications. The effective moduli space for $t^1$ thus degenerates into the fundamental domain $\mathcal{H}/SL(2,\mathbb{Z})$ as $t^2\to\infty$.

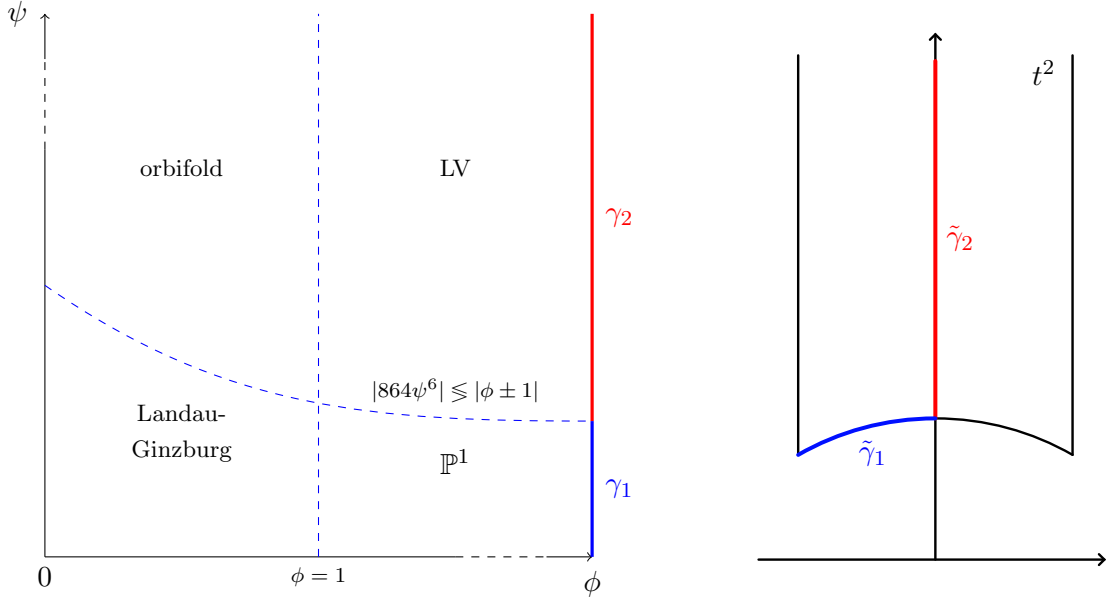
\begin{figure}
    \centering
    \raisebox{-0.5\height}{
    \begin{tikzpicture}[xscale=1.2,yscale=1.2]
        \node at (-3.3,6) {$\psi$};
        \node at (3,-0.3) {$\phi$};
        \node[align=center,below] at (0,0) {\scriptsize $\phi=1$};
        \node[align=left,below] at (-3,0) {$0$};
        
        \node[align=center,above] at (1.5,1.6) {\scriptsize $|864 \psi^6| \lessgtr |\phi \pm 1|$};
        \draw[dashed] (-3,4.5) -> (-3,5.5);
        \draw[-] (-3,0) -> (-3,4.5);
        
        \draw[->] (-3,5.5) -> (-3,6);
        \draw[dashed] (1.5,0) -> (2.5,0);
        \draw (-3,0) -> (1.5,0);
        
        \draw[->] (2.5,0) -> (3,0);
        \draw[dashed,blue] (0,0) -> (0,6);
        \coordinate (A) at (-3,3);
        \coordinate (B) at (3,1.5);
        \draw[dashed, blue]    (A) to[out=-35,in=180] (B);

        \node[align=center,below] at (1.5,4.5) {\footnotesize LV};
        \node[align=center,below] at (1.5,1.3) {$\mathbb P^1$};
        \node[align=center,below] at (-1.5,1.8) {\footnotesize Landau- \\ \footnotesize{Ginzburg}};
        \node[align=center,below] at (-1.5,4.5) {\footnotesize orbifold};

        \draw[red,very thick,-] (3,1.5) to (3,6);
        \node[red] at (3.3,3.75) {$\gamma_2$};
        \draw[blue,very thick,-] (3,0) to (3,1.5);
        \node[blue] at (3.3,0.75) {$\gamma_1$};
    \end{tikzpicture}
    }
    \hfill
    \raisebox{-0.5\height}{
    \begin{tikzpicture}
        \node at (0,0) {\includegraphics[width=4.7cm]{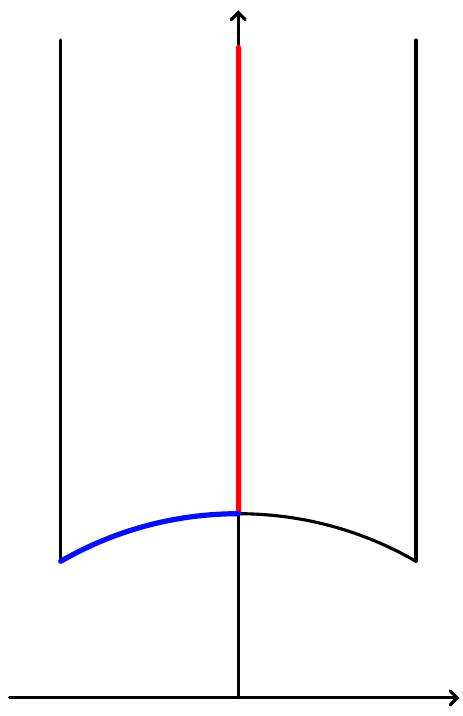}};
        \node at (1.5,3) {$t^2$};
        \node[red] at (0.4,0.8) {$\tilde{\gamma}_2$};
        \node[blue] at (-0.75,-2.05) {$\tilde{\gamma}_1$};
    \end{tikzpicture}
    }
    \caption{Mapping of type II and heterotic moduli spaces in the weak coupling limit.}
    \label{fig:11226AsymptoticModuliSpace}
\end{figure}

Given this information, let us now reproduce the result~\eqref{eq:11226FiniteDistanceApprox} for the asymptotic diameter of the $\mathbb{P}^1$ hybrid phase. In~\cite{Kachru:1995fv} the map between the type IIB and heterotic moduli was determined to be
\begin{equation}
    \label{eq:11226HeteroticCoordinateRelation}
    x=\frac{1728}{j(T)}+\dots\;,\qquad y=e^{-S}+\dots\;,
\end{equation}
where we use again the coordinates~\eqref{eq:11226SWCoordinates} and corrections away from the weak coupling limit are neglected. The fact that the mirror map of $X$ is related to the $j$ function was first pointed out by~\cite{Candelas:1993dm}. In the $(x,y)$ coordinates, the conifold component of the discriminant locus of the IIB compactification on $\check{X}$ takes the form
\begin{equation}
    \label{eq:11226Conifold}
    \Delta_\textup{coni}=(1-x)^2-x^2 y\;.
\end{equation}
As $y\to 0$, the developing double singularity at $x=1$ is mapped via~\eqref{eq:11226HeteroticCoordinateRelation} to the $\mathbb{Z}_2$ orbifold point $T=i$ of the fundamental domain. Furthermore, it is easy to see from the explicit form of the map~\eqref{eq:11226HeteroticCoordinateRelation} that the point $(x,y)=(0,\infty)$ is mapped to $(T,S)=(t^1,t^2)=(-\bar{\rho},\infty)$, where $\rho=e^{i \pi/3}$ is the $\mathbb{Z}_3$ orbifold point of the fundamental domain. As a result, the asymptotic trajectory $\gamma_1$ in the $\psi$ direction along which the distance $\Delta\phi_c\approx 0.27$ was computed reveals itself as the geodesic arc $\tilde{\gamma}_1$ joining the two points $T=i$ and $T=-\bar{\rho}$. The map is depicted in Figure~\ref{fig:11226AsymptoticModuliSpace}.

With this information, we can now parametrize $\tilde{\gamma}_2$ by $T=\exp(i\varphi)$ and evaluate the diameter $\Delta\phi_c$ exactly as
\begin{equation}
    \label{eq:11226DeltaPhiExact}
    \Delta\phi_c=\int_{\pi/2}^{2\pi/3}d\varphi \sqrt{g_{T\bar{T}}\frac{d T(\varphi)}{d\varphi}\frac{d \bar{T}(\varphi)}{d\varphi}}=\frac{1}{\sqrt{2}}\int_{\pi/2}^{2\pi/3} \frac{d\varphi}{\sin\varphi}=\frac{\log(3)}{2\sqrt{2}}\approx 0.39\;.
\end{equation}
We can recover our approximate result as follows. To leading order around $T=-\bar{\rho}$, the $j$ function can be expanded as~\cite{BayerTravesa:2007}
\begin{equation}
    \label{eq:JFunctionExpansionLeadingOrder}
    j(T)=1728\left(k\frac{T+\bar{\rho}}{\rho-\bar{\rho}}\right)^3+\dots\;,\qquad k=e^{i\pi/3}\frac{12}{\sqrt{\pi}}\frac{\Gamma(\tfrac76)^2}{\Gamma(\tfrac56)}\;.
\end{equation}
By inverting this and using the relations~\eqref{eq:11226HeteroticCoordinateRelation}, we can compute
\begin{equation}
    g_{\psi\bar{\psi}}=\left|\frac{\partial T}{\partial\psi}\right|^2g_{T\bar{T}}=\left|\frac{\partial T}{\partial\psi}\right|^2\frac{1}{2(\textup{Im}T)^2}\approx\frac{5184\pi\cdot 2^{1/3}\cdot\Gamma(\tfrac56)^2}{\Gamma(\tfrac16)^2}\frac{|\psi|^2}{|\phi|^{2/3}}\approx 27.23\frac{|\psi|^2}{|\phi|^{2/3}}\;,
\end{equation}
which recovers~\eqref{eq:11226MetricP1Phase}. Similarly, one can compute the leading term of $g_{\phi\bar{\phi}}$. Finally, we note that along the geodesic arc $\tilde{\gamma}_1$ the $j$ function is excellently approximated by
\begin{equation}
    \label{eq:JFunctionArcApproximation}
    j(T=e^{i\varphi})\approx 1728\left(\tfrac12-\tfrac12\cos(6\varphi)\right)^{3/2}\;,
\end{equation}
which can be used to interpolate between the leading~\eqref{eq:11226FiniteDistanceApprox} and exact~\eqref{eq:11226DeltaPhiExact} results. We find that the resulting series expansion of the metric converges very slowly, which explains the discrepancy\footnote{Around 100 terms are required to achieve an accuracy of 10\%.}.

The above discussion should carry over to other K3 fibered Calabi-Yau manifolds with $h^{11}=2$ with generic fiber a sextic K3 surface $\mathbb{P}^3_{1113}[6]$. These may differ from the manifold $\mathbb{P}^4_{11226}[12]$ via the structure of the fibration, which will be reflected in data such as the triple intersection numbers or the second Chern class. In the large base limit these details should not be important and it is expected that the moduli space always degenerates into the $SL(2,\mathbb{Z})$ fundamental domain. We will discuss this base/fiber decoupling more extensively in the next section.

\subsection{Fibration by \texorpdfstring{$\mathbb{P}^3[4]$}{quartic K3} - \texorpdfstring{$\Gamma_0(2)^+$}{Gamma0+(2)}}
\label{subsec:11222}

As a first example of a heterotic/IIA dual pair that realizes a non-trivial modular curve in its moduli space we will consider the resolution $X$ of the degree eight hypersurface $\mathbb{P}_{11222}^4[8]$ with Hodge numbers $(h^{11},h^{21})=(2,86)$. The vector multiplet moduli space of this Calabi-Yau has been determined using mirror symmetry in~\cite{Candelas:1993dm}. The type IIA geometry is more conveniently described in terms of an equivalent CICY construction~\cite{Schimmrigk:1989tz}
\begin{equation}
    X=\mathbb{P}_{11222}^4[8]\cong
    \left(
    \begin{array}{c|cc}
        \mathbb{P}^1 & 0 & 2\\
        \mathbb{P}^4 & 4 & 1\\
    \end{array}
    \right)\;,
\end{equation}
which is manifestly fibered by quartic K3 surfaces $\mathbb{P}^3[4]$. In terms of the Kähler cone generators of the ambient space, the geometry is characterized by the numerical invariants
\begin{equation}
    \label{eq:11222Intersections}
    J_2^3=8\;,\qquad J_2^2\cdot J_1=4\;,\qquad c_2\cdot J_1=24\;,\qquad c_2\cdot J_2=56\;.
\end{equation}
The fact that $c_2\cdot J_1=24=\chi(K3)$ confirms that $J_1=\left[\mathbb{P}^4[4,1]\right]$ is the class of the K3 fiber. 

The mirror $\check{X}$ has a simple description as a (Greene-Plesser) quotient of the original manifold $X$~\cite{Candelas:1993dm}
\begin{equation}
    \label{eq:11222MirrorEquation}
    V\left(x_0^8+x_1^8+x_2^4+x_3^4+x_4^4-8\psi\prod_{i} x_i-2\phi\, x_1^4x_2^4\right)\Big/\mathbb{Z}_4^3\;.
\end{equation}
In the $\mathbb{P}^4_{11222}[8]$ description of $X$, a linear system of fibral K3 divisors is again given by $x_0=\lambda x_1$, where $\lambda\in\mathbb{P}^1$~\cite{Candelas:1993dm,Klemm:1995tj}. On this locus, the two-parameter family~\eqref{eq:11222MirrorEquation} degenerates to the one-parameter Dwork family of K3 surfaces
\begin{equation}
\label{eq:MirrorQuarticK3}
    V\left(y_0^4+y_1^4+y_2^4+y_3^4-4\tilde{\psi}\,y_0y_1y_2y_3\right)\big/\mathbb{Z}_4^2\;.
\end{equation}

The moduli space of $X$ is qualitatively similar to the one of $\mathbb{P}^4_{11226}[12]$~\cite{Candelas:1993dm,Blumenhagen:2018nts}, so we will not reproduce it here. As in the last section, it is useful to introduce also the coordinates
\begin{equation}
    \label{eq:11222SWCoordinates}
    x=-\frac{1}{8}\frac{\phi}{\psi^4}\;,\qquad y=\frac{1}{\phi^2}\;,
\end{equation}
in which the conifold locus takes the same form as~\eqref{eq:11226Conifold}. Again, we have a $\mathbb{P}^1$ hybrid phase, in which the metric asymptotes to
\begin{equation}
    \label{eq:11222MetricP1Phase}
    g_{\alpha\bar{\beta}}=
    \begin{pmatrix}
        \frac{1}{4|\phi|^2(\log|\phi|)^2} & 0\\
        0 & \frac{0.5905}{\sqrt{|\phi|}}
    \end{pmatrix}\;.
\end{equation}
The asymptotic diameter in the $y\to 0$ limit was computed in~\cite{Blumenhagen:2018nts} as
\begin{equation}
    \label{eq:11222FiniteDistanceApprox}
    \Delta\phi_c\approx
    \int_{0}^{|\psi_\textup{max}|}d|\psi|\sqrt{g_{\psi\bar{\psi}}}\Bigg|_{|\phi|\to\infty}
    \approx \sqrt{0.5905}\int_{0}^{|\psi_\textup{max}|}d|\psi|\frac{1}{|\phi|^{1/4}}
    \approx 0.46\;,
\end{equation}
where $|8\psi_\textup{max}^4|=|\phi|$.

We would like to reproduce this result and complete it into an exact expression, in an analogous manner to the previous section. The asymptotic moduli space metric in type IIA coordinates is of course the same as~\eqref{eq:K3FibrationAsymptoticModuliMetric}. The duality group acting on the upper half-plane was determined in~\cite{Lian:1995js,Klemm:1995tj} to be the congruence subgroup $\Gamma_0(2)^+$\footnote{While the group $\Gamma_0(2)$ is an honest subgroup of $SL(2,\mathbb{Z})$, the group $\Gamma_0(2)^+$ corresponds to its normalizer in $SL(2,\mathbb{R})$. Thus, it is not a subgroup of $SL(2,\mathbb{Z})$. We will nonetheless use the term ``congruence subgroup'' also for (partial) normalizers of honest congruence subgroups.}. The group $\Gamma_0(N)^+$ consists of matrices of the form
\begin{equation}
    \begin{pmatrix}
        a&b\\c&d\\
    \end{pmatrix}
    \in SL(2,\mathbb{Z})\;, \qquad c\equiv 0\,\textup{mod}\,N\;,
\end{equation}
which form the congruence subgroup $\Gamma_0(N)$, together with the Fricke involution $T\to-1/(NT)$. The modularity constraint on $c$ removes the usual T-duality $T\to -1/T$ from the group, thus adding an additional cusp at $T=0$. The Fricke involution acts as a generalized T-duality and folds the cusp at $T=0$ back to the cusp at $i\infty$. The self-dual radius is now reduced to $T_\textup{sd}=i/\sqrt{N}$. The resulting modular curve $\mathcal{H}/\Gamma_0(2)^+$ is depicted in Figure~\ref{fig:gamma2fundomain}. More details on congruence subgroups and modular curves can be found in Appendices~\ref{app:CongruenceSubgroups} and~\ref{app:FunDomains}.
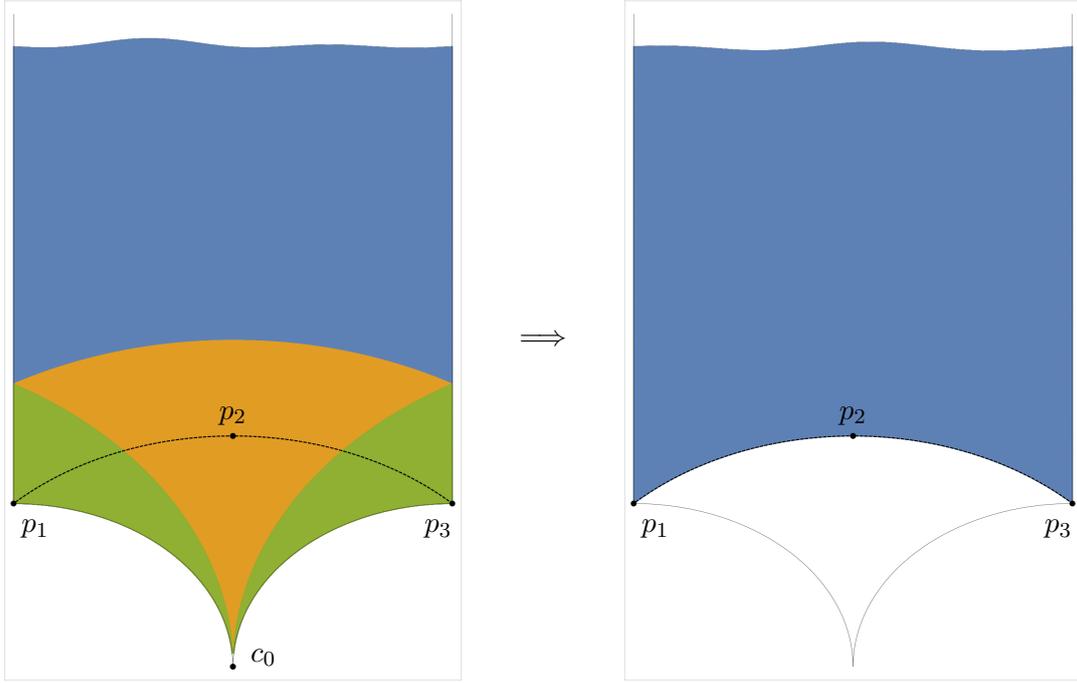
\begin{figure}
    \centering
    \raisebox{-0.5\height}{
    \begin{tikzpicture}
        \node at (0,0) {\includegraphics[width=6cm]{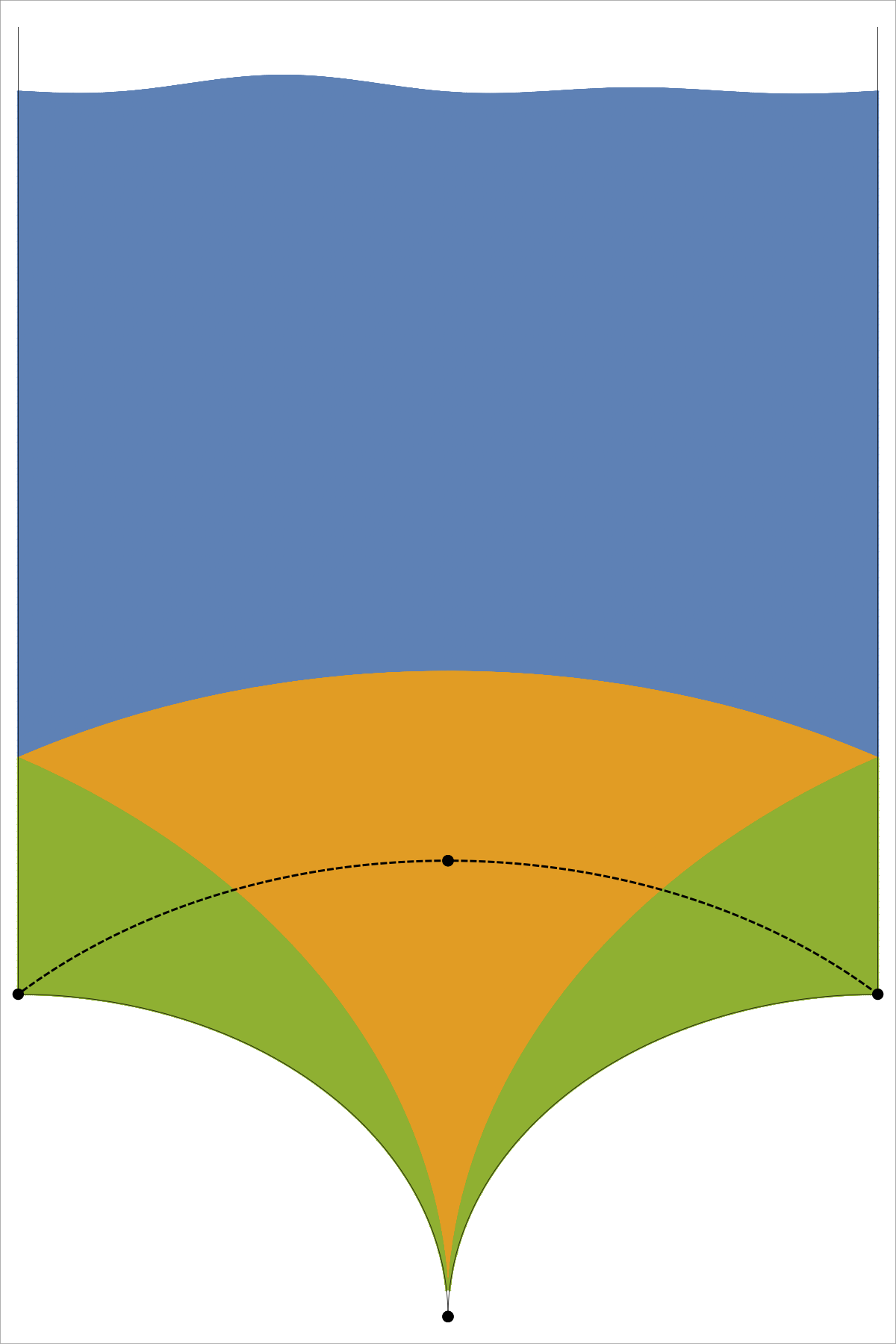}};
        \node at (-2.6,-2.5) {$p_1$};
        \node at (0,-1) {$p_2$};
        \node at (2.7,-2.5) {$p_3$};
        \node at (0.4,-4.2) {$c_0$};
    \end{tikzpicture}
    }
    \quad
    \raisebox{-0.5\height}{$\Longrightarrow$}
    \quad
    \raisebox{-0.5\height}{
    \begin{tikzpicture}
        \node at (0,0) {\includegraphics[width=6cm]{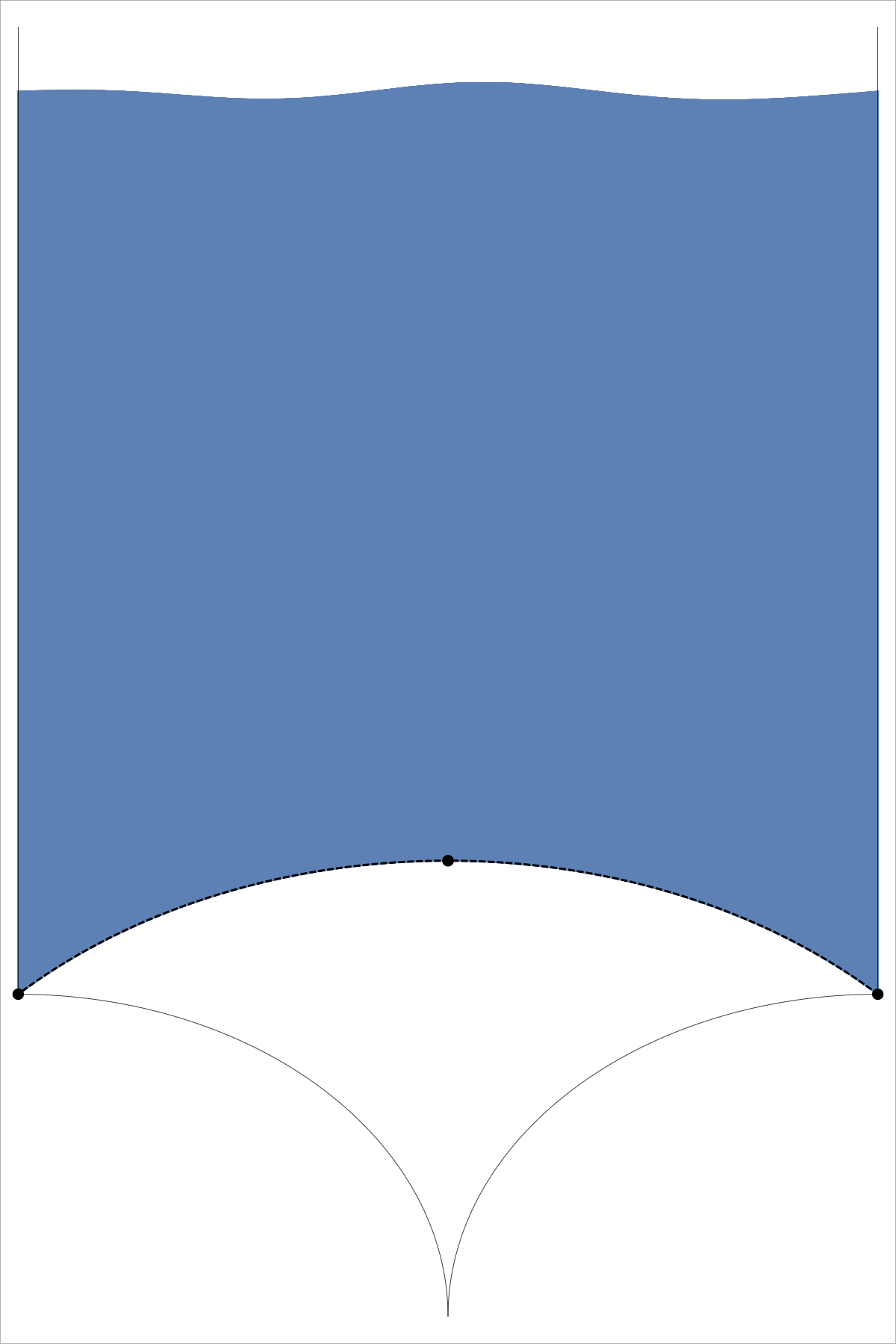}};
        \node at (-2.6,-2.5) {$p_1$};
        \node at (0,-1) {$p_2$};
        \node at (2.7,-2.5) {$p_3$};
    \end{tikzpicture}
    }
    \caption{Left: A fundamental domain $X_0(2)$ for $\Gamma_0(2)$ that is symmetric under the Fricke involution $F_2$. Right: A fundamental domain for $\Gamma_0(2)^+$ is obtained as $X_0(2)/F_2$.\\The marked points are the following: $p_3=e^{i\pi/4}/\sqrt{2}$ is an orbifold point of order $4$ and $p_1=-\bar{p}_3$. $p_2=i/\sqrt{2}$ is the fixed point of $F_2$ and $c_0=0$ is the second cusp of $X_0(2)$.}
    \label{fig:gamma2fundomain}
\end{figure}

Although, to the best knowledge of the author, a precise heterotic dual has not been identified in the literature\footnote{See appendix~\ref{app:HetDuals} for some speculations.}, we can write down an analogous map between the type IIB complex structure coordinates and the would-be heterotic coordinates $T=t^1$ and $S=t^2$ as in~\eqref{eq:11226HeteroticCoordinateRelation}
\begin{equation}
    \label{eq:11222HeteroticCoordinateRelation}
    x=\frac{256}{j_2^+(T)}+\dots\;,\qquad y=e^{-S}+\dots\;.
\end{equation}
The so-called Hauptmodul $j_2^+$ plays the same role for $X_0(2)^+\equiv\mathcal{H}/\Gamma_0(2)^+$ as the $j$ function plays for the $SL(2,\mathbb{Z})$ fundamental domain. It is a generator of the function field of $X_0(2)^+$. An explicit expression is given by~\cite{Harnad:1998hh}
\begin{equation}
    j_2^+(T)=2^4
    \frac
    {\left(\theta_3(q_T)^4   +\theta_4(q_T)^4\right)^4}
    {\,\theta_2(q_T)^8\theta_3(q_T)^4\theta_4(q_T)^4}\;,
\end{equation}
where $\theta_i$ are the Jacobi theta functions and $q_T=\exp(i\pi T)$ is the nome of $T$.

Just as in section~\ref{subsec:11226}, we can now see that the asymptotic trajectory in the $\psi$ direction is identified with the geodesic arc that connects the orbifold points $p_1$ and $p_2$ in Figure~\ref{fig:gamma2fundomain}. The resulting exact expression for the distance between these two points is
\begin{equation}
    \label{eq:11222DeltaPhiExact}
    \Delta\phi_c=\int_{\pi/2}^{3\pi/4}d\varphi \sqrt{g_{T\bar{T}}\frac{d T(\varphi)}{d\varphi}\frac{d \bar{T}(\varphi)}{d\varphi}}=\frac{1}{\sqrt{2}}\int_{\pi/2}^{3\pi/4} \frac{d\varphi}{\sin\varphi}=\frac{\log(\cot(\tfrac{\pi}{8}))}{\sqrt{2}}\approx 0.62\;.
\end{equation}

It is again possible to expand the Hauptmodul $j_2^+$ to leading order around $p_1$\cite{BayerTravesa:2007}
\begin{equation}
    \label{eq:J2PlusExpansionLeadingOrder}
    j_2^+(T)=256\left(k\frac{T-T_0}{T_0+\overline{T_0}}\right)^4+\dots\;,\qquad k=4\sqrt{2}\frac{\Gamma(\tfrac54)^2}{\Gamma(\tfrac34)^2}\;, \qquad T_0=\tfrac{1}{\sqrt{2}}e^{3\pi i/4}\;,
\end{equation}
in order to recover the approximate result~\eqref{eq:11222FiniteDistanceApprox}
\begin{equation}
    g_{\psi\bar{\psi}}=\left|\frac{\partial T}{\partial\psi}\right|^2g_{T\bar{T}}=\left|\frac{\partial T}{\partial\psi}\right|^2\frac{1}{2(\textup{Im}T)^2}\approx\frac{32\sqrt{2}\Gamma(\tfrac34)^4}{\Gamma(\tfrac14)^4}\frac{1}{\sqrt{|\phi|}}\approx \frac{0.5905}{\sqrt{|\phi|}}\;,
\end{equation}
and similarly for $g_{\psi\bar{\psi}}$. An analogue of the approximation~\eqref{eq:JFunctionArcApproximation} is
\begin{equation}
    j_2^+(T=\tfrac{1}{\sqrt{2}}e^{i\varphi})\approx 256\left(\tfrac12+\tfrac12\cos(4\varphi)\right)^2\;.
\end{equation}

Before moving on to the next example, we would like to stress that the distance $\Delta\phi_c$ is a property of the fiber rather than that of the whole Calabi-Yau. This can be seen explicitly by looking at another K3 fibered Calabi-Yau with the same fiber but different twist. By scanning the CICY list of~\cite{Anderson:2017aux}, we find the following examples
\begin{equation}
    \begin{aligned}
    \left(
    \begin{array}{c|ccc}
        \mathbb{P}^1 & 2\\
        \mathbb{P}^3 & 4\\
    \end{array}
    \right)\;&:\quad
    J_2^3=2\;,\quad J_2^2\cdot J_1=4\;,\quad
    c_2\cdot J_i=(24,44)\;,\quad\chi=-168
    \\
    \left(
    \begin{array}{c|ccc}
        \mathbb{P}^1 & 1 & 1\\
        \mathbb{P}^4 & 4 & 1\\
    \end{array}
    \right)\;&:\quad
    J_2^3=5\;,\quad J_2^2\cdot J_1=4\;,\quad
    c_2\cdot J_i=(24,50)\;,\quad\chi=-168
    \end{aligned}\;.
\end{equation}
We can compute the Picard-Fuchs system for these examples using standard methods~\cite{Hosono:1994ax}. In the large base limit, in each case one finds that the system of differential equations degenerates into the PF operator of the fiber
\begin{equation}
    \mathcal{D}=\theta^3-4x(4\theta+3)(4\theta+2)(4\theta+1)\;,\qquad \theta=x\frac{d}{dx}\;,
\end{equation}
which is known to have solutions associated with $\Gamma_0(2)^+$~\cite{Lian:1995js,Klemm:1995tj}. We expect that many more inequivalent fibrations can be constructed using full-blown toric methods\footnote{Several examples can be found in~\cite{Klemm:2004km}.}.

\subsection{Fibration by \texorpdfstring{$\mathbb{P}^4[2,3]$}{codimension 2 K3} - \texorpdfstring{$\Gamma_0(3)^+$}{Gamma0+(3)}}
\label{subsec:112222}

The previous two examples have shown that in the large base limit the diameter $\Delta\phi_c$ of the $\mathbb{P}^1$ hybrid phase is dictated by the duality group acting on the Kähler modulus of the K3 fiber, which was either $SL(2,\mathbb{Z})$ or the congruence subgroup $\Gamma_0(N)^+$. Compactifications that feature $\Gamma_0(N)^+$ duality groups in this way have been constructed also for $N=3,4$. For $N=3$, a possible Calabi-Yau realization is given by~\cite{Klemm:1995tj}
\begin{equation}
    X=\mathbb{P}_{1^2 2^4}^5[4,6]\cong
    \left(
    \begin{array}{c|ccc}
        \mathbb{P}^1 & 0 & 0 & 2\\
        \mathbb{P}^5 & 3 & 2 & 1\\
    \end{array}
    \right)\;,
\end{equation}
with Hodge numbers $(h^{11},h^{21})=(2,68)$ and intersection data
\begin{equation}
    J_2^3=12\;,\qquad J_2^2\cdot J_1=6\;,\qquad c_2\cdot J_1=24\;,\qquad c_2\cdot J_2=60\;.
\end{equation}
The manifold $X$ clearly admits a fibration by the K3 surface $\mathbb{P}^4[2,3]$. Other ways to fiber this K3 over $\mathbb{P}^1$ can be found in the CICY list of~\cite{Anderson:2017aux}
\begin{equation}
    \begin{aligned}
        \left(
        \begin{array}{c|cc}
            \mathbb{P}^1 & 0 & 2\\
            \mathbb{P}^4 & 2 & 3\\
        \end{array}
        \right)\;&:\quad
        J_2^3=4\;,\quad J_2^2\cdot J_1=6\;,\quad
        c_2\cdot J_i=(24,52)\;,\quad
        \chi=-148
        \\
        \left(
        \begin{array}{c|cc}
            \mathbb{P}^1 & 1 & 1\\
            \mathbb{P}^4 & 2 & 3\\
        \end{array}
        \right)\;&:\quad
        J_2^3=5\;,\quad J_2^2\cdot J_1=6\;,\quad
        c_2\cdot J_i=(24,50)\;,\quad
        \chi=-128
        \\
        \left(
        \begin{array}{c|cc}
            \mathbb{P}^1 & 2 & 0\\
            \mathbb{P}^4 & 2 & 3\\
        \end{array}
        \right)\;&:\quad
        J_2^3=6\;,\quad J_2^2\cdot J_1=6\;,\quad
        c_2\cdot J_i=(24,72)\;,\quad
        \chi=-108
        \\
        \left(
        \begin{array}{c|ccc}
            \mathbb{P}^1 & 0 & 1 & 1\\
            \mathbb{P}^4 & 2 & 3 & 1\\
        \end{array}
        \right)\;&:\quad
        J_2^3=8\;,\quad J_2^2\cdot J_1=6\;,\quad
        c_2\cdot J_i=(24,56)\;,\quad
        \chi=-140
        \\
        \left(
        \begin{array}{c|ccc}
            \mathbb{P}^1 & 1 & 0 & 1\\
            \mathbb{P}^4 & 2 & 3 & 1\\
        \end{array}
        \right)\;&:\quad
        J_2^3=9\;,\quad J_2^2\cdot J_1=6\;,\quad
        c_2\cdot J_i=(24,54)\;,\quad
        \chi=-120
        \end{aligned}\;.
\end{equation}

\begin{figure}
    \centering
    \raisebox{-0.5\height}{
    \begin{tikzpicture}
        \node at (0,0) {\includegraphics[width=6cm]{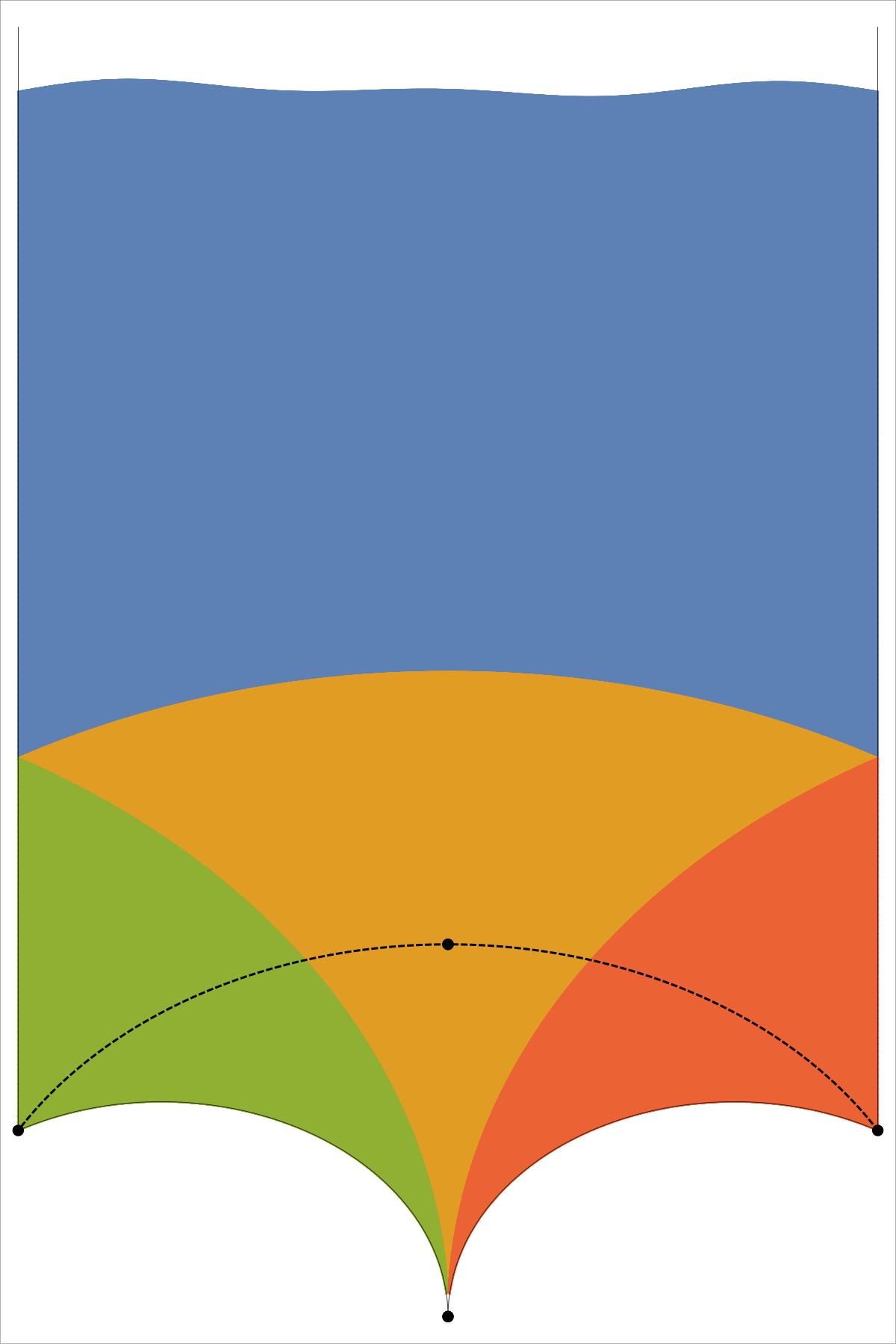}};
        \node at (-2.6,-3.3) {$p_4$};
        \node at (0,-1.6) {$p_5$};
        \node at (2.7,-3.3) {$p_6$};
        \node at (0.4,-4.2) {$c_0$};
    \end{tikzpicture}
    }
    \quad
    \raisebox{-0.5\height}{$\Longrightarrow$}
    \quad
    \raisebox{-0.5\height}{
    \begin{tikzpicture}
        \node at (0,0) {\includegraphics[width=6cm]{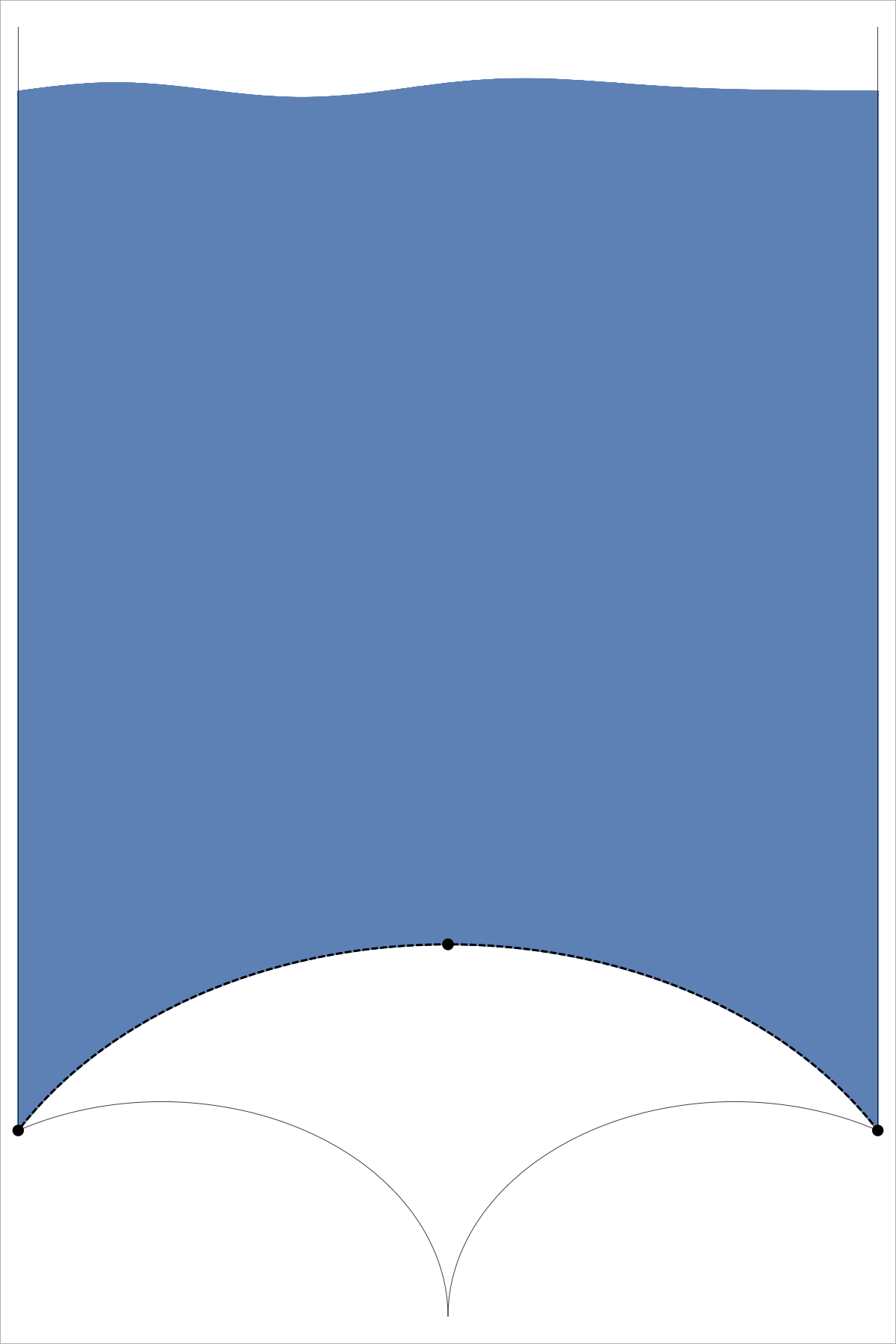}};
        \node at (-2.6,-3.3) {$p_4$};
        \node at (0,-1.6) {$p_5$};
        \node at (2.7,-3.3) {$p_6$};
    \end{tikzpicture}
    }
    \caption{Left: A fundamental domain $X_0(3)$ for $\Gamma_0(3)$ that is symmetric under the Fricke involution $F_3$. Right: A fundamental domain for $\Gamma_0(3)^+$ is obtained as $X_0(3)/F_3$.\\The marked points are the following: $p_6=e^{i\pi/6}/\sqrt{3}$ is an orbifold point of order $6$ and $p_4=-\bar{p}_6$. $p_5=i/\sqrt{3}$ is the fixed point of $F_3$ and $c_0=0$ is the second cusp of $X_0(3)$.}
    \label{fig:gamma3fundomain}
\end{figure}
The resulting modular curve $X_0(3)^+=\mathcal{H}/\Gamma_0(3)^+$ is depicted in Figure~\ref{fig:gamma3fundomain}. We see that it shares the qualitative properties of the previous two examples. The Hauptmodul for $X_0(3)^+$ is given by~\cite{Harnad:1998hh}
\begin{equation}
    j_3^+(T)=\frac
{\left(\eta(T)^{12}+27\eta(3T)^{12}\right)^2}
{\eta(T)^{12}\eta(3T)^{12}}\;,
\end{equation}
where $\eta(T)$ is the Dedekind eta function.

The moduli space of $\mathbb{P}^5_{1^22^4}[4,6]$ has the same structure as before. We can choose coordinates $(x,y)$ on the complex structure moduli space of $\check{X}$, such that the conifold part of the discriminant locus is again given by~\ref{eq:11226Conifold}. The map to heterotic variables is then given by
\begin{equation}
    \label{eq:112222HeteroticCoordinateRelation}
    x=\frac{108}{j_3^+(T)}+\dots\;,\qquad y=e^{-S}+\dots\;.
\end{equation}

The exact asymptotic diameter of the $\mathbb{P}^1$ hybrid phase is obtained by integrating along the geodesic arc connecting the points $p_4$ and $p_5$ in Figure~\ref{fig:gamma3fundomain}. We compute
\begin{equation}
    \label{eq:112222DeltaPhiExact}
    \Delta\phi_c=\int_{\pi/2}^{5\pi/6}d\varphi \sqrt{g_{T\bar{T}}\frac{d T(\varphi)}{d\varphi}\frac{d \bar{T}(\varphi)}{d\varphi}}=\frac{1}{\sqrt{2}}\int_{\pi/2}^{5\pi/6} \frac{d\varphi}{\sin\varphi}=\sqrt{2}\coth^{-1}\left(\sqrt{3}\right)\approx 0.93\;.
\end{equation}
We note that approximations similar to~\eqref{eq:JFunctionExpansionLeadingOrder} and~\eqref{eq:JFunctionArcApproximation} can be found for $j_3^+$ and may in principle be used to match a perturbative mirror symmetry computation of $\Delta\phi_c$.

\subsection{Fibration by \texorpdfstring{$\mathbb{P}^5[2,2,2]$}{codimension 3 K3} - \texorpdfstring{$\Gamma_0(4)^+$}{Gamma0+(4)}}
\label{subsec:1122222}

The final toric example realizes the modular curve $\mathcal{H}/\Gamma_0(4)^+$. It is given as a codimension three complete intersection~\cite{Klemm:1995tj}
\begin{equation}
    X=\mathbb{P}_{1^2 2^5}^6[4,4,4]\cong
    \left(
    \begin{array}{c|cccc}
        \mathbb{P}^1 & 0 & 0 & 0 & 2\\
        \mathbb{P}^6 & 2 & 2 & 2 & 1\\
    \end{array}
    \right)\;,
\end{equation}
with Hodge numbers $(h^{11},h^{21})=(2,58)$ and numerical data
\begin{equation}
    J_2^3=16\;,\qquad J_2^2\cdot J_1=8\;,\qquad c_2\cdot J_1=24\;,\qquad c_2\cdot J_2=64\;.
\end{equation}
This manifold admits a fibration by $\mathbb{P}^5[2,2,2]$. Other complete intersection Calabi-Yau threefolds with the same fiber are
\begin{equation}
    \begin{aligned}
        \left(
        \begin{array}{c|ccc}
            \mathbb{P}^1 & 0 & 0 & 2\\
            \mathbb{P}^4 & 2 & 2 & 2\\
        \end{array}
        \right)\;&:\quad
        J_2^3=8\;,\quad J_2^2\cdot J_1=8\;,\quad
        c_2\cdot J_i=(24,56)\;,\quad
        \chi=-112
        \\
        \left(
        \begin{array}{c|cccc}
            \mathbb{P}^1 & 0 & 0 & 1 & 1\\
            \mathbb{P}^4 & 2 & 2 & 2 & 1\\
        \end{array}
        \right)\;&:\quad
        J_2^3=12\;,\quad J_2^2\cdot J_1=8\;,\quad
        c_2\cdot J_i=(24,60)\;,\quad
        \chi=-112
        \end{aligned}\;.
\end{equation}

\begin{figure}
    \centering
    \raisebox{-0.5\height}{
    \begin{tikzpicture}
        \node at (0,0) {\includegraphics[width=6cm]{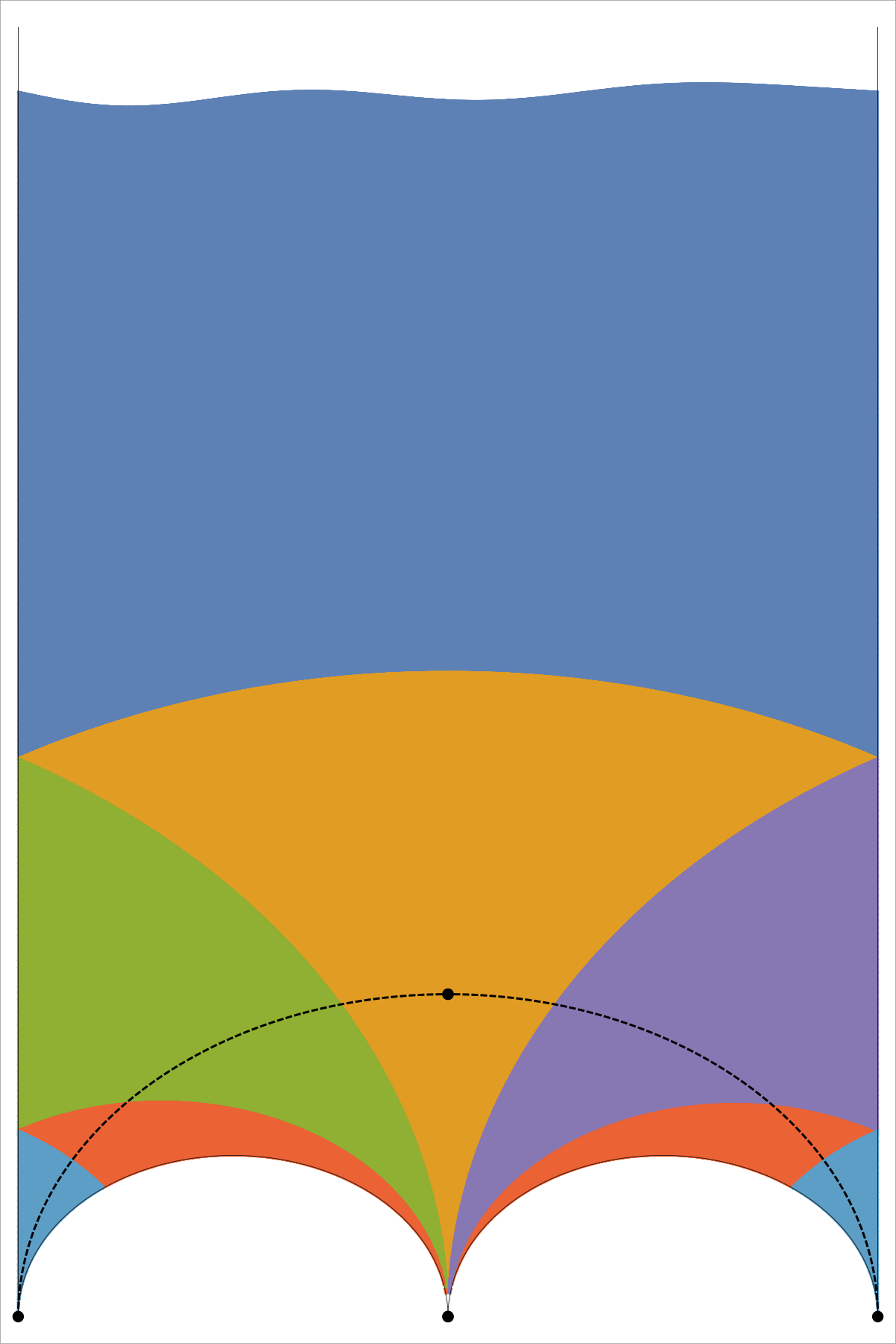}};
        \node at (-2.5,-4.25) {$p_7$};
        \node at (0,-1.9) {$p_8$};
        \node at (2.6,-4.25) {$p_9$};
        \node at (0.4,-4.25) {$c_0$};
    \end{tikzpicture}
    }
    \quad
    \raisebox{-0.5\height}{$\Longrightarrow$}
    \quad
    \raisebox{-0.5\height}{
    \begin{tikzpicture}
        \node at (0,0) {\includegraphics[width=6cm]{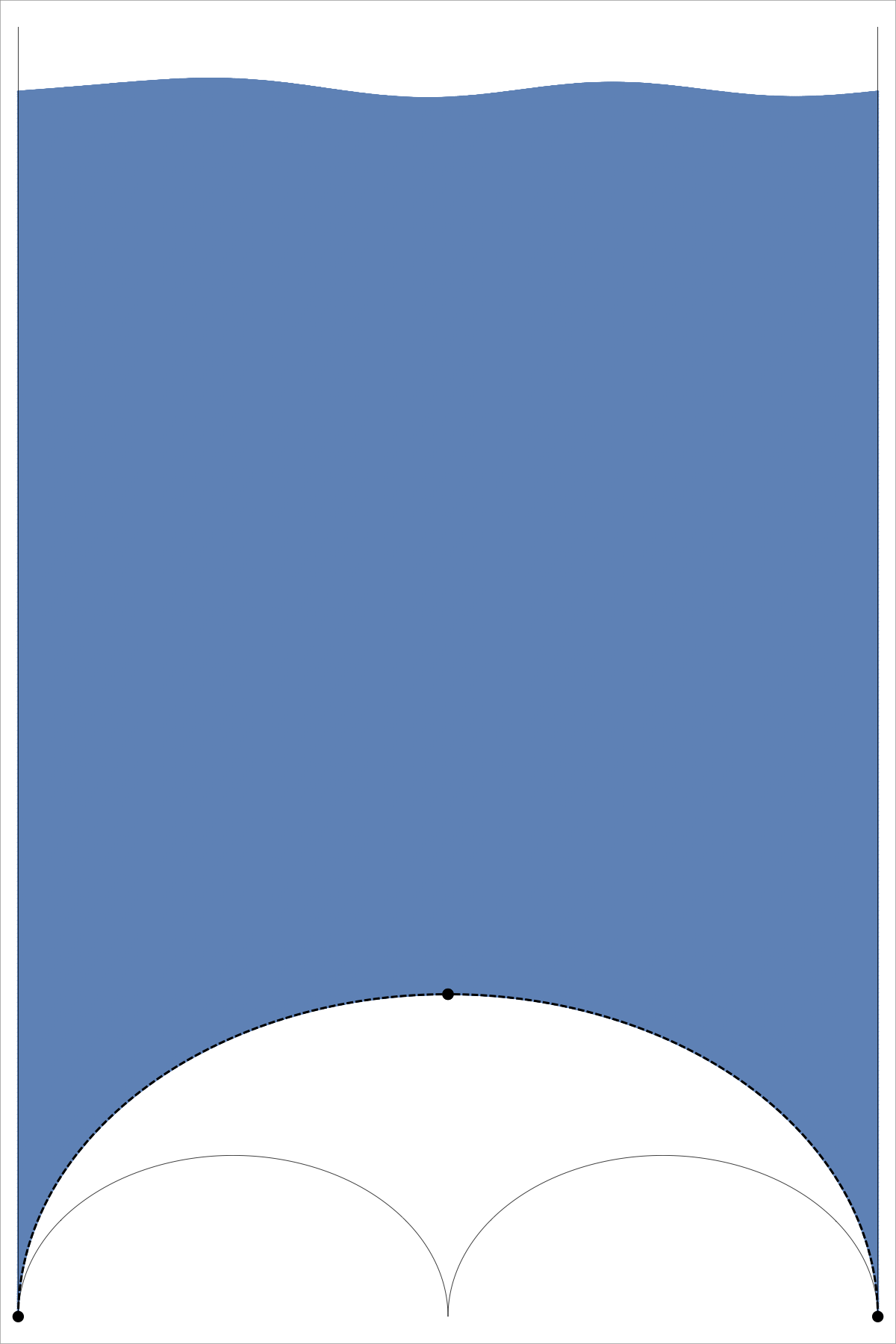}};
        \node at (-2.5,-4.25) {$p_7$};
        \node at (0,-1.9) {$p_8$};
        \node at (2.6,-4.25) {$p_9$};
    \end{tikzpicture}
    }
    \caption{Left: A fundamental domain $X_0(4)$ for $\Gamma_0(4)$ that is symmetric under the Fricke involution $F_4$. Right: A fundamental domain for $\Gamma_0(4)^+$ is obtained as $X_0(4)/F_4$.\\The marked points are the following: $p_9=1/2$ is a cusp and $p_7=-\bar{p}_9$. $p_8=i/2$ is the fixed point of $F_4$ and $c_0=0$ is the third cusp of $X_0(4)$.}
    \label{fig:gamma4fundomain}
\end{figure}
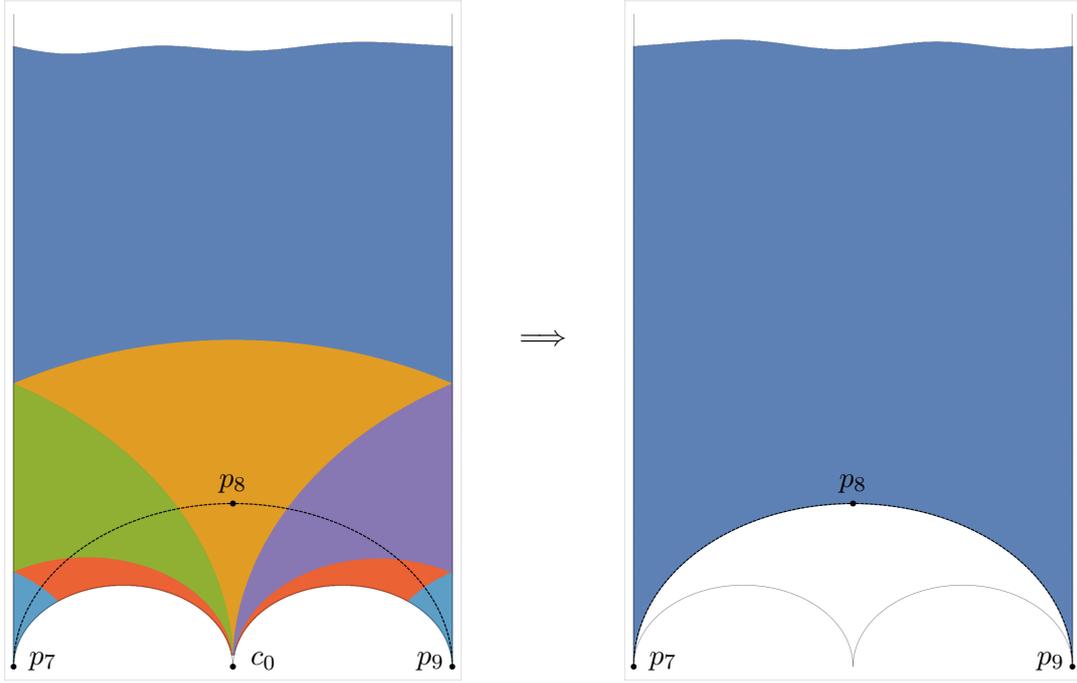
From Figure~\ref{fig:gamma4fundomain}, we see that the modular curve $X_0(4)^+=\mathcal{H}/\Gamma_0(4)^+$ is now qualitatively different because the special points at $\textup{Re}(T)=\pm\tfrac12$ that were previously orbifold points have now turned into cusps and are at infinite distance. As a result, the diameter of the $\mathbb{P}^1$ hybrid phase is now infinite
\begin{equation}
    \label{eq:1122222DeltaPhiExact}
    \Delta\phi_c=\int_{\pi/2}^{\pi}d\varphi \sqrt{g_{T\bar{T}}\frac{d T(\varphi)}{d\varphi}\frac{d \bar{T}(\varphi)}{d\varphi}}=\frac{1}{\sqrt{2}}\int_{\pi/2}^{\pi} \frac{d\varphi}{\sin\varphi}=\infty\;.
\end{equation}
This means that we will now expect an infinite tower also when approaching $T=p_7$ in the hybrid phase. As a result, the RDC is expected to hold trivially in this part of the moduli space. We should point out that this is an artifact of $N=4$ and that $N>4$ will again allow for an interesting challenge for the distance conjecture. For example, for $N=5$ there are no cusps other than $i\infty$ in $X_0(5)^+$. Additional cusps will generally appear if $N$ is not a prime number.

\section{RDC for CY Threefolds Fibered by Degree \texorpdfstring{$2N$}{2N} K3 Surfaces}
\label{sec:PolarizedK3Fibrations}

In the previous section, we have studied the diameter of a certain hybrid phase in the moduli space of type IIA compactified on K3 fibered Calabi-Yau threefolds. In the large base limit, the moduli space reduces to a modular curve $X_0(N)^+$, given as a quotient of the upper half-plane by a congruence subgroup $\Gamma_0(N)^+$. The diameter could then be computed as the length of a geodesic arc connecting two orbifold points in $X_0(N)^+$. The resulting diameters of the $\mathbb{P}^1$ hybrid phase are summarized in table~\ref{tab:FiniteDistances1to4}
\begin{table}
    \renewcommand{\arraystretch}{1.3}
    \centering
    \begin{tabular}{llc}
        \toprule
        Congruence Subgroup & Calabi-Yau Example & $\Delta\phi_c$\\
        \midrule
        $SL(2,\mathbb{Z})$ & $\mathbb{P}^4_{11226}[12]$ & $0.39$\\
        $\Gamma_0(2)^+$ & $\mathbb{P}^4_{1^22^3}[8]$ & $0.62$\\
        $\Gamma_0(3)^+$ & $\mathbb{P}^4_{1^22^4}[4,6]$ & $0.93$\\
        $\Gamma_0(4)^+$ & $\mathbb{P}^4_{1^22^5}[4,4,4]$ & $\infty$\\
        \bottomrule
    \end{tabular}
    \caption{Diameter of the $\mathbb{P}^1$ hybrid phase for Calabi-Yau threefolds fibered by K3 fibers with moduli space $\mathcal{H}/\Gamma$, where $\Gamma\subset SL(2,\mathbb{R})$.}
    \label{tab:FiniteDistances1to4}
\end{table}
We see that the diameter of the phase is finite, except for $N=4$, and grows with $N$. Already for $N=3$ we get a distance of almost one. In order to potentially break the refined distance conjecture, we should thus aim to find models that realize $N\gg 1$ and confirm that this growth continues. Unfortunately, this is not so easy as we will discuss momentarily.

The even integer $2N$ is the \emph{degree} of a maximally generic family of \emph{polarized K3 surfaces}. This means that the generic fiber of our Calabi-Yau threefold has a primitive ample divisor $L$ such that $L^2=2N$. The role of $L$ in the context of the explicit models of section~\ref{sec:RDC_LowDegree} was played by the Kähler cone generator $J_2$. Such K3 surfaces are indeed known to exist for any $N$, but toric constructions of very general families of degree $2N$ polarized K3 surfaces are limited to $N\leq 4$~\cite{Huybrechts:2016uxh}. General polarized families of degree $2N>8$ have been constructed by Mukai~\cite{Mukai1,Mukai2,Mukai3,Mukai4,Mukai5,Mukai6}. Many of them can be obtained as complete intersections specified by vector bundles over a Grassmannian.

In the following, we will discuss the relevance of large $N$ for testing the refined distance conjecture. Section~\ref{subsec:Generalities} discusses general properties of K3 fibrations. In section~\ref{subsec:ViolatingRDC} we argue that large $N$ leads to logarithmic tension with the RDC. Finally, section~\ref{subsec:Grassmannian Constructions} is concerned with constructing Calabi-Yau threefolds that are fibered by general polarized degree $2N>8$ K3 surfaces.

\subsection{K3 Fibrations with \texorpdfstring{$h^{11}=2$}{h11=2}: Generalities}
\label{subsec:Generalities}

In order to systematize the results of section~\ref{sec:RDC_LowDegree}, we want to consider a smooth, projective Calabi-Yau three-fold $X$ with $h^{11}=2$ that admits a K3 fibration morphism $p:X\to\mathbb{P}^1$. The generic fiber is required to be a general, smooth degree $2N$ polarized K3 surface with one-dimensional Picard lattice generated by the primitive ample divisor class with self-intersection $2N$
\begin{equation}
    \textup{Pic}=\langle +2N\rangle\;.
\end{equation}
As reviewed in appendix~\ref{app:K3SurfacesAndFibrations}, the mirror moduli space for such a K3 is indeed the modular curve $X_0(N)^+=\mathcal{H}/\Gamma_0(N)^+$.

Since $h^{11}=2$, the Kähler cone will be simplicial and has only two generators, which we will call $J_1$ and $J_2$. Due to a theorem of Ooguiso~\cite{Oguiso1,Oguiso2}, the existence of a K3 fibration implies the existence of a nef divisor $D$ satisfying $D^2=0$ and $c_2(X)\cdot D=24$. The nef divisor $D$ actually has to be one of the Kähler cone generators, as shown in~\cite{Lee:2019wij}. Without loss of generality, we can assume $J_1=D$. Thus, our manifold will be characterized by the following numerical data
\begin{equation}
    \label{eq:IntersectionsGeneral}
    J_2^3=d_2\;,\quad J_2^2\cdot J_1=d_1\;,\quad c_2\cdot J_1=24\;,\quad c_2\cdot J_2=c\;,\quad \chi=k\;,
\end{equation}
where $d_1,d_2$ as well as $c,k$ are some integers\footnote{See~\cite{Kanazawa2014} for some interesting constraints on these numbers.} that depend on the fibration structure and determine our Calabi-Yau threefold $X$ up to diffeomorphism if it is furthermore assumed to be simply connected~\cite{Wall1966}. Since $J_1$ is the class of the K3 fiber and $J_2$ descends to an ample divisor on it, we can see that
\begin{equation}
    2N=\left(J_2\big|_\textup{K3}\right)^2=J_2^2\cdot J_1=d_1\;.
\end{equation}

All of the models discussed in section~\ref{sec:RDC_LowDegree} fit into the structure described above.

\subsection{Violating the Refined Distance Conjecture?}
\label{subsec:ViolatingRDC}

Let us now see whether we can put the RDC to the test by constructing models with large $N$. Before discussing actual Calabi-Yau fibrations, we may compute the length of various finite distance geodesics that connect orbifold points in $X_0(N)^+$. The results for $5\leq N \leq 12$ are tabulated in~\ref{tab:finiteDistance5to12}. These have been computed in explicit models for the fundamental domain of $\Gamma_0(N)^+$, which can be found in figures~\ref{fig:gamma5678Fundomain} and~\ref{fig:gamma9101112Fundomain} in the appendix. The maximal distance is obtained for $N=11$ and is $\Delta\phi\approx 2.12$. By inspecting tables~\ref{tab:FiniteDistances1to4} and~\ref{tab:finiteDistance5to12}, we find empirically that the growth of the length of realizable finite distance geodesics is scattered around $\tfrac{1}{\sqrt{2}}\log(\kappa_1+\kappa_2 N)$, where $\kappa_i$ are order one numbers\footnote{Similar growth with $\log(N)$ was observed in~\cite{Hebecker:2017lxm}, in this case for $\Gamma^0(N)$.}.

The logarithmic scaling can be intuitively explained by observing that the boundary circles of the hyperbolic polygon that marks the boundary of $X_0(N)^+$ are pushed to smaller and smaller $\textup{Im}(T)$ and by the fact that the metric diverges logarithmically when approaching the real axis. To be more precise, due to the action of the Fricke involution $X_0(N)^+$ will always contain a circular arc of radius $1/\sqrt{N}$, with $\textup{Arg}(T)\in [\varphi_0,\pi/2]$. Integrating along this arc, we obtain
\begin{equation}
    \Delta\phi=\int_{\varphi_0}^{\pi/2}\frac{1}{\sqrt{2}\sin(\varphi)}=\frac{1}{\sqrt{2}}\log\cot\left(\frac{\varphi_0}{2}\right)\overset{\varphi_0\to 0}{\sim}\frac{1}{\sqrt{2}}\log\left(\frac{1}{\varphi_0}\right)\;,
\end{equation}
which explains the observed behavior if $\varphi_0\sim 1/N$ as $N\to \infty$.

Thus, for exponentially large $N$, we may expect some tension with the RDC. We would have to study the moduli spaces of actual Calabi-Yau fibrations in order to confirm that the computed distances correspond to the diameter of some phase of the moduli space.
\begin{table}
    \centering
    \renewcommand{\arraystretch}{1.5}
    \begin{tabular}{lcr}
        \toprule
        Congruence Subgroup & Geodesic & Distance\\
        \midrule
        $\Gamma_0(5)^+$ & $\overline{p_{13}\,p_{11}}$ & $\frac{1}{\sqrt{2}}\sinh^{-1}(2)\approx 1.02$\\
        &$\overline{p_{11}\,p_{12}}$ & $\frac{1}{\sqrt{2}}\ln(\cot(\tfrac12\sec^{-1}(\tfrac{3}{\sqrt{5}})))\approx 0.68$\\
        $\Gamma_0(6)^+$ & $\overline{p_{20}\,p_{18}}$ & $\frac{1}{2\sqrt{2}}\ln(49+20\sqrt{6})\approx 1.62$\\
        $\Gamma_0(7)^+$ & $\overline{p_{27}\,p_{25}}$ & $\frac{1}{\sqrt{2}}\ln(\cot(\tfrac12\cos^{-1}(-\tfrac{5}{2\sqrt{7}})))\approx 1.26$\\
        & $\overline{p_{25}\,p_{26}}$ & $\frac{1}{\sqrt{2}}\ln(\cot(\tfrac12\cos^{-1}(\tfrac{2}{\sqrt{7}})))\approx 0.70$\\
        $\Gamma_0(8)^+$ & $\overline{p_{34}\,p_{32}}$ & $\frac{1}{2\sqrt{2}}\log(17+12\sqrt{2})\approx 1.25$\\
        $\Gamma_0(10)^+$ & $\overline{p_{49}\,p_{47}}$ & $\frac{1}{2\sqrt{2}}\log(19+6\sqrt{10})\approx 1.29$\\
        $\Gamma_0(11)^+$ & $\overline{p_{57}\,p_{55}}$ & $\frac{1}{2\sqrt{2}}\log(199+60\sqrt{11})\approx 2.12$\\
        & $\overline{p_{55}\,p_{56}}$ & $\frac{1}{2\sqrt{2}}\log(97+56\sqrt{3})\approx 1.86$\\
        $\Gamma_0(12)^+$ & $\overline{p_{66}\,p_{63}}$ & $\frac{1}{2\sqrt{2}}\log(97+56\sqrt{3})\approx 1.86$\\
        \bottomrule
    \end{tabular}
    \caption{Finite distance geodesics in $X_0(N)^+$ for $5\leq N\leq 12$.}
    \label{tab:finiteDistance5to12}
\end{table}

A crucial question is the maximal value of the parameter $N$ that is realized in the Calabi-Yau landscape. An upper bound would be expected if the number of Calabi-Yau threefolds were finite. There are indications that this could be true from the investigation of genus one fibered Calabi-Yau threefolds, of which there are known to be only a finite number of families~\cite{Gross1993AFT}. Although it is expected that most CY threefolds do admit a genus one fibration~\cite{Anderson:2017aux}, an additional genus one fibration is impossible for the $h^{11}=2$ K3 fibrations that we discuss here. Currently, there seems to be no such powerful finiteness result for K3 fibered CYs~\cite{Wilson2017}.

In relation to the question of finiteness of $N$, an interesting observation about $\Gamma_0(N)^+$ is that the genus of $X_0(N)^+$ is equal to zero only for a finite number of values of $N$, listed for example in~\cite{BayerTravesa:2007}. This is relevant because conjecture 4 of~\cite{Ooguri:2006in} states that the (compactified) moduli space of quantum gravity should be simply connected. Naively, this would suggest that the values of $N$ that appear for K3 fibered CY threefolds are at most those that lead to genus zero curves $X_0(N)^+$\footnote{See \cite{Hajouji:2019vxs,Dierigl:2020lai} for related observations in the context of allowed Mordell-Weil torsion groups in F-theory.}. The main problem with such an argument seems to be that non-trivial one-cycles in the weak coupling slice of moduli space that we are looking at might trivialize in the strong coupling region.

The concrete determination of an upper bound on $N$ will be left for future work. As a first step in this direction, we will propose a geometric construction of $N>4$ K3 fibrations as complete intersections in Grassmann bundles over $\mathbb{P}^1$ in the next section.

\subsection{Explicit Construction of Degree \texorpdfstring{$d>8$}{d>8} K3 Fibrations}
\label{subsec:Grassmannian Constructions}

Let us now study how we can construct Calabi-Yau threefolds fibered by general K3 surfaces of degree $2N>8$. To do this, we will utilize the language of vector bundles on Grassmannians.

\subsubsection*{Warmup: Projective Bundles}
To motivate the construction, let us consider the manifold $\mathbb{P}^4_{11222}[8]$ from section~\ref{subsec:11222}. The fact that this is fibered by quartic K3 surfaces $\mathbb{P}^3[4]$ can be made manifest by first fibering the ambient space $\mathbb{P}^3$ of the K3 over $\mathbb{P}^1$ and then considering a hypersurface in this projective bundle. The bundle we will consider is
\begin{equation}
    Y=\mathbb{P}\left(\mathcal{O}_{\mathbb{P}^1}\oplus\mathcal{O}_{\mathbb{P}^1}\oplus\mathcal{O}_{\mathbb{P}^1}\oplus\mathcal{O}_{\mathbb{P}^1}(2)\right)\;,\qquad p:Y\to\mathbb{P}^1\;.
\end{equation}
Standard methods can be used to compute the canonical bundle of $Y$, which is given by
\begin{equation}
    \omega_Y=\mathcal{O}_Y(-4)\otimes p^* \mathcal{O}_{\mathbb{P}^1}(-4)\;,
\end{equation}
where $\mathcal{O}_Y(1)$ is the dual of the tautological bundle of Y. By adjunction, the vanishing locus $X$ of a general section of the line bundle
\begin{equation}
    \label{eq:Degree4VectorBundle}
    V = \omega_Y^{-1} = \mathcal{O}_Y(4)\otimes p^* \mathcal{O}_{\mathbb{P}^1}(4)
\end{equation}
will have trivial canonical bundle. Over a generic point in $\mathbb{P}^1$, it is manifest that a section of $V$ reduces to a section of $\mathcal{O}_{\mathbb{P}^3}(4)$ and produces a quartic K3 surface. To see that the resulting manifold is diffeomorphic to $\mathbb{P}^4_{11222}[8]$, we can compute the assiciated numerical data~\ref{eq:IntersectionsGeneral} and see that it coincides.

It is well known that the Chow ring $A(\mathbb{P}(E))$ of the projectivization of a vector bundle $E$ of rank $r$ over some smooth base $B$ is generated as an $A(B)$ algebra by the class $r=c_1(\mathcal{O}_{\mathbb{P}(E)}(1))$ with the following relation~\cite{eisenbud_harris_2016}
\begin{equation}
    \label{eq:ChowRingProjectiveBundle}
    A(\mathbb{P}(E))=A(B)[r]\Big/\left(\sum_{i=0}^{r}r^ip^*c_{r-i}(E)\right)\;.
\end{equation}

In this light, let $h=c_1(p^* \mathcal{O}_{\mathbb{P}^1}(1))=p^*c_1(\mathcal{O}_{\mathbb{P}^1}(1))$ and $r=c_1(\mathcal{O}_Y(1))$, so that~\eqref{eq:ChowRingProjectiveBundle} reduces to 
\begin{equation}
    A(Y)=A(\mathbb{P}^1)[r]/\left(r^4+2hr^3\right)=\mathbb{Z}[h,r]/\left(h^2,r^4+2hr^3\right)\;.
\end{equation}
Furthermore, we set $J_1=h,J_2=r+h$. Since the Poincaré dual of the homology class corresponding to the vanishing locus of a generic section of $V$ is the Euler class $e(V)=c_\textup{top}(V)$, we can compute intersections on $X$ via
\begin{equation}
    \label{eq:IntersectionFromAmbietEulerClass}
    D_\alpha\cdot D_\beta\cdot D_\gamma|_X=D_\alpha\cdot D_\beta\cdot D_\gamma\cdot e(V)\;.
\end{equation}
Using this, we can compute
\begin{equation}
    \begin{aligned}
        J_2^3|_X&=(r+h)^3\cdot c_1(V)=(r+h)^3\cdot(4r+4h)=8\\
        J_2^2\cdot J_1|_X&=(r+h)^2\cdot h\cdot c_1(V)=(r+h)^2\cdot h\cdot(4r+4h)=4
    \end{aligned}\;,
\end{equation}
precisely matching~\eqref{eq:11222Intersections}. Using $c(X)=c(Y)/c(V)$, one can see that also the data associated with $c_2(X)$ and $c_3(X)$ coincides.

In the same way, the examples of sections~\ref{subsec:112222} and~\ref{subsec:1122222} can be reproduced in projective bundles by considering
\begin{equation}
    V=\left(\mathcal{O}_Y(2)\otimes p^* \mathcal{O}_{\mathbb{P}^1}(2)\right)\oplus\left(\mathcal{O}_Y(3)\otimes p^* \mathcal{O}_{\mathbb{P}^1}(3)\right)\qquad\textup{on}\qquad\mathbb{P}\left(\mathcal{O}_{\mathbb{P}^1}^{\oplus 4}\oplus\mathcal{O}_{\mathbb{P}^1}(3)\right)\;,
\end{equation}
and
\begin{equation}
    V=\left(\mathcal{O}_Y(2)\otimes p^* \mathcal{O}_{\mathbb{P}^1}(2)\right)^{\oplus 3}\qquad\textup{on}\qquad\mathbb{P}\left(\mathcal{O}_{\mathbb{P}^1}^{\oplus 5}\oplus\mathcal{O}_{\mathbb{P}^1}(4)\right)\;.
\end{equation}
Many more examples of toric K3 fibrations can be constructed by generalizing to weighted projective bundles~\cite{mullet2006toric,mullet_thesis}.

\subsubsection*{Grassmann Bundles}
Turning now to higher degree K3 fibrations, we want to make the fibration manifest by considering an ambient space that is itself fibered over $\mathbb{P}^1$. Since higher degree K3 fibers can be constructed in Grassmannians, it is thus natural to consider Grassmannians of vector bundles over $\mathbb{P}^1$. Let us review some basic facts about these objects in order to fix notation~\cite{eisenbud_harris_2016,fulton2016intersection}.

The ordinary Grassmannian $\textup{Gr}(k,n)$ is the space of $k$-planes in an $n$-dimensional vector space $V\cong \mathbb{C}^n$ and has dimension $k(n-k)$. This definition includes projective space as the special case $k=1$. There are two natural vector bundles on the Grassmannian. The tautological (sub-)bundle $S$ associates to a point $P$ in $\textup{Gr}(k,n)$ the $k$-dimensional subspace $P\subset V$ that it represents, whereas the tautological quotient bundle $Q$ associates the $(n-k)$-dimensional quotient space $V/P$ to it. The line bundle $O_G(1)\equiv \textup{det}(S)^\vee$ determines the Plücker embedding into $\textup{Gr}(k,n)\hookrightarrow \mathbb{P}(\Lambda^k V)\cong\mathbb{P}^{\binom{n}{k}-1}$.

In the same way, given a vector bundle $E$ over some manifold $B$, we can consider the Grassmann bundle $p:\textup{Gr}(k,E)\to B$ of $k$-dimensional linear subspaces of the fibers of $E$. Let us now record some important facts about this construction. The space $\textup{Gr}(k,E)$ is equipped with the obvious tautological bundle $S$ and the quotient bundle $Q=p^*E/S$. These satisfy the tautological exact sequence
\begin{equation}
    \label{eq:GrassmannTautologicalSequence}
    0\longrightarrow S\longrightarrow p^* E\longrightarrow Q\longrightarrow 0\;.
\end{equation}
The tangent bundle of $\textup{Gr}(k,E)$ fits into the exact sequence\footnote{The bundle $S^\vee \otimes Q$ is the \emph{relative} tangent bundle with respect to the projection $p$~\cite{fulton2016intersection}.}
\begin{equation}
    0\longrightarrow S^\vee \otimes Q\longrightarrow T\textup{Gr}(k,E)\longrightarrow p^* T_B\longrightarrow 0\;.
\end{equation}
By dualising and taking determinants, we determine the canonical bundle
\begin{equation}
    \omega_{\textup{Gr}(k,E)}=\textup{det}(Q^\vee)^k\otimes \textup{det}(S)^{n-k}\otimes p^*\omega_B\;.
\end{equation}
The Chow ring of $\textup{Gr}(k,E)$ is generated over that of the base $B$ by the Chern classes of $c_i(S)$ and $c_i(Q)$
\begin{equation}
    A(\textup{Gr}(k,E))=A(B)\left[c_i(S),c_i(Q)\right]\big/ \left(c(E)-c(S)c(Q)\right)\;.
\end{equation}
The relation $c(E)=c(S)c(Q)$ follows from the tautological sequence~\eqref{eq:GrassmannTautologicalSequence} and can be solved in order to express either the $c_i(Q)$ in terms of the $c_i(S)$ or vice versa. There is also a nice description of the Chow ring in terms of Schubert cycles, the intersection of which is governed by Schubert calculus~\cite{eisenbud_harris_2016}. In the present work, the calculation of intersection numbers was carried out using the \texttt{Schubert2} package~\cite{Schubert2Source} for \texttt{Macaulay2}~\cite{M2}.

The following examples exhaust the list of polarized K3 surfaces that can be constructed in Grassmannian ambient spaces given in~\cite{debarre2020hyperkahler}, which includes examples up to degree $38$. It is important to note that we have not checked, but rather assumed, that the given complete intersections actually give rise to \emph{smooth} Calabi-Yau threefolds with $h^{11}=2$.

We note that all of the computed $\chi(X)$ appear in~\cite{Batyrev:2008rp}, which studied the Hodge numbers that arise from conifold transitions out of the Kreuzer-Skarke list. Although~\cite{Batyrev:2008rp} only computed more refined data for $h^{11}=1$ manifolds, the matching of the Euler character is a first idication that our constructions may be connected to the toric landscape via extremal transitions. For a similar relation in the context of genus one fibrations, see~\cite{Knapp:2021vkm}.

\subsubsection*{Degree 10}

A general degree 10 K3 surface can be constructed as the complete intersection of the Plücker embedding $\textup{Gr}(2,5)\hookrightarrow\mathbb{P}^9$ with a quadric and a $\mathbb{P}^6\subset \mathbb{P}^9$~\cite{debarre2020hyperkahler}. The latter are given by sections of $\mathcal{O}_{\mathbb{P}^9}(2)$ and $\mathcal{O}_{\mathbb{P}^9}(1)^{\oplus 3}$, respectively. By pulling back we can equivalently describe the K3 as the zero locus of a generic section of the rank four bundle
\begin{equation}
    \label{eq:Degree10FiberDefiningBundle}
    V_G=\mathcal{O}_G(1)^{\oplus 3}\oplus \mathcal{O}_G(2)
\end{equation}
on the six-dimensional Grassmannian $G=\textup{Gr}(2,5)$.

We would now like to promote the Grassmannian to a Grassmann bundle
\begin{equation}
    Y = \textup{Gr}(2,E)\;,\qquad E = \mathcal{O}_{\mathbb{P}^1}^{\oplus 4}\oplus \mathcal{O}_{\mathbb{P}^1}(t)\;,
\end{equation}
where $t\in \mathbb{Z}$ is a twist which is to be specified. We now want to twist the bundle~\eqref{eq:Degree10FiberDefiningBundle} by powers of $p^*\mathcal{O}_{\mathbb{P}^1}(1)$ such that the zero locus $X$ of a generic section has trivial canonical bundle. We make the ansatz
\begin{equation}
    V_Y = \left(\bigoplus_{i=1}^{3}\mathcal{O}_Y(1)\otimes p^*\mathcal{O}_{\mathbb{P}^1}(s_i)\right)\oplus\left(\mathcal{O}_Y(2)\otimes p^*\mathcal{O}_{\mathbb{P}^1}(s_4)\right)\;.
\end{equation}
By adjunction, $\omega_X=\omega_Y\otimes p^*\textup{det}(E)|_X$, we compute
\begin{equation}
    \omega_X=p^*\mathcal{O}_{\mathbb{P}^1}(-2-2t+\textstyle{\sum}_i s_i)\;.
\end{equation}
Here and in the following examples, we will denote the generators of the Chow group $A^1(Y)$ of divisors of $Y$ by $h=p^*c_1(\mathcal{O}_{\mathbb{P}^1}(1))$ and $r=c_1(\mathcal{O}_Y(1))$. Defining $J_1=h|_X$ and $J_2=r|_X$, for small values of the twist $t=0,1$, we find the examples\footnote{The last example is also mentioned in~\cite{Knapp:2021vkm}.}:
\begin{equation}
    \begin{array}{ccccc}
        t & s_i & J_2^3 & c_2\cdot J_2 & \chi\\
        \hline
        0 & (2,0,0,0) & 20 & 68 & -100\\
        0 & (1,0,0,1) & 15 & 66 & -110\\
        0 & (0,0,0,2) & 10 & 64 & -120\\
        1 & (1,1,1,1) & 7 & 58 & -102\\
    \end{array}
\end{equation}
where here and in the following it is understood that $J_2^2\cdot J_1=2N$ and $c_2\cdot J_1=24$.

\subsubsection*{Degree 12}

A general K3 surface of degree 12 is realized as the zero locus of a generic section of the vector bundle~\cite{Mukai5}
\begin{equation}
    \label{eq:Degree12FiberDefiningBundle}
    V_G=\mathcal{O}_G(1)^{\oplus 2}\oplus S(2)\cong \mathcal{O}_G(1)^{\oplus 2}\oplus \left(S^\vee \otimes \textup{det}(Q)\right)
\end{equation}
over the six-dimensional Grassmannian $\textup{Gr}(2,5)$. We promote the ambient space and bundles to
\begin{equation}
    \begin{gathered}
    Y = \textup{Gr}(2,E)\;,\qquad E = \mathcal{O}_{\mathbb{P}^1}^{\oplus 4}\oplus \mathcal{O}_{\mathbb{P}^1}(t)\;,\\
    V_Y=\left(\bigoplus_{i=1}^2\mathcal{O}_Y(1)\otimes p^*\mathcal{O}_{\mathbb{P}^1}(s_i)\right)\oplus \left(S^\vee \otimes \textup{det}(Q)\otimes p^*\mathcal{O}_{\mathbb{P}^1}(s_3)\right)\;.
    \end{gathered}
\end{equation}
The resulting canonical bundle is
\begin{equation}
    \omega_X=p^*\mathcal{O}_{\mathbb{P}^1}(-2+s_1+s_2+2s_2)\;,
\end{equation}
which leads to a finite number of possibilities for $s_i$. For small twist, we find:
\begin{equation}
    \begin{array}{cccccc}
        t & \textstyle{\sum_{i=1}^2}s_i & s_3 & J_2^3 & c_2\cdot J_2 & \chi\\
        \hline
        0 & 2 & 0 & 24 & 72 & -92\\
        0 & 0 & 1 & 15 & 66 & -92\\
        1 & 2 & 0 &  6 & 60 & -92\\
        1 & 0 & 1 & 33 & 78 & -92\\
    \end{array}
\end{equation}

\subsubsection*{Degree 14}

A general K3 surface of degree 14 is realized as the intersection of the Plücker embedding $\textup{Gr}(2,6)\hookrightarrow\mathbb{P}^{14}$ with a $\mathbb{P}^8\subset\mathbb{P}^{14}$ or equivalently as the zero locus of a generic section of the rank six vector bundle~\cite{debarre2020hyperkahler}
\begin{equation}
    \label{eq:Degree14FiberDefiningBundle}
    V_G=\mathcal{O}_G(1)^{\oplus 6}
\end{equation}
over the eight-dimensional Grassmannian $\textup{Gr}(2,6)$. We promote the ambient space and bundles to
\begin{equation}
    \begin{gathered}
        Y = \textup{Gr}(2,E)\;,\qquad E = \mathcal{O}_{\mathbb{P}^1}^{\oplus 5}\oplus \mathcal{O}_{\mathbb{P}^1}(t)\;,\\
    V_Y=\bigoplus_{i=1}^{6}\mathcal{O}_Y(1)\otimes p^*\mathcal{O}_{\mathbb{P}^1}(s_i)\;.
    \end{gathered}
\end{equation}
The resulting canonical bundle is
\begin{equation}
    \omega_X=p^*\mathcal{O}_{\mathbb{P}^1}(-2-2t+\textstyle{\sum_i}s_i)\;.
\end{equation}
For small twist, we find:
\begin{equation}
    \begin{array}{ccccc}
        t & \textstyle{\sum}_{i}s_i & J_2^3 & c_2\cdot J_2 & \chi\\
        \hline
        0 & 2 & 28 & 76 & -88\\
        1 & 4 & 14 & 68 & -88\\
        2 & 6 & 42 & 84 & -88\\
    \end{array}
\end{equation}

\subsubsection*{Degree 16}

A general K3 surface of degree 16 is realized as the zero locus of a generic section of the vector bundle~\cite{Mukai2}
\begin{equation}
    \label{eq:Degree16FiberDefiningBundle}
    V_G=\mathcal{O}_G(1)^{\oplus 4}\oplus S(1)\cong\mathcal{O}_G(1)^{\oplus 4}\oplus\Lambda^2 S^\vee
\end{equation}
over the nine-dimensional Grassmannian $\textup{Gr}(3,6)$. We promote the ambient space and bundles to
\begin{equation}
    \begin{gathered}
    Y = \textup{Gr}(3,E)\;,\qquad E = \mathcal{O}_{\mathbb{P}^1}^{\oplus 5}\oplus \mathcal{O}_{\mathbb{P}^1}(t)\;,\\
    V_Y=\left(\bigoplus_{i=1}^{4}\mathcal{O}_Y(1)\otimes p^*\mathcal{O}_{\mathbb{P}^1}(s_i)\right)\oplus \left(\Lambda^2 S^\vee\otimes p^*\mathcal{O}_{\mathbb{P}^1}(s_5)\right)\;.
    \end{gathered}
\end{equation}
The resulting canonical bundle is
\begin{equation}
    \omega_X=p^*\mathcal{O}_{\mathbb{P}^1}(-2-3t+\textstyle{\sum}_{i=1}^4s_i+3s_5)\;.
\end{equation}
For small twist, we find:
\begin{equation}
    \begin{array}{cccccc}
        t & \textstyle{\sum}_{i=1}^4s_i & s_5 & J_2^3 & c_2\cdot J_2 & \chi\\
        \hline
        0 & 2 & 0 & 32 & 80 & -84\\
        1 & 5 & 0 &  3 & 66 & -86\\
        1 & 2 & 1 & 18 & 72 & -80\\
    \end{array}
\end{equation}

\subsubsection*{Degree 18}

A general K3 surface of degree 18 is realized as the zero locus of a generic section of the vector bundle~\cite{Mukai2}
\begin{equation}
    \label{eq:Degree18FiberDefiningBundle}
    V_G=\mathcal{O}_G(1)^{\oplus 3}\oplus Q^\vee (1)\cong \mathcal{O}_G(1)^{\oplus 3}\oplus\Lambda^4 Q
\end{equation}
over the 10-dimensional Grassmannian $\textup{Gr}(2,7)$. We promote the ambient space and bundles to
\begin{equation}
    \begin{gathered}
    Y = \textup{Gr}(2,E)\;,\qquad E = \mathcal{O}_{\mathbb{P}^1}^{\oplus 6}\oplus \mathcal{O}_{\mathbb{P}^1}(t)\;,\\
    V_Y=\left(\bigoplus_{i=1}^3\mathcal{O}_Y(1)\otimes p^*\mathcal{O}_{\mathbb{P}^1}(s_i)\right)\oplus \left(\Lambda^4 Q\otimes p^*\mathcal{O}_{\mathbb{P}^1}(s_4)\right)\;.
    \end{gathered}
\end{equation}
The resulting canonical bundle is
\begin{equation}
    \omega_X=p^*\mathcal{O}_{\mathbb{P}^1}(-2+2t+\textstyle{\sum}_{i=1}^3s_i+5s_4)\;,
\end{equation}
which leads to a finite number of possibilities for $t,s_i$. For small twist, we find:
\begin{equation}
    \begin{array}{cccccc}
        t & \textstyle{\sum}_{i=1}^3s_i & s_4 & J_2^3 & c_2\cdot J_2 & \chi\\
        \hline
        0 & 2 & 0 & 36 & 84 & -84\\
        1 & 0 & 0 & 28 & 76 & -68\\
    \end{array}
\end{equation}

\subsubsection*{Degree 22}

A general K3 surface of degree 22 is realized as the zero locus of a generic section of the vector bundle~\cite{Mukai5}
\begin{equation}
    \label{eq:Degree22FiberDefiningBundle}
    V_G=\mathcal{O}_G(1)\oplus S(1)^{\oplus 3}\cong \mathcal{O}_G(1)\oplus \left(\Lambda^2 S^\vee\right)^{\oplus 3}
\end{equation}
over the 12-dimensional Grassmannian $\textup{Gr}(3,7)$. We promote the ambient space and bundles to
\begin{equation}
    \begin{gathered}
        Y = \textup{Gr}(3,E)\;,\qquad E = \mathcal{O}_{\mathbb{P}^1}^{\oplus 6}\oplus \mathcal{O}_{\mathbb{P}^1}(t)\;,\\
    V_Y=\left(\mathcal{O}_Y(1)\otimes p^*\mathcal{O}_{\mathbb{P}^1}(s_1)\right)\oplus \left(\bigoplus_{i=2}^4 \Lambda^2 S^\vee\otimes p^*\mathcal{O}_{\mathbb{P}^1}(s_i)\right)\;.
    \end{gathered}
\end{equation}
The resulting canonical bundle is
\begin{equation}
    \omega_X=p^*\mathcal{O}_{\mathbb{P}^1}(-2-3t+s_1+3\textstyle{\sum}_{i=2}^4s_i)\;.
\end{equation}
For small twist, we find:
\begin{equation}
    \begin{array}{cccccc}
        t & s_1 & \textstyle{\sum}_{i=2}^4s_i & J_2^3 & c_2\cdot J_2 & \chi\\
        \hline
        0 & 2 & 0 & 44 & 92 & -84\\
        1 & 5 & 0 & 62 & 104 & -108\\
        1 & 2 & 1 & 19 & 82 & -80\\
    \end{array}
\end{equation}

\subsubsection*{Degree 24}

A general K3 surface of degree 24 is realized as the zero locus of a generic section of the vector bundle~\cite{Mukai5}
\begin{equation}
    \label{eq:Degree24FiberDefiningBundle}
    V_G=S(1)^{\oplus 2}\oplus Q^\vee (1)\cong \left(\Lambda^2 S^\vee\right)^{\oplus 2}\oplus \Lambda^3 Q
\end{equation}
over the 12-dimensional Grassmannian $\textup{Gr}(3,7)$. We promote the ambient space and bundles to
\begin{equation}
    \begin{gathered}
        Y = \textup{Gr}(3,E)\;,\qquad E = \mathcal{O}_{\mathbb{P}^1}^{\oplus 6}\oplus \mathcal{O}_{\mathbb{P}^1}(t)\;,\\
    V_Y=\left(\bigoplus_{i=1}^2\Lambda^2 S^\vee\otimes p^*\mathcal{O}_{\mathbb{P}^1}(s_i)\right)\oplus \left(\Lambda^3 Q\otimes p^*\mathcal{O}_{\mathbb{P}^1}(s_3)\right)\;.
    \end{gathered}
\end{equation}
The resulting canonical bundle is
\begin{equation}
    \omega_X=p^*\mathcal{O}_{\mathbb{P}^1}(-2+3s_1+3s_2+4s_3)\;.
\end{equation}
There are no solutions to $\omega_X=\mathcal{O}_X$.

\subsubsection*{Degree 34}

A general K3 surface of degree 34 is realized as the zero locus of a generic section of the vector bundle~\cite{Mukai5}
\begin{equation}
    \label{eq:Degree34FiberDefiningBundle}
    V_G=\Lambda^3S^\vee\oplus (\Lambda^2Q)^{\oplus 2}
\end{equation}
over the 12-dimensional Grassmannian $\textup{Gr}(4,7)$. We promote the ambient space and bundles to
\begin{equation}
    \begin{gathered}
    Y = \textup{Gr}(4,E)\;,\qquad E = \mathcal{O}_{\mathbb{P}^1}^{\oplus 6}\oplus \mathcal{O}_{\mathbb{P}^1}(t)\;,\\
    V_Y=\left(\Lambda^3S^\vee\otimes p^*\mathcal{O}_{\mathbb{P}^1}(s_1)\right)\oplus \left(\Lambda^2Q\otimes p^*\mathcal{O}_{\mathbb{P}^1}(s_2)\right)\oplus \left(\Lambda^2Q\otimes p^*\mathcal{O}_{\mathbb{P}^1}(s_3)\right)\;.
    \end{gathered}
\end{equation}
The resulting canonical bundle is
\begin{equation}
    \omega_X=p^*\mathcal{O}_{\mathbb{P}^1}(-2+4s_1+3s_2+3s_3)\;.
\end{equation}
There are no solutions to $\omega_X=\mathcal{O}_X$.

\subsubsection*{Degree 38}

A general K3 surface of degree 38 is realized as the zero locus of a generic section of the vector bundle~\cite{debarre2020hyperkahler}
\begin{equation}
    \label{eq:Degree38FiberDefiningBundle}
    V_G=(\Lambda^2S^\vee)^{\oplus 3}
\end{equation}
over the 20-dimensional Grassmannian $\textup{Gr}(4,9)$. We promote the ambient space and bundles to
\begin{equation}
    \begin{gathered}
    Y = \textup{Gr}(4,E)\;,\qquad E = \mathcal{O}_{\mathbb{P}^1}^{\oplus 8}\oplus \mathcal{O}_{\mathbb{P}^1}(t)\;,\\
    V_Y=\bigoplus_{i=1}^3\Lambda^2S^\vee\otimes p^*\mathcal{O}_{\mathbb{P}^1}(s_i)\;.
    \end{gathered}
\end{equation}
The resulting canonical bundle is
\begin{equation}
    \omega_X=p^*\mathcal{O}_{\mathbb{P}^1}(-2-4t+6\textstyle{\sum}_i s_i)\;.
\end{equation}
For small twist, we find:
\begin{equation}
    \begin{array}{ccccc}
        t & \textstyle{\sum}_{i=1}^3s_i & J_2^3 & c_2\cdot J_2 & \chi\\
        \hline
        1 & 1 & 83 & 110 & -52\\
        4 & 3 & 45 & 102 & -52\\
    \end{array}
\end{equation}

\section{Discussion and Outlook}
\label{sec:Conclusions}

In this work we have tested the validity of the refined distance conjecture in the 4D $N=2$ vector multiplet moduli space. We have identified a series of Calabi-Yau compactifications of the type IIA string, indexed by a discrete parameter $N$, which is in tension with the conjecture because the diameter of certain hybrid phases of the moduli space is expected to grow as $\sim\log(N)$. This means that the onset of the exponential mass decay of the distance conjecture tower could be delayed by this amount. The Calabi-Yau manifolds under consideration have $h^{11}=2$ and admit a fibration by a general degree $2N$ polarized K3 surface. In the large base limit, the mirror map is expected to reduce to the Hauptmodul for the group $\Gamma_0(N)^+$, which consists of the congruence subgroup $\Gamma_0(N)$ and the Fricke involution. Orbifold points under the action of $\Gamma_0(N)^+$ on the upper half-plane correspond to components of the discriminant of the mirror IIB compactification and hence to phase boundaries. The computation of the diameter of these phases is thus reduced to integrals along geodesics in $\mathcal{H}/\Gamma_0(N)^+$, which can be evaluated exactly. We have explicitly performed these computations for $N=1,\dots,4$, which can be realized as complete intersections in toric ambient spaces. Furthermore, we have computed distances between orbifold points in $\mathcal{H}/\Gamma_0(N)^+$ for $N=5,\dots,12$, which is expected to arise as the large base limit of the moduli space of degree $d=10,\dots 24$ K3 fibrations. The largest distance found in this way was $\Delta\phi\approx 2.12$. We made first steps towards explicitly realizing these moduli spaces by suggesting a construction of degrees $2N=2,\dots 38$ in terms of Grassmann bundles.

Because of the logarithmic growth of distances\footnote{We point out that log corrections to other swampland conjectures have been considered in~\cite{Blumenhagen:2019vgj}.}, the fate of the refined distance conjecture depends on whether large values of $N$ are actually realized in the Calabi-Yau landscape. Clearly, the present work is only a first step towards addressing this question, which is related to the question of finiteness of families of K3 fibered Calabi-Yau threefolds. In particular, we would like to address the Grassmannian constructions in more detail in future work. A first step would be to address the question of smoothness and compute actual Hodge numbers instead of $\chi$ using the techniques of~\cite{Knapp:2021vkm}. It is also desirable to have a GLSM description~\cite{Hori:2013gga,Caldararu:2017usq,Knapp:2019cih} of the manifolds in order to compute the periods and analyse the moduli space away from the large base limit. Furthermore, since Grassmannians are known to admit toric degenerations and due to the matching values of $\chi$ in~\cite{Batyrev:2008rp}, it will be interesting to study extremal transitions to toric Calabi-Yau threefolds. Finally, heterotic/IIA duality for the Calabi-Yau threefolds with $h^{11}=2$ deserves to be better understood.

\subsection*{Acknowledgements}
It is a pleasure to thank Timo Weigand, Lorenz Schlechter, Johanna Knapp, Emanuel Scheidegger and Thorsten Schimannek for useful discussions. The work of Daniel Kläwer is supported in part by Deutsche Forschungsgemeinschaft under Germany's Excellence Strategy EXC 2121 Quantum Universe 390833306.

{\bfseries Note added:} After this work appeared as a preprint, the paper~\cite{Alvarez-Garcia:2021mzv} was posted to arXiv, which employs different methods in order to obtain analytic expressions for periods. Using these results, the authors confirm the computations of sections~\ref{subsec:11226} and~\ref{subsec:11222}.

\appendix

\section{Congruence Subgroups of \texorpdfstring{$SL(2,\mathbb{Z})$}{SL(2,Z)}}
\label{app:CongruenceSubgroups}

In this appendix we give a brief review of the various congruence subgroups of the modular group and fix the relevant notation. Given the group
\begin{equation}
    SL(2,\mathbb{Z})=\left\{\left.\begin{pmatrix}a&b\\c&d\end{pmatrix}\right|ad-bc=1\right\}\;,
\end{equation}
it is natural to consider subgroups $G\subset SL(2,\mathbb{Z})$ defined by congruences modulo some integer $N\in\mathbb{N}$ of the matrix entries $a,b,c,d$.

A prominent example is the principal congruence subgroup
\begin{equation}
    \Gamma(N)=\left\{\left.\begin{pmatrix}a&b\\c&d\end{pmatrix}\in SL(2,\mathbb{Z})\;\right|\;a,d\equiv 1 \textup{ mod }N\;,\quad b,c\equiv 0\textup{ mod }N\right\}\;.
\end{equation}

In this paper, we will mostly be interested in the Hecke congruence subgroups
\begin{equation}
    \begin{aligned}
        \Gamma_0(N)&=\left\{\left.\begin{pmatrix}a&b\\c&d\end{pmatrix}\in SL(2,\mathbb{Z})\;\right|\;c\equiv 0 \textup{ mod } N\right\}\;,\\
        \Gamma^0(N)&=\left\{\left.\begin{pmatrix}a&b\\c&d\end{pmatrix}\in SL(2,\mathbb{Z})\;\right|\;b\equiv 0 \textup{ mod } N\right\}\;.
    \end{aligned}
\end{equation}
An important property of these is that they map, under the obvious linear action, the lattices $\mathbb{Z}\oplus N\mathbb{Z}$ and $N\mathbb{Z}\oplus \mathbb{Z}$ respectively onto themselves. We note that the two groups are equivalent to each other up to a modular $S$-transformation and hence restrict attention to $\Gamma_0(N)$.

The normalizer of the Hecke congruence subgroups $\Gamma_0(N)$ in $SL(2,\mathbb{R})$ contains the so-called Atkin-Lehner involutions
\begin{equation}
    A_k=\frac{1}{\sqrt{k}}\begin{pmatrix}
        a\cdot k & b\\
        c\cdot N & d\cdot k
    \end{pmatrix}\;,\qquad \textup{det}(A_k)\overset{!}{=}1\;,
\end{equation}
where $k|N$ is a Hall divisor\footnote{This means that, in addition to $k$ being a divisor, $k$ and $N/k$ are coprime.}. Here it is understood that two possible choices of $a,b,c,d$ are equivalent if they are related by $\Gamma_0(N)$ transformations. The Atkin-Lehner involutions can be chosen such that they square to $\boldsymbol{1}_2$ and hence they act as involutions when considered as modular transformations on the upper half-plane. A special case is the Fricke involution
\begin{equation}
    F_N=A_N=\frac{1}{\sqrt{N}}
        \begin{pmatrix}
            0 & -1\\
            N & 0
        \end{pmatrix}\;.
\end{equation}
As $F_N$ induces the transformation $T\to-\frac{1}{NT}$, it can be thought of as a T-duality transformation where the self-dual radius is lowered by a factor of $1/\sqrt{N}$.

The Fricke modular group is the group generated by $\Gamma_0(N)$ together with the Fricke involution. We will denote it by
\begin{equation}
    \Gamma_0(N)^+=\left<\Gamma_0(N),F_N\right>\;.
\end{equation}

Furthermore, the group which is obtained by including all of the Atkin-Lehner involutions will be denoted by\footnote{Sometimes this is denoted as $\Gamma_0(N)+$ in the literature, whereas the Fricke modular group is denoted by $\Gamma_0(N)+N$, see for example~\cite{Conway:1979qga,Lian:1995js}.}
\begin{equation}
    \Gamma_0(N)^*=\left<\Gamma_0(N),A_k\right>\;.
\end{equation}
We note that $\Gamma_0(N)^+=\Gamma_0(N)^*$ for all $N\leq 5$ and in fact for all prime powers $N=p^k$.

Fundamental domains for the finite index subgroups $H$ of $G=SL(2,\mathbb{Z})$ can be determined using group theory. For this, we choose representatives for the cosets $G/H=\{g_iH\}_i$. Each $g_i$ defines a transformed copy of the $SL(2,\mathbb{Z})$ fundamental domain $\mathcal{F}$
\begin{equation}
    \mathcal{F}_{g_i}=\left\{ T\in\mathcal{H} \;\left|\;-\frac12\leq\textup{Re}(T_{g_i})\leq\frac12\;,\quad |T_{g_i}|\geq1\right.\right\}\;,
\end{equation}
where $\mathcal{H}$ is the upper half-plane and $T_{g_i}$ denotes the modular action of $g_i$ on $T$. One can then choose the representatives $g_i$ judiciously such that the $\mathcal{F}_i$ glue together to form a connected fundamental domain
\begin{equation}
\label{eq:fundomain_cosets}
    \mathcal{F}_H=\bigcup_i \mathcal{F}_{g_i}\subset\mathcal{H}
\end{equation}
for the subgroup $H$. Together with the appropriate boundary identifications, the region $\mathcal{F}_H$ is a complex curve, the modular curve of $H$. It can be compactified by adding a finite number of cusp points. The compactified modular curves corresponding to the groups $\Gamma(N)$ and $\Gamma_0(N)$ are conventionally denoted by $X(N)$ an $X_0(N)$.

\section{Fundamental Domains For \texorpdfstring{$\Gamma_0(N)^+$}{Gamma0+(N)}}
\label{app:FunDomains}

Fundamental domains for some of the groups $\Gamma_0(N)$ and $\Gamma_0(N)^+$ have been determined already by Fricke~\cite{fricke1922elliptischen,FrickeSpringer1,FrickeSpringer2,FrickeSpringer3}. The latter can be constructed as a quotient of the former if we can find a representation that is symmetric with respect to the Fricke involution $F_N$. Unfortunately, it is not possible for all $N\in\mathbb{N}$ to find such a representation of the form~\eqref{eq:fundomain_cosets}. One rather has to allow for the additional freedom of cutting the $SL(2,\mathbb{Z})$ fundamental domain into two hyperbolic triangles and then transforming both halves by different representatives from the same coset. This procedure of cutting and glueing by $\Gamma_0(N)$ transformations is described in~\cite{BayerTravesa:2007} for $N=2,3,4$ and in the following we will illustrate it for $N=2$.

In many cases, including $N\in\{2,3,4,6,8,9,10\}$, it is sufficient to cut the $SL(2,\mathbb{Z})$ fundamental domain along the imaginary axis $\textup{Re}(T)=0$. We will denote the transformed left and right parts of the triangulation of $\mathcal{F}$ by
\begin{equation}
    \begin{aligned}
        \mathcal{F}^L_{g_i}&=\left\{ T\in\mathcal{H} \;\left|\;-\frac12\leq\textup{Re}(T_{g_i})\leq 0\;,\quad |T_{g_i}|\geq1\right.\right\}\;,\\
        \mathcal{F}^R_{g_i}&=\left\{ T\in\mathcal{H} \;\left|\;0\leq\textup{Re}(T_{g_i})\leq \frac12\;,\quad |T_{g_i}|\geq1\right.\right\}\;.
    \end{aligned}
\end{equation}
We are looking for a decomposition of the $\Gamma_0(N)$ fundamental domain of the form
\begin{equation}
    \mathcal{F}_H=\bigcup_i\left( \mathcal{F}^L_{g_i} \cup \mathcal{F}^R_{g_i\cdot h_i} \right)\;,
\end{equation}
where $g_i$ are a set of coset representatives and $h_i\in H=\Gamma_0(N)$, such that $(g_i\cdot h_i) H=g_iH$.

\begin{figure}
    \centering
    \begin{tikzpicture}
        \node at (0,0) {\includegraphics[width=0.4\textwidth]{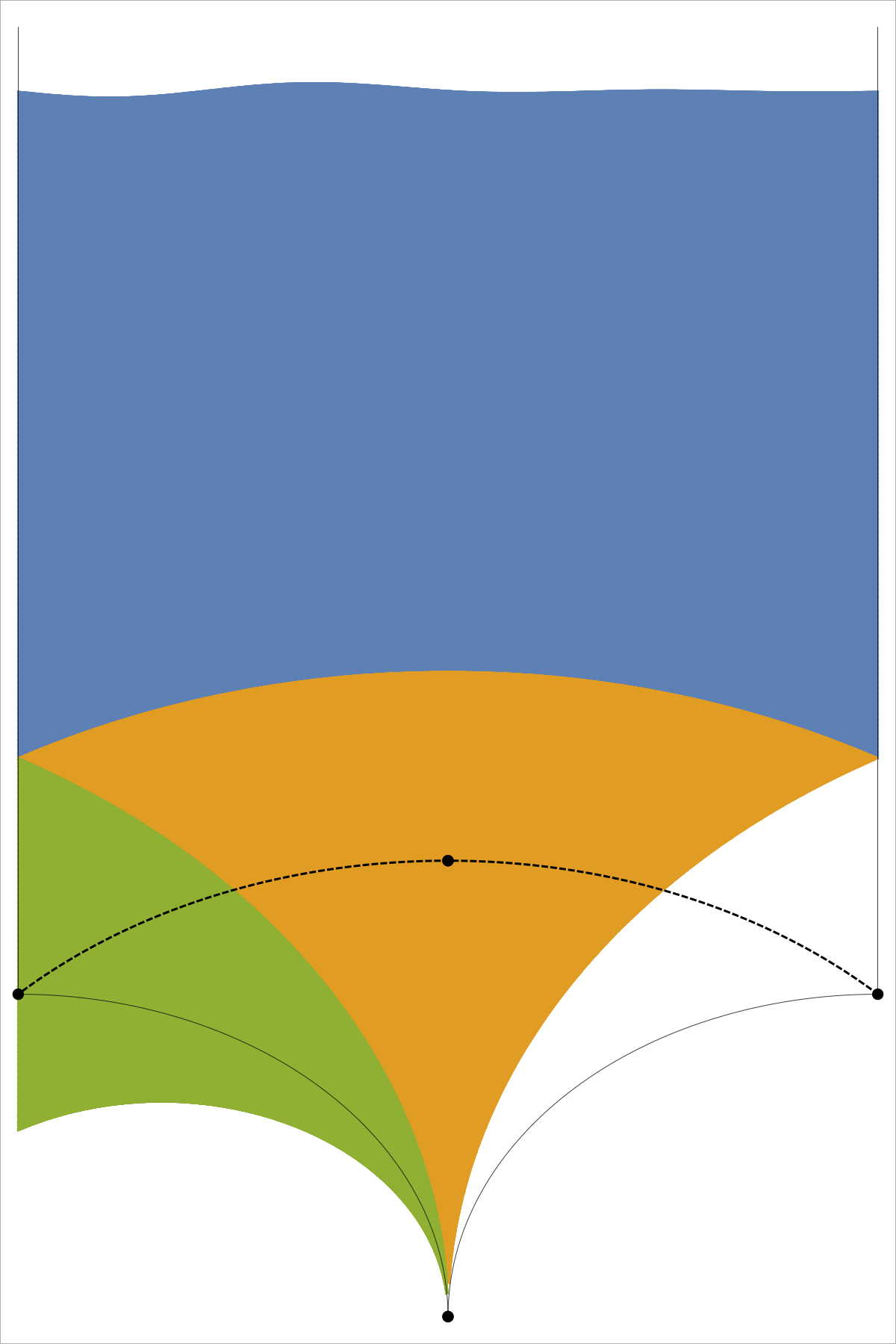}};
        \node at (8,0) {\includegraphics[width=0.4\textwidth]{img/gamma2.png}};
        \draw[->, line width = 0.8pt] (-2.2,-2.8) .. controls (4,-4.5) .. (9.55,-2.15);
        \node at (4,-3.7) {$h_3$};
    \end{tikzpicture}
    \caption{Constructing a $F_2$-symmetric fundamental domain for $\Gamma_0(2)$.}
    \label{fig:cutAndGlueIllustration}
\end{figure}

A convenient set of coset representatives for $\Gamma_0(2)$ is given by
\begin{equation}
    g_1=
    \begin{pmatrix}
        1&0\\
        0&1
    \end{pmatrix}\;,\quad
    g_2=
    \begin{pmatrix}
        0&-1\\
        1&0
    \end{pmatrix}\;,\quad
    g_3=
    \begin{pmatrix}
        1&1\\
        -1&0
    \end{pmatrix}\;.\quad
\end{equation}
The resulting set $\mathcal{F}_{g_1}\cup\mathcal{F}_{g_2}\cup\mathcal{F}_{g_3}$ is shown on the left-hand side of Figure~\ref{fig:cutAndGlueIllustration}. It is clearly not invariant under $T\to-\tfrac{1}{2T}$. This can be remedied by cutting the green region, corresponding to the coset $g_3H$, and glueing the lower half to the right using the $\Gamma_0(2)$ element
\begin{equation}
    h_3=
    \begin{pmatrix}
        1&0\\
        -2&1
    \end{pmatrix}
    \in \Gamma_0(2)\;,
\end{equation}
such that we obtain the symmetric fundamental domain
\begin{equation}
    X_0(2)^+=\mathcal{F}_{g_1}\cup\mathcal{F}_{g_2}\cup\mathcal{F}^L_{g_3}\cup\mathcal{F}^R_{g_3\cdot h_3}\;.
\end{equation}
The process is depicted in Figure~\ref{fig:cutAndGlueIllustration}.

We record the coset representatives $g_i$, glueing transformations $h_i$ and fundamental domains for $N=2,3,4,6,8,9,10$ in table~\ref{tab:cosetReps}.

\begin{table}
\centering
    \begin{tabular}{lcc}
    \toprule
        Group & $g_i=(\cdots)_i$ & $h_i=(\cdots)_i$\\
    \midrule

    \addlinespace[1ex]
        $\Gamma_0(2)$ &
        $\begin{pmatrix}
            1&0\\
            0&1
        \end{pmatrix}_1
        \begin{pmatrix}
            0&-1\\
            1&0
        \end{pmatrix}_2
        \begin{pmatrix}
            1&1\\
            -1&0
        \end{pmatrix}_3$&
        $\begin{pmatrix}
            1&0\\
            -2&1
        \end{pmatrix}_3$\\

    \addlinespace[1ex]
    \midrule
    \addlinespace[1ex]
        $\Gamma_0(3)$ &
        $\begin{pmatrix}
            1&0\\
            0&1
        \end{pmatrix}_1
        \begin{pmatrix}
            0&-1\\
            1&0
        \end{pmatrix}_2
        \begin{pmatrix}
            1&1\\
            -1&0
        \end{pmatrix}_3
        \begin{pmatrix}
            -1&1\\
            -1&0
        \end{pmatrix}_4$&
        $\emptyset$\\

    \addlinespace[1ex]
    \midrule
    \addlinespace[1ex]
        \multirow{4}{*}[1.5pt]{$\Gamma_0(4)$} &
        $\begin{pmatrix}
            1&0\\
            0&1
        \end{pmatrix}_1
        \begin{pmatrix}
            0&-1\\
            1&0
        \end{pmatrix}_2
        \begin{pmatrix}
            1&1\\
            -1&0
        \end{pmatrix}_3
        \begin{pmatrix}
            2&1\\
            -1&0
        \end{pmatrix}_4$&
        \multirow{4}{*}[1.5pt]{
        $\begin{pmatrix}
            1&0\\
            -4&1
        \end{pmatrix}_4
        \begin{pmatrix}
            1&0\\
            4&1
        \end{pmatrix}_6$}\\
    \addlinespace[1ex]
        &$\begin{pmatrix}
            -1&1\\
            -1&0
        \end{pmatrix}_5
        \begin{pmatrix}
            -1&0\\
            2&-1
        \end{pmatrix}_6$&\\

    \addlinespace[1ex]
    \midrule
    \addlinespace[1ex]
        \multirow{6}{*}[1.5pt]{$\Gamma_0(6)$}&
        $\begin{pmatrix}
            1&0\\
            0&1
        \end{pmatrix}_1
        \begin{pmatrix}
            0&-1\\
            1&0
        \end{pmatrix}_2
        \begin{pmatrix}
            1&1\\
            -1&0
        \end{pmatrix}_3
        \begin{pmatrix}
            2&1\\
            -1&0
        \end{pmatrix}_4$
        &
        \multirow{3}{*}{
        $\begin{pmatrix}
            1&0\\
            -6&1
        \end{pmatrix}_5
        \begin{pmatrix}
            5&2\\
            12&5
        \end{pmatrix}_9$}\\
        \addlinespace[1ex]
        &$\begin{pmatrix}
            3&1\\
            -1&0
        \end{pmatrix}_5
        \begin{pmatrix}
            -2&1\\
            -1&0
        \end{pmatrix}_6
        \begin{pmatrix}
            -1&1\\
            -1&0
        \end{pmatrix}_7
        \begin{pmatrix}
            1&0\\
            -2&1
        \end{pmatrix}_8$
        &
        \multirow{3}{*}[-1ex]{
        $\begin{pmatrix}
            1&0\\
            6&1
        \end{pmatrix}_{11}
        \begin{pmatrix}
            -5&2\\
            12&-5
        \end{pmatrix}_{12}$}\\
        \addlinespace[1ex]
        &$\begin{pmatrix}
            3&-1\\
            -2&1
        \end{pmatrix}_9
        \begin{pmatrix}
            1&0\\
            2&1
        \end{pmatrix}_{10}
        \begin{pmatrix}
            1&0\\
            -3&1
        \end{pmatrix}_{11}
        \begin{pmatrix}
            2&1\\
            -3&-1
        \end{pmatrix}_{12}$
        &\\

    \addlinespace[1ex]
    \midrule
    \addlinespace[1ex]
        \multirow{6}{*}[1.5pt]{$\Gamma_0(8)$}&
        $\begin{pmatrix}
            1&0\\
            0&1
        \end{pmatrix}_1
        \begin{pmatrix}
            0&-1\\
            1&0
        \end{pmatrix}_2
        \begin{pmatrix}
            1&1\\
            -1&0
        \end{pmatrix}_3
        \begin{pmatrix}
            2&1\\
            -1&0
        \end{pmatrix}_4$
        &
        \multirow{6}{*}[1.5pt]{$\begin{pmatrix}
            1&0\\
            -8&1
        \end{pmatrix}_6
        \begin{pmatrix}
            1&0\\
            8&1
        \end{pmatrix}_{12}$}\\
        \addlinespace[1ex]
        &$\begin{pmatrix}
            3&1\\
            -1&0
        \end{pmatrix}_5
        \begin{pmatrix}
            4&1\\
            -1&0
        \end{pmatrix}_6
        \begin{pmatrix}
            -3&1\\
            -1&0
        \end{pmatrix}_7
        \begin{pmatrix}
            -2&1\\
            -1&0
        \end{pmatrix}_8$
        &\\
        \addlinespace[1ex]
        &$\begin{pmatrix}
            -1&1\\
            -1&0
        \end{pmatrix}_9
        \begin{pmatrix}
            1&0\\
            -2&1
        \end{pmatrix}_{10}
        \begin{pmatrix}
            1&0\\
            2&1
        \end{pmatrix}_{11}
        \begin{pmatrix}
            1&0\\
            -4&1
        \end{pmatrix}_{12}$
        &\\
    \addlinespace[1ex]
    \bottomrule
    \end{tabular}
\caption{Coset representatives $g_i$ and glueing transformations $h_i$ for some selected $\Gamma_0(N)$. The table is continued at~\ref{tab:cosetReps2}.}
\label{tab:cosetReps}
\end{table}

\begin{table}
\centering
    \begin{tabular}{lcc}
    \toprule
        Group & $g_i=(\cdots)_i$ & $h_i=(\cdots)_i$\\
    \midrule
    \addlinespace[1ex]
        \multirow{6}{*}[1.5pt]{$\Gamma_0(9)$}&
        $\begin{pmatrix}
            1&0\\
            0&1
        \end{pmatrix}_1
        \begin{pmatrix}
            0&-1\\
            1&0
        \end{pmatrix}_2
        \begin{pmatrix}
            1&1\\
            -1&0
        \end{pmatrix}_3
        \begin{pmatrix}
            2&1\\
            -1&0
        \end{pmatrix}_4$
        &\multirow{6}{*}[1.5pt]{$\emptyset$}\\
        \addlinespace[1ex]
        &$\begin{pmatrix}
            3&1\\
            -1&0
        \end{pmatrix}_5
        \begin{pmatrix}
            4&1\\
            -1&0
        \end{pmatrix}_6
        \begin{pmatrix}
            -4&1\\
            -1&0
        \end{pmatrix}_7
        \begin{pmatrix}
            -3&1\\
            -1&0
        \end{pmatrix}_8$
        &\\
        \addlinespace[1ex]
        &$\begin{pmatrix}
            -2&1\\
            -1&0
        \end{pmatrix}_9
        \begin{pmatrix}
            -1&1\\
            -1&0
        \end{pmatrix}_{10}
        \begin{pmatrix}
            1&0\\
            -3&1
        \end{pmatrix}_{11}
        \begin{pmatrix}
            1&1\\
            3&1
        \end{pmatrix}_{12}$
        &\\

    \addlinespace[1ex]
    \midrule
    \addlinespace[1ex]
        \multirow{8}{*}[1.5pt]{$\Gamma_0(10)$}&
        $\begin{pmatrix}
            1&0\\
            0&1
        \end{pmatrix}_1
        \begin{pmatrix}
            0&-1\\
            1&0
        \end{pmatrix}_2
        \begin{pmatrix}
            1&1\\
            -1&0
        \end{pmatrix}_3
        \begin{pmatrix}
            2&1\\
            -1&0
        \end{pmatrix}_4$
        &\\
        \addlinespace[1ex]
        &
        $\begin{pmatrix}
            3&1\\
            -1&0
        \end{pmatrix}_5
        \begin{pmatrix}
            4&1\\
            -1&0
        \end{pmatrix}_6
        \begin{pmatrix}
            5&1\\
            -1&0
        \end{pmatrix}_7
        \begin{pmatrix}
            -4&1\\
            -1&0
        \end{pmatrix}_8$
        &
        $\begin{pmatrix}
            1&0\\
            -10&1
        \end{pmatrix}_7
        \begin{pmatrix}
            9&2\\
            40&9
        \end{pmatrix}_{14}$\\
        \addlinespace[1ex]
        &
        $\begin{pmatrix}
            -3&1\\
            -1&0
        \end{pmatrix}_9
        \begin{pmatrix}
            -2&1\\
            -1&0
        \end{pmatrix}_{10}
        \begin{pmatrix}
            -1&1\\
            -1&0
        \end{pmatrix}_{11}
        \begin{pmatrix}
            1&0\\
            -2&1
        \end{pmatrix}_{12}$
        &
        $\begin{pmatrix}
            1&0\\
            10&1
        \end{pmatrix}_{17}
        \begin{pmatrix}
            9&-2\\
            -40&9
        \end{pmatrix}_{18}$\\
        \addlinespace[1ex]
        &
        $\begin{pmatrix}
            1&0\\
            -4&1
        \end{pmatrix}_{13}
        \begin{pmatrix}
            5&-1\\
            -4&1
        \end{pmatrix}_{14}$
        $\begin{pmatrix}
            1&0\\
            4&1
        \end{pmatrix}_{15}
        \begin{pmatrix}
            1&0\\
            2&1
        \end{pmatrix}_{16}$
        &\\
        \addlinespace[1ex]
        &
        $\begin{pmatrix}
            1&0\\
            -5&1
        \end{pmatrix}_{17}
        \begin{pmatrix}
            4&1\\
            -5&-1
        \end{pmatrix}_{18}$
        &\\
        \addlinespace[1ex]

    \midrule
    \addlinespace[1ex]
        \multirow{8}{*}[1.5pt]{$\Gamma_0(12)$}&
        $\begin{pmatrix}
            1&0\\
            0&1
        \end{pmatrix}_1
        \begin{pmatrix}
            0&-1\\
            1&0
        \end{pmatrix}_2
        \begin{pmatrix}
            1&1\\
            -1&0
        \end{pmatrix}_3
        \begin{pmatrix}
            2&1\\
            -1&0
        \end{pmatrix}_4$
        &\\
        \addlinespace[1ex]
        &$
        \begin{pmatrix}
            3&1\\
            -1&0
        \end{pmatrix}_5
        \begin{pmatrix}
            4&1\\
            -1&0
        \end{pmatrix}_6
        \begin{pmatrix}
            5&1\\
            -1&0
        \end{pmatrix}_7
        \begin{pmatrix}
            6&1\\
            -1&0
        \end{pmatrix}_8$
        &
        $\begin{pmatrix}
            1&0\\
            -12&1
        \end{pmatrix}_8
        \begin{pmatrix}
            5&2\\
            12&5
        \end{pmatrix}_{15}$\\
        \addlinespace[1ex]
        &$
        \begin{pmatrix}
            -5&1\\
            -1&0
        \end{pmatrix}_9
        \begin{pmatrix}
            -4&1\\
            -1&0
        \end{pmatrix}_{10}
        \begin{pmatrix}
            -3&1\\
            -1&0
        \end{pmatrix}_{11}
        \begin{pmatrix}
            -2&1\\
            -1&0
        \end{pmatrix}_{12}$
        &
        $\begin{pmatrix}
            5&-2\\
            -12&5
        \end{pmatrix}_{18}
        \begin{pmatrix}
            7&2\\
            24&7
        \end{pmatrix}_{19}$\\
        \addlinespace[1ex]
        &$
        \begin{pmatrix}
            -1&1\\
            -1&0
        \end{pmatrix}_{13}
        \begin{pmatrix}
            1&0\\
            -2&1
        \end{pmatrix}_{14}
        \begin{pmatrix}
            3&-1\\
            -2&1
        \end{pmatrix}_{15}
        \begin{pmatrix}
            1&0\\
            2&1
        \end{pmatrix}_{16}$
        &
        $\begin{pmatrix}
            7&-2\\
            -24&7
        \end{pmatrix}_{22}
        \begin{pmatrix}
            1&0\\
            12&1
        \end{pmatrix}_{24}$\\
        \addlinespace[1ex]
        &$
        \begin{pmatrix}
            1&0\\
            -3&1
        \end{pmatrix}_{17}
        \begin{pmatrix}
            2&1\\
            -3&-1
        \end{pmatrix}_{18}
        \begin{pmatrix}
            4&-1\\
            -3&1
        \end{pmatrix}_{19}
        \begin{pmatrix}
            1&0\\
            3&1
        \end{pmatrix}_{20}$
        &\\
    \addlinespace[1ex]
        &$
        \begin{pmatrix}
            1&0\\
            -4&1
        \end{pmatrix}_{21}
        \begin{pmatrix}
            3&1\\
            -4&-1
        \end{pmatrix}_{22}
        \begin{pmatrix}
            1&0\\
            4&1
        \end{pmatrix}_{23}
        \begin{pmatrix}
            1&0\\
            -6&1
        \end{pmatrix}_{24}$
        &\\
        \addlinespace[1ex]
    \bottomrule
    \end{tabular}
\caption{Coset representatives $g_i$ and glueing transformations $h_i$ for some selected $\Gamma_0(N)$, continued.}
\label{tab:cosetReps2}
\end{table}

\begin{table}
\centering
    \begin{tabular}{ll}
    \toprule
    Group & Fundamental Domain\\
    \midrule
    \addlinespace[1ex]
    $\Gamma_0(2)$ &
    $\mathcal{F}_{g_1}\cup\mathcal{F}_{g_2}\cup\mathcal{F}^L_{g_3}\cup\mathcal{F}^R_{g_3\cdot h_3}$\\
    \addlinespace[1ex]
    $\Gamma_0(3)$ &
    $\bigcup\limits_{i=1}^{4}\mathcal{F}_{g_i}$\\
    \addlinespace[1ex]
    $\Gamma_0(4)$ &
    $\left(\bigcup\limits_{i\notin\mathcal{I}}\mathcal{F}_{g_i}\right)\cup\left(\bigcup\limits_{i\in\mathcal{I}}\mathcal{F}^L_{g_i}\cup\mathcal{F}^R_{g_i\cdot h_i}\right)\;,\qquad \mathcal{I}=\{4,6\}$\\
    \addlinespace[1ex]
    $\Gamma_0(6)$ & 
    $\left(\bigcup\limits_{i\notin\mathcal{I}}\mathcal{F}_{g_i}\right)\cup\left(\bigcup\limits_{i\in\mathcal{I}}\mathcal{F}^L_{g_i}\cup\mathcal{F}^R_{g_i\cdot h_i}\right)\;,\qquad \mathcal{I}=\{5,9,11,12\}$\\
    \addlinespace[1ex]
    $\Gamma_0(8)$ &
    $\left(\bigcup\limits_{i\notin\mathcal{I}}\mathcal{F}_{g_i}\right)\cup\left(\bigcup\limits_{i\in\mathcal{I}}\mathcal{F}^L_{g_i}\cup\mathcal{F}^R_{g_i\cdot h_i}\right)\;,\qquad \mathcal{I}=\{6,12\}$\\
    \addlinespace[1ex]
    $\Gamma_0(9)$ &
    $\bigcup\limits_{i=1}^{12}\mathcal{F}_{g_i}$\\
    \addlinespace[1ex]
    $\Gamma_0(10)$ &
    $\left(\bigcup\limits_{i\notin\mathcal{I}}\mathcal{F}_{g_i}\right)\cup\left(\bigcup\limits_{i\in\mathcal{I}}\mathcal{F}^L_{g_i}\cup\mathcal{F}^R_{g_i\cdot h_i}\right)\;,\qquad \mathcal{I}=\{7,14,17,18\}$\\
    \addlinespace[1ex]
    $\Gamma_0(12)$ &
    $\left(\bigcup\limits_{i\notin\mathcal{I}}\mathcal{F}_{g_i}\right)\cup\left(\bigcup\limits_{i\in\mathcal{I}}\mathcal{F}^L_{g_i}\cup\mathcal{F}^R_{g_i\cdot h_i}\right)\;,\qquad \mathcal{I}=\{8,15,18,19,22,24\}$\\
    \addlinespace[1ex]
    \bottomrule
    \end{tabular}
    \caption{Selected $F_N$-symmetric fundamental domains for $\Gamma_0(N)$. The group elements $g_i$ and $h_i$ are listed in Table~\ref{tab:cosetReps}. The resulting fundamental domains are depicted in Figures~\ref{fig:gamma2fundomain},~\ref{fig:gamma3fundomain},~\ref{fig:gamma4fundomain},~\ref{fig:gamma5678Fundomain} and~\ref{fig:gamma9101112Fundomain}}
    \label{tab:fundomainsCutAndGlue}
\end{table}

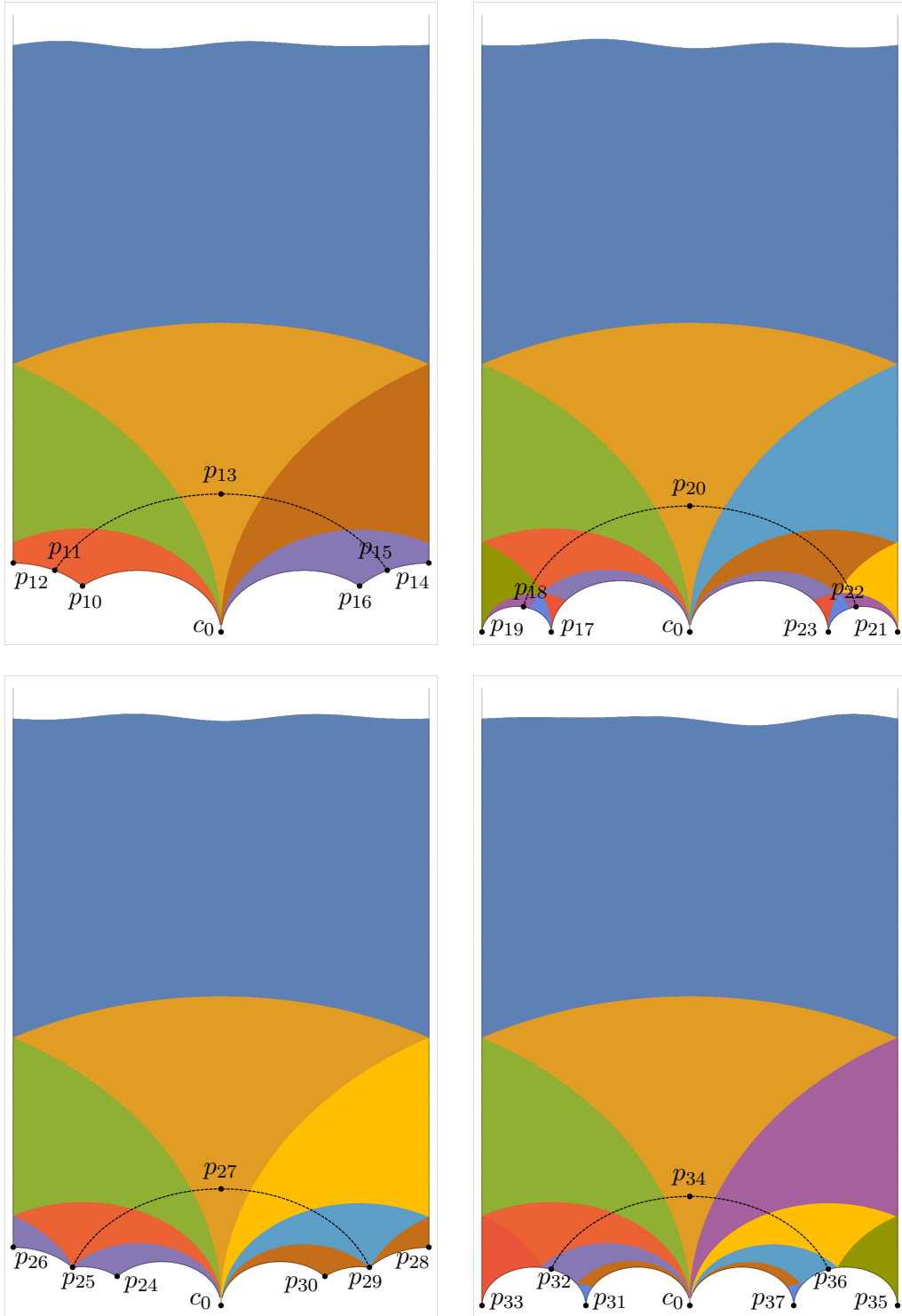
\begin{figure}
    \centering
    \begin{tikzpicture}
        \node at (0,0) {\includegraphics[width=6.4cm]{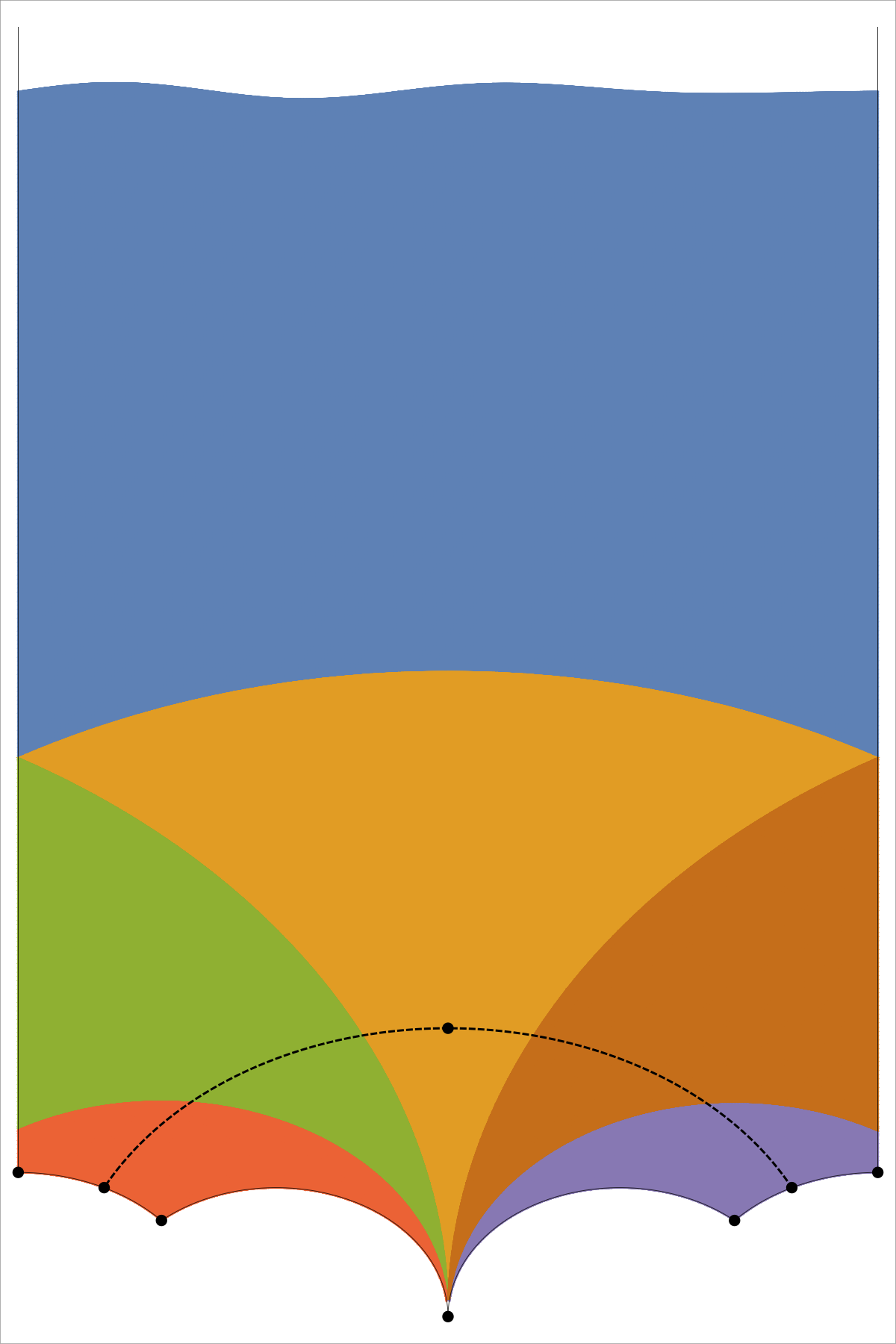}};
        \node at (-0.25,-4.55) {$c_0$};
        \node at (-2,-4.15) {$p_{10}$};
        \node at (-2.3,-3.4) {$p_{11}$};
        \node at (-2.8,-3.85) {$p_{12}$};
        \node at (0,-2.25) {$p_{13}$};
        \node at (2.85,-3.85) {$p_{14}$};
        \node at (2.3,-3.4) {$p_{15}$};
        \node at (2,-4.15) {$p_{16}$};
    \end{tikzpicture}
    \hspace{1pt}
    \begin{tikzpicture}
        \node at (0,0) {\includegraphics[width=6.4cm]{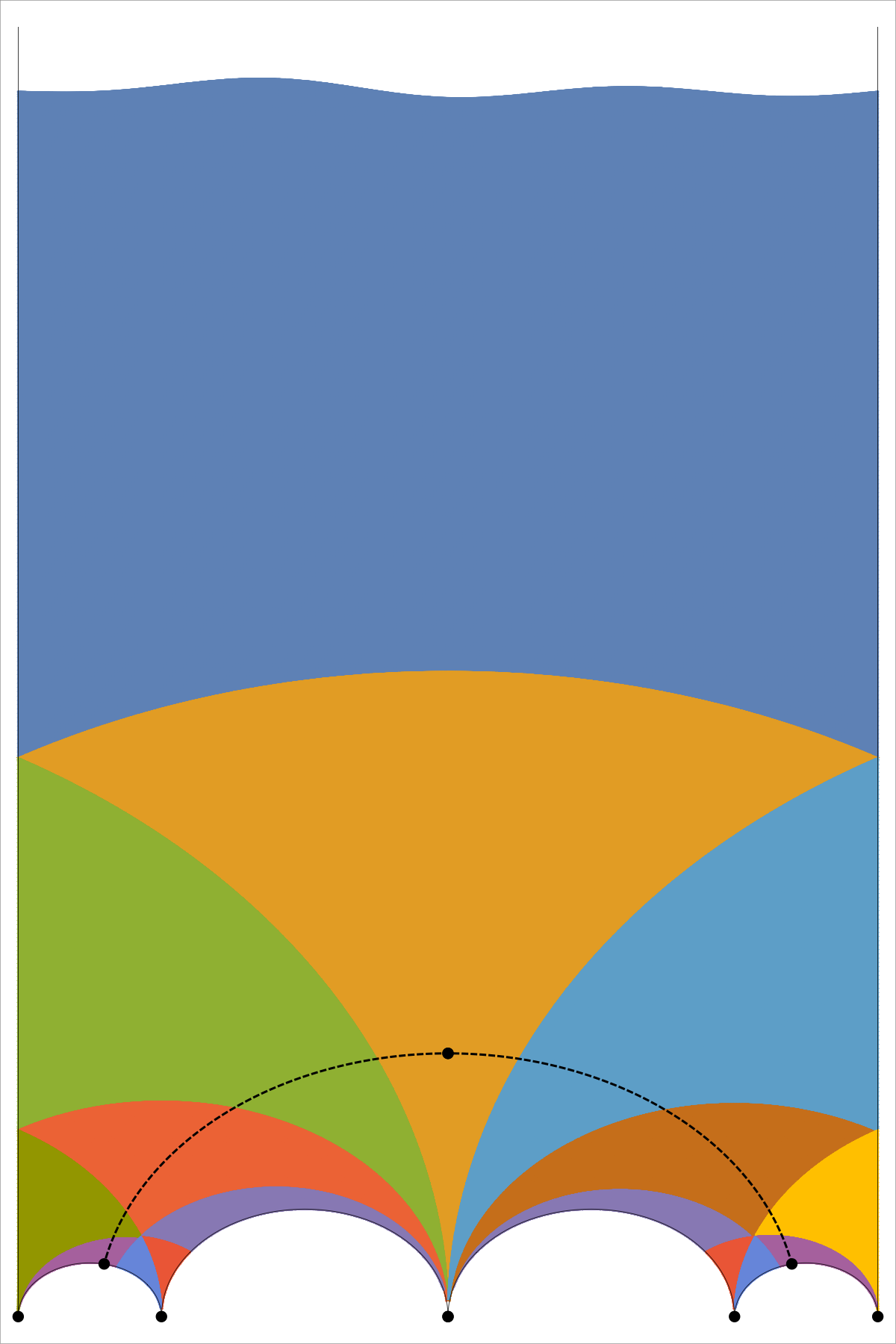}};
        \node at (-0.25,-4.55) {$c_0$};
        \node at (-1.65,-4.55) {$p_{17}$};
        \node at (-2.35,-4) {$p_{18}$};
        \node at (-2.7,-4.55) {$p_{19}$};
        \node at (0,-2.45) {$p_{20}$};
        \node at (2.7,-4.55) {$p_{21}$};
        \node at (2.35,-4) {$p_{22}$};
        \node at (1.65,-4.55) {$p_{23}$};
    \end{tikzpicture}\\[1ex]
    \begin{tikzpicture}
        \node at (0,0) {\includegraphics[width=6.4cm]{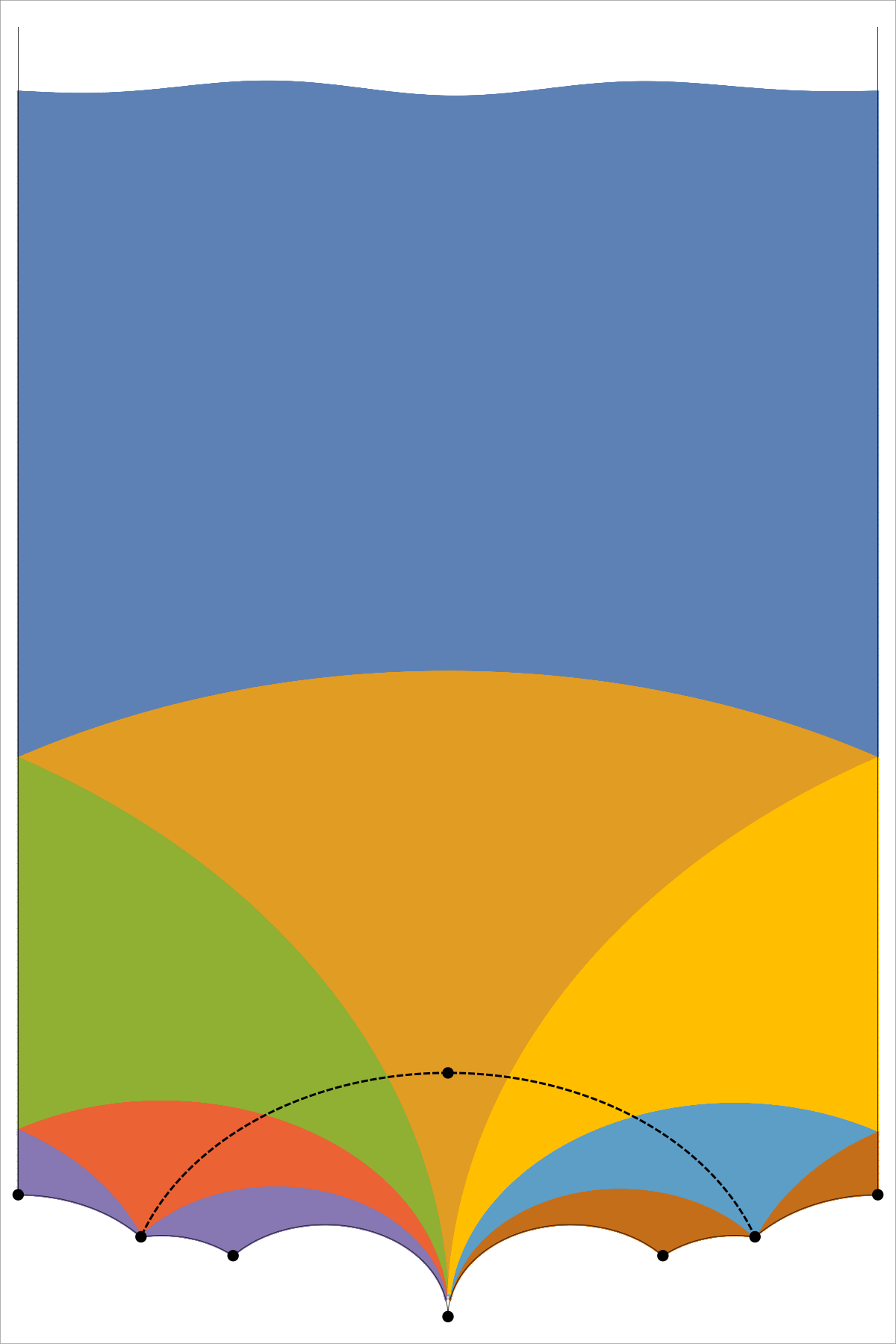}};
        \node at (-0.3,-4.55) {$c_0$};
        \node at (-1.175,-4.3) {$p_{24}$};
        \node at (-2.1,-4.25) {$p_{25}$};
        \node at (-2.8,-3.95) {$p_{26}$};
        \node at (0,-2.6) {$p_{27}$};
        \node at (2.85,-3.95) {$p_{28}$};
        \node at (2.15,-4.25) {$p_{29}$};
        \node at (1.2,-4.3) {$p_{30}$};
    \end{tikzpicture}
    \hspace{1pt}
    \begin{tikzpicture}
        \node at (0,0) {\includegraphics[width=6.4cm]{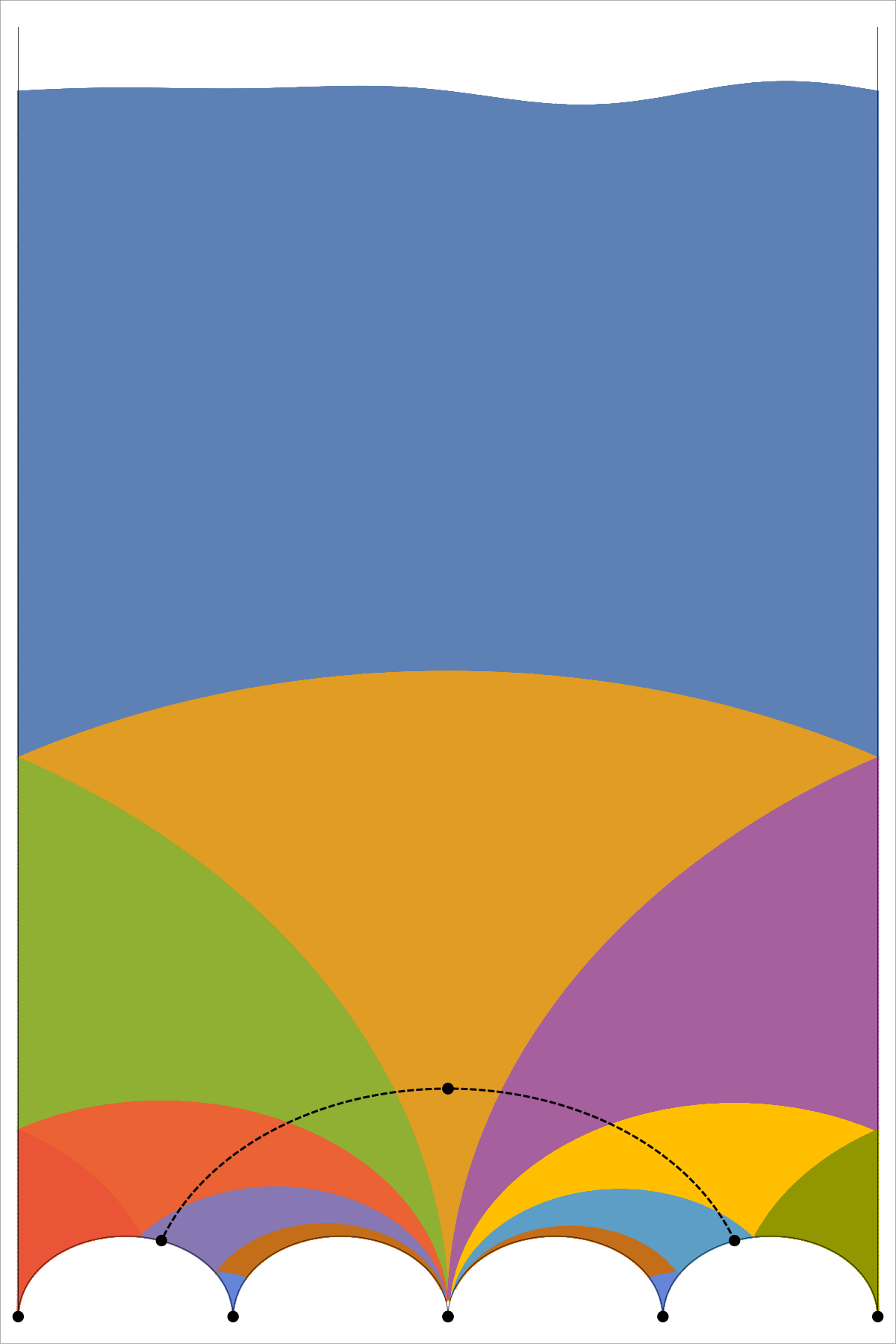}};
        \node at (-0.25,-4.55) {$c_0$};
        \node at (-1.175,-4.55) {$p_{31}$};
        \node at (-2,-4.25) {$p_{32}$};
        \node at (-2.7,-4.55) {$p_{33}$};
        \node at (0,-2.7) {$p_{34}$};
        \node at (2.7,-4.55) {$p_{35}$};
        \node at (2.1,-4.25) {$p_{36}$};
        \node at (1.175,-4.55) {$p_{37}$};
    \end{tikzpicture}
    \caption{Symmetric fundamental domains for $\Gamma_0(N)$ with $N=5,6,7,8$. In each case, a fundamental domain for $\Gamma_0(N)^+$ is obtained by taking only the part above the dashed line, corresponding to quotienting by $F_N$. The labelled points are listed in table~\ref{tab:specialPoints}.}
    \label{fig:gamma5678Fundomain}
\end{figure}

More generally, as it occurs for example in the cases $N\in\{5,7,11\}$, one has to apply cuts along hyperbolic lines emanating from the points $\rho=e^{i\pi/3}$ and $-1/\rho$. We will not describe this in detail. The resulting fundamental domains are shown in Figures~\ref{fig:gamma5678Fundomain} and~\ref{fig:gamma9101112Fundomain}.

\begin{figure}
    \centering
    \begin{tikzpicture}
        \node at (0,0) {\includegraphics[width=6.4cm]{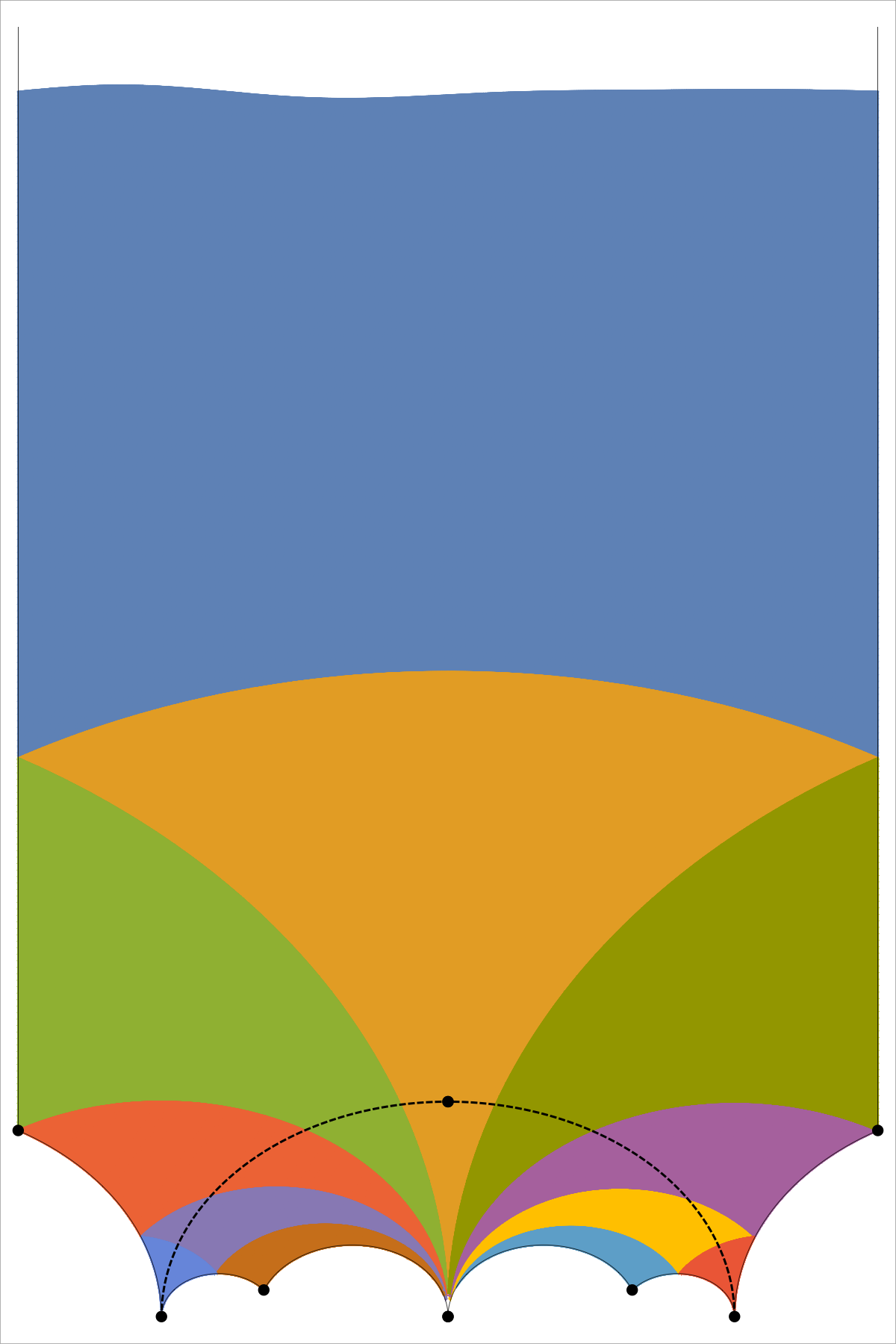}};
        \node at (-0.25,-4.55) {$c_0$};
        \node at (-1,-4.6) {$p_{38}$};
        \node at (-2.4,-4.55) {$p_{39}$};
        \node at (-2.8,-3.7) {$p_{40}$};
        \node at (0,-2.8) {$p_{41}$};
        \node at (2.8,-3.7) {$p_{42}$};
        \node at (2.4,-4.55) {$p_{43}$};
        \node at (1.1,-4.6) {$p_{44}$};
    \end{tikzpicture}
    \hspace{1pt}
    \begin{tikzpicture}
        \node at (0,0) {\includegraphics[width=6.4cm]{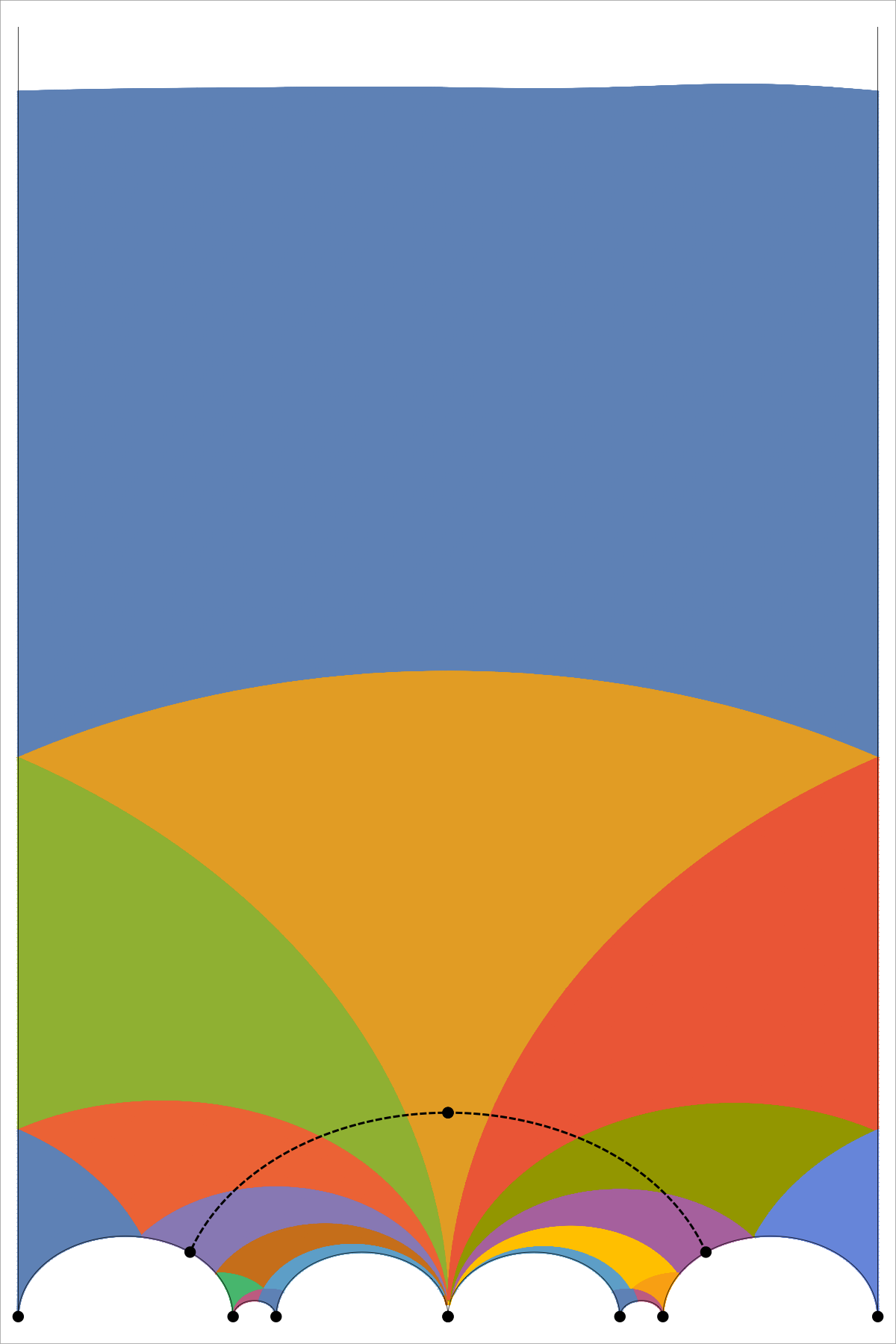}};
        \node at (-0.25,-4.55) {$c_0$};
        \node at (-0.85,-4.6) {$p_{45}$};
        \node at (-1.85,-4.6) {$p_{46}$};
        \node at (-1.95,-4.3) {$p_{47}$};
        \node at (-2.7,-4.55) {$p_{48}$};
        \node at (0,-2.9) {$p_{49}$};
        \node at (2.7,-4.55) {$p_{50}$};
        \node at (2.05,-4.3) {$p_{51}$};
        \node at (1.9,-4.6) {$p_{52}$};
        \node at (0.85,-4.6) {$p_{53}$};
    \end{tikzpicture}\\[1ex]
    \begin{tikzpicture}
        \node at (0,0) {\includegraphics[width=6.4cm]{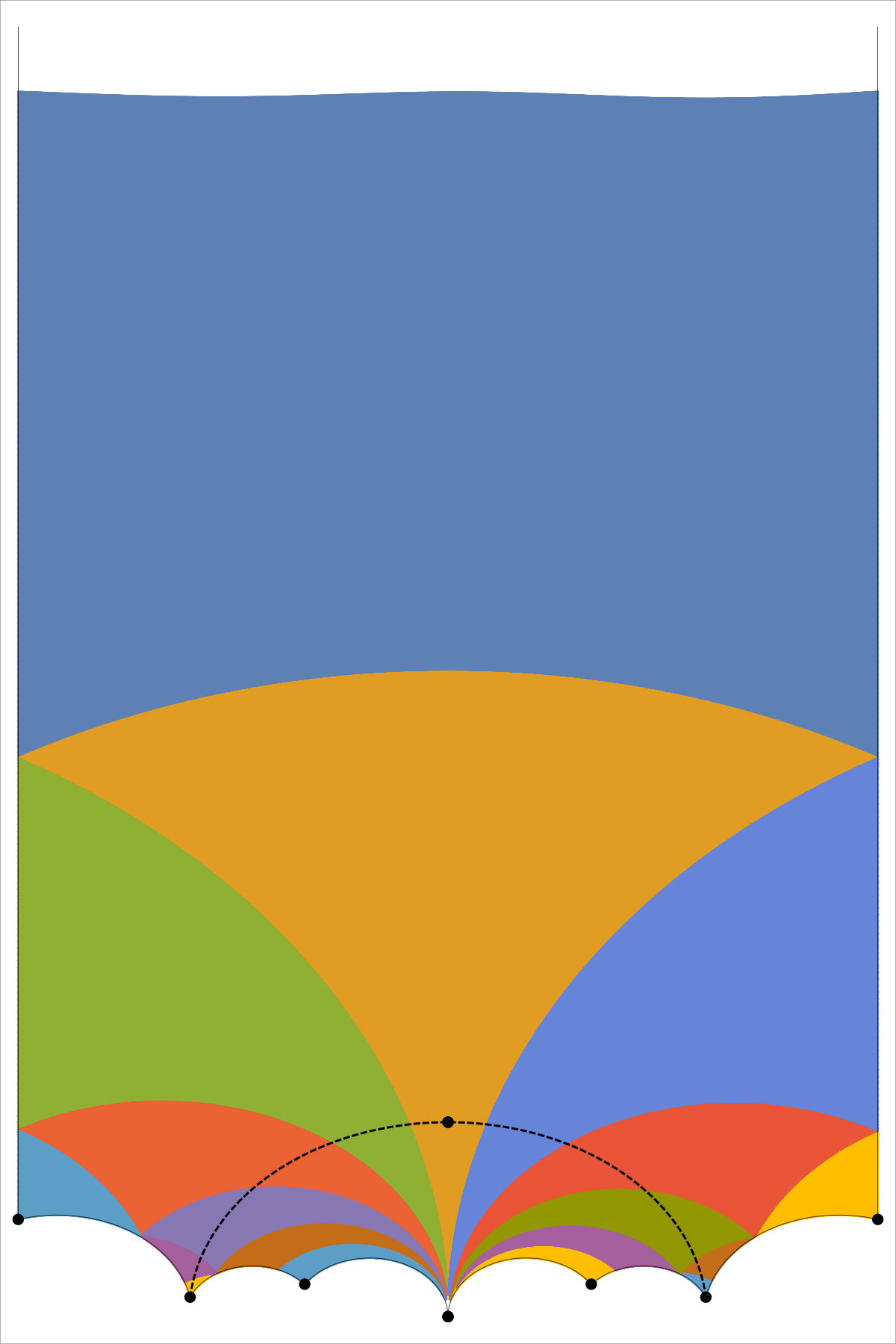}};
        \node at (-0.25,-4.55) {$c_0$};
        \node at (-0.9,-4.55) {$p_{54}$};
        \node at (-2.1,-4.55) {$p_{55}$};
        \node at (-2.8,-4.1) {$p_{56}$};
        \node at (0,-2.95) {$p_{57}$};
        \node at (2.85,-4.1) {$p_{58}$};
        \node at (2.15,-4.55) {$p_{59}$};
        \node at (0.95,-4.55) {$p_{60}$};
    \end{tikzpicture}
    \begin{tikzpicture}
        \node at (0,0) {\includegraphics[width=6.4cm]{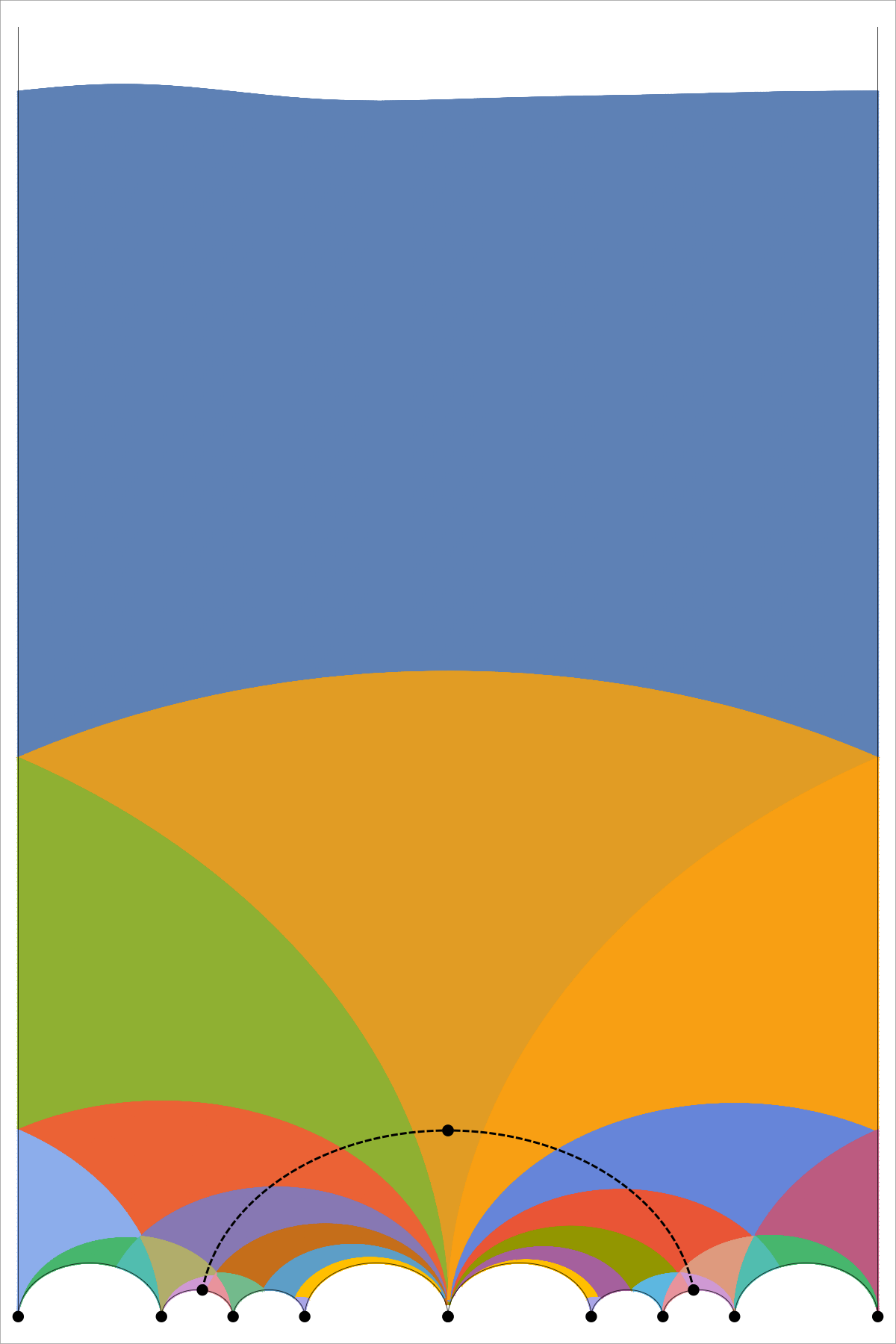}};
        \node at (-0.25,-4.55) {$c_0$};
        \node at (-0.85,-4.8) {$p_{61}$};
        \node at (-1.5,-4.8) {$p_{62}$};
        \node at (-1.8,-4.1) {$p_{63}$};
        \node at (-2.1,-4.8) {$p_{64}$};
        \node at (-2.8,-4.8) {$p_{65}$};
        \node at (0,-2.95) {$p_{66}$};
        \node at (2.9,-4.8) {$p_{67}$};
        \node at (2.1,-4.8) {$p_{68}$};
        \node at (1.8,-4.1) {$p_{69}$};
        \node at (1.5,-4.8) {$p_{70}$};
        \node at (0.85,-4.8) {$p_{71}$};
    \end{tikzpicture}
    \caption{Symmetric fundamental domains for $\Gamma_0(N)$ with $N=9,10,11,12$. The labelled points are listed in table~\ref{tab:specialPoints}.}
    \label{fig:gamma9101112Fundomain}
\end{figure}
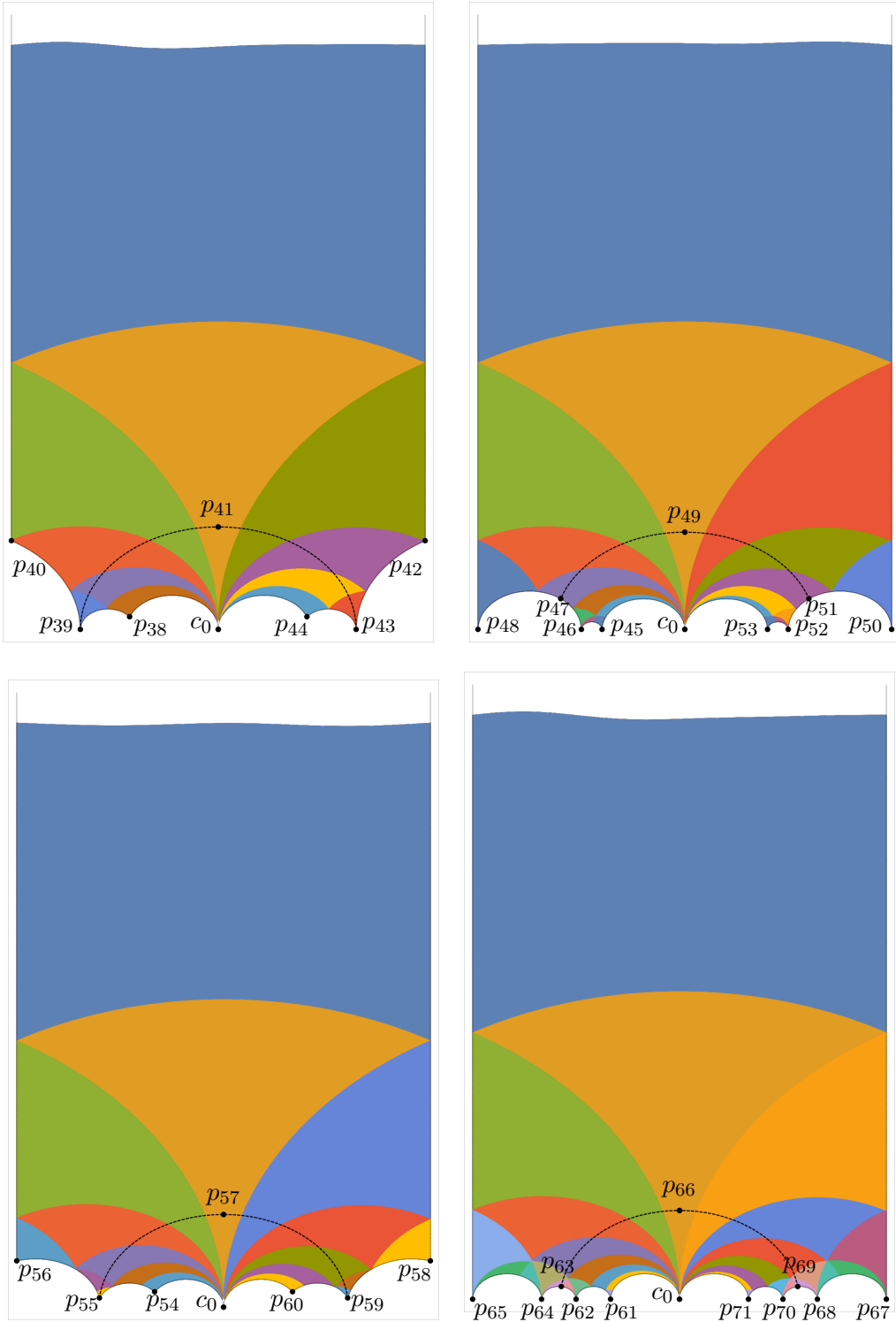
\begin{table}
\centering
\renewcommand{\arraystretch}{1.5}
    \begin{tabular}{lc}
    \toprule
    Point & Location\\
    \midrule
    $c_0$ & $0$\\
    $c_\infty$ & $i\infty$\\
    $p_1$ & $-\frac{1}{2}+\frac{i}{2}$\\
    $p_2$ & $\frac{i}{\sqrt{2}}$\\
    $p_3$ & $\frac{1}{2}+\frac{i}{2}$\\
    $p_4$ & $-\frac{1}{2}+\frac{i}{2\sqrt{3}}$\\
    $p_5$ & $\frac{i}{\sqrt{3}}$\\
    $p_6$ & $\frac{1}{2}+\frac{i}{2\sqrt{3}}$\\
    $p_7$ & $-\frac{1}{2}$\\
    $p_8$ & $\frac{i}{2}$\\
    $p_9$ & $\frac{1}{2}$\\
    $p_{10}$ & $-\frac{1}{3}+\frac{i}{3\sqrt{5}}$\\
    $p_{11}$ & $-\frac{2}{5}+\frac{i}{5}$\\
    $p_{12}$ & $-\frac{1}{2}+\frac{i\sqrt{5}}{10}$\\
    $p_{13}$ & $\frac{i}{\sqrt{5}}$\\
    $p_{14}$ & $\frac{1}{2}+\frac{i\sqrt{5}}{10}$\\
    $p_{15}$ & $\frac{2}{5}+\frac{i}{5}$\\
    $p_{16}$ & $\frac{1}{3}+\frac{i}{3\sqrt{5}}$\\
    $p_{17}$ & $-\frac13$\\
    $p_{18}$ & $-\frac{2}{5}+\frac{i}{5\sqrt{6}}$\\
    $p_{19}$ & $-\frac12$\\
    $p_{20}$ & $\frac{i}{\sqrt{6}}$\\
    $p_{21}$ & $\frac12$\\
    $p_{22}$ & $\frac{2}{5}+\frac{i}{5\sqrt{6}}$\\
    $p_{23}$ & $\frac13$\\
    \bottomrule
    \end{tabular}\hspace{1ex}
    \begin{tabular}{lc}
    \toprule
    Point & Location\\
    \midrule
    $p_{24}$ & $-\frac{1}{4}+\frac{i}{4\sqrt{7}}$\\
    $p_{25}$ & $-\frac{5}{14}+\frac{i\sqrt{3}}{14}$\\
    $p_{26}$ & $-\frac{1}{2}+\frac{i\sqrt{7}}{14}$\\
    $p_{27}$ & $\frac{i}{\sqrt{7}}$\\
    $p_{28}$ & $\frac{1}{2}+\frac{i\sqrt{7}}{14}$\\
    $p_{29}$ & $\frac{5}{14}+\frac{i\sqrt{3}}{14}$\\
    $p_{30}$ & $\frac{1}{4}+\frac{i}{4\sqrt{7}}$\\
    $p_{31}$ & $-\frac14$\\
    $p_{32}$ & $-\frac13+\frac{i}{6\sqrt{2}}$\\
    $p_{33}$ & $-\frac12$\\
    $p_{34}$ & $\frac{i}{\sqrt{8}}$\\
    $p_{35}$ & $\frac12$\\
    $p_{36}$ & $\frac13+\frac{i}{6\sqrt{2}}$\\
    $p_{37}$ & $\frac14$\\
    $p_{38}$ & $-\frac{3}{14}+\frac{i}{14\sqrt{3}}$\\
    $p_{39}$ & $-\frac13$\\
    $p_{40}$ & $-\frac12+\frac{i}{2\sqrt{3}}$\\
    $p_{41}$ & $\frac{i}{3}$\\
    $p_{42}$ & $\frac12+\frac{i}{2\sqrt{3}}$\\
    $p_{43}$ & $\frac13$\\
    $p_{44}$ & $\frac{3}{14}+\frac{i}{14\sqrt{3}}$\\
    $p_{45}$ & $-\frac15$\\
    $p_{46}$ & $-\frac14$\\
    $p_{47}$ & $-\frac{3}{10}+\frac{i}{10}$\\
    \bottomrule
    \end{tabular}\hspace{1ex}
    \begin{tabular}{lc}
    \toprule
    Point & Location\\
    \midrule
    $p_{48}$ & $-\frac12$\\
    $p_{49}$ & $\frac{i}{\sqrt{10}}$\\
    $p_{50}$ & $\frac12$\\
    $p_{51}$ & $\frac{3}{10}+\frac{i}{10}$\\
    $p_{52}$ & $\frac14$\\
    $p_{53}$ & $\frac15$\\
    $p_{54}$ & $-\frac{1}{6}+\frac{i}{6\sqrt{11}}$\\
    $p_{55}$ & $-\frac{3}{10}+\frac{i}{10\sqrt{11}}$\\
    $p_{56}$ & $-\frac12+\frac{i}{2\sqrt{11}}$\\
    $p_{57}$ & $\frac{i}{\sqrt{11}}$\\
    $p_{58}$ & $\frac12+\frac{i}{2\sqrt{11}}$\\
    $p_{59}$ & $\frac{3}{10}+\frac{i}{10\sqrt{11}}$\\
    $p_{60}$ & $\frac{1}{6}+\frac{i}{6\sqrt{11}}$\\
    $p_{61}$ & $-\frac{1}{6}$\\
    $p_{62}$ & $-\frac{1}{4}$\\
    $p_{63}$ & $-\frac{2}{7}+\frac{i}{14\sqrt{3}}$\\
    $p_{64}$ & $-\frac{1}{3}$\\
    $p_{65}$ & $-\frac{1}{2}$\\
    $p_{66}$ & $\frac{i}{\sqrt{12}}$\\
    $p_{67}$ & $\frac{1}{2}$\\
    $p_{68}$ & $\frac{1}{3}$\\
    $p_{69}$ & $\frac{2}{7}+\frac{i}{14\sqrt{3}}$\\
    $p_{70}$ & $\frac{1}{4}$\\
    $p_{71}$ & $\frac{1}{6}$\\
    \bottomrule
    \end{tabular}\hspace{1ex}
    \caption{Special points in the $\Gamma_0(N)$ fundamental domains.}
    \label{tab:specialPoints}
\end{table}

\newpage
\section{K3 Surfaces and Fibrations}
\label{app:K3SurfacesAndFibrations}

There are many articles, reviews and books discussing the geometry of K3 surfaces and the resulting physics of string compactifications, see for example~\cite{Huybrechts:2016uxh,debarre2020hyperkahler,Aspinwall:1996mn,Braun:2016sks,Enoki:2019deb}. In this paper we are concerned with complex analytic K3 surfaces, that is, simply connected, compact complex manifolds with a nowhere-vanishing $(2,0)$-form $\Omega$. When the K3 surface is furthermore projective, we can consider it as an algebraic variety. An algebraic surface is a K3 surface if $\omega_X=\mathcal{O}_X$ and $H^1(X,\mathcal{O}_X)=0$. All complex K3 surfaces are diffeomorphic and the second cohomology forms the lattice
\begin{equation}
	H^2(X,\mathbb{Z})\cong \Gamma^{3,19} = E_8(-1)\oplus E_8(-1) \oplus U \oplus U \oplus U\;,
\end{equation}
where $E_8(-1)$ is the $E_8$ lattice with negative definite inner product and $U$ is the hyperbolic lattice in two dimensions generated by $f,g$ with $e^2=f^2=0$ and $e\cdot f=1$. The signature of this lattice is $(3,19)$. By the global Torelli theorem, the moduli space of complex structures on a complex K3 surface is given by the space of periods of the holomorphic $(2,0)$-form $\Omega$. Specifying such a $(2,0)$-form $\Omega$ turns out to be the same as fixing a spacelike, oriented 2-plane in $\mathbb{R}^{3,19}=H^2(X,\mathbb{R})\supset H^2(X,\mathbb{Z})$. The space of such 2-planes is the Grassmannian $O^+(3,19)/(O(2)\times O(1,19))^+$. Hence, after modding out lattice isomorphisms, the complex structure moduli space of K3 surfaces is given by the double quotient
\begin{equation}
	\mathcal{M}= O^+(\Gamma^{3,19})\!\setminus\! O^+(3,19)/(O(2)\times O(1,19))^+\;.
\end{equation}
The action of $O^+(\Gamma^{3,19})$ is not properly discontinuous and the resulting complex 20-dimensional space is not Hausdorff.

An important role in the theory of K3 surfaces is played by the Picard group of isomorphism classes of complex analytic line bundles $L$ over $X$. For K3 surfaces, this is the same as the Néron-Severi group of divisors modulo algebraic equivalence. It embeds as a lattice into the K3 lattice
\begin{equation}
	\textup{NS}(X)\cong \textup{Pic}(X)\overset{c_1}{\cong}H^{1,1}(X,\mathbb{C})\cap H^2(X,\mathbb{Z})\hookrightarrow \Gamma^{3,19}\;.
\end{equation}
The rank $\rho$ of the Picard lattice jumps under complex structure deformations. A general complex analytic K3 surface has $\rho=0$, while the maximal Picard rank is $\rho=20$. The signature of the Picard lattice is $(1,\rho-1)$. The orthogonal complement of $\textup{Pic}(X)$ in $H^2(X,\mathbb{Z})$ is the transcendental lattice
\begin{equation}
	T(X)=\textup{Pic}(x)^\perp\subset H^2(X,\mathbb{Z})
\end{equation}

In the projective setting, it is natural to consider ample line bundles on $X$. A K3 surface together with a primitive ample line bundle $(X,L)$ is called a polarized K3 surface. The degree of $(X,L)$ is the degree of the line bundle $d=c_1(L)^2$. For smooth K3 surfaces, the degree is related to the genus by the formula $d=2-2g$. Fixing an element of the Picard group $[L]$ is the same as fixing a class $c_1(L)\in H^2(X,\mathbb{Z})$. The (coarse) moduli space of smooth degree $d$ K3-surfaces is an irreducible, quasi-projective, 19-dimensional complex variety. For a generic polarized K3 surface of degree $d=2n$ we have
\begin{equation}
	\textup{Pic}(X)=\langle +2n\rangle\;,\qquad\qquad T(X)=E_8(-1)\oplus E_8(-1)\oplus U \oplus U \oplus \langle -2n\rangle \;,
\end{equation}
where $\langle\pm 2n\rangle$ is the one-dimensional lattice with generator $e^2=\pm 2n$.

Fixing a primitive ample line-bundle can be viewed as fixing the embedding of a one-dimensional lattice, generated by $c_1(L)$, into $\textup{Pic}(X)$. Generalizing this idea leads to the notion of lattice polarized K3 surfaces. An M-polarized K3 surface is a pair $(X,j)$, where $X$ is a K3 surface, M is an even lattice of signature $(1,r-1)$ and $i:M\hookrightarrow \textup{Pic}(X)$ is a primitive lattice embedding such that $j(M)$ contains a pseudo-ample divisor class~\cite{Dolgachev:1996xw,Dolgachev:2005abc}.

The notion of a lattice polarized K3 appears naturally when we consider non-singular Calabi-Yau threefolds $\mathcal{M}$ fibered by K3 surfaces. That is, we have an algebraic fiber space structure
\begin{equation}
	\pi:\mathcal{M}\to \mathbb{P}^1\;,
\end{equation}
with generic fiber a $K3$ surface $X_t=\pi^{-1}(t)$, where $t\in \mathbb{P}^1$ is a generic point. We obtain a lattice embedding into $\textup{NS}(X_t)=\textup{Pic}(X_t)$ by restricting divisors of $\mathcal{M}$
\begin{equation}
	\big\langle n_\alpha D_\alpha\big|_{X_t}\big\rangle_{n_\alpha\in\mathbb{Z}} \lhook\joinrel\longrightarrow \textup{NS}(X_t)\cong\textup{Pic}(X_t)\;.
\end{equation}

An interesting class of lattice polarized K3 surfaces consists of those that are polarized by near-maximal rank lattices with $r=19$. These are extremely rigid and have only a one-dimensional space of complex structure deformations. In particular, we consider K3 surfaces with
\begin{equation}
	\textup{Pic}(X)=M_n=E_8(-1)\oplus E_8(-1)\oplus U \oplus \langle -2n \rangle\;,\qquad T(X)=U\oplus \langle +2n\rangle\;.
\end{equation}
These are known as $M_n$-polarized K3 surfaces and are mirror to degree $d=2n$ K3 surfaces as discussed above. The complex structure moduli space of $M_n$-polarized K3 surfaces is given by $X_0(n)^+$~\cite{Dolgachev:1996xw,Doran_2019}.

\section{Heterotic Duals}
\label{app:HetDuals}
In this appendix we speculate about possible heterotic duals for the type IIA compactifications discussed in the main text. We recall from section~\ref{subsec:11226} that type IIA on $\mathbb{P}^4_{11226}[12]$ is dual to the heterotic string on $K3\times T^2$, where the radius of the $T^2$ is frozen at the self-dual radius by introducing an $SU(2)^3$ instanton background $(10,10,4)\hookrightarrow E_8\times E_8\times SU(2)$. It is tempting to speculate that the reduced self-dual radius $T=-1/\sqrt{N}$ in the manifolds with a degree $2N$ K3 fibration could be associated with gauge enhancements at these radii. Reduced self-dual radii occur in heterotic compactifications with discrete Wilson lines~\cite{Narain:1985jj,Narain:1986am,Polchinski:1995df} and can lead to interesting gauge enhancements~\cite{Fraiman:2018ebo,Font:2020rsk,Font:2021uyw}.

Let us recall the computation of the heterotic spectrum from~\cite{Kachru:1995wm}. After first compactifying to 8D on the $T^2$ factor, the massless spectrum consists of the SUGRA and vector multiplets, so all charged fields are in the adjoint. Compactifying on K3, the gravitational sector contributes $N_h^\textup{grav}=20$ neutral hypermultiplets. An $SU(N)$ bundle $V$ over K3 with instanton number $\textstyle{\int}_{K3}c_2(V)=k$ gives another
\begin{equation}
    N_h^\textup{bundle}=kN-(N^2-1)\;.
\end{equation}
Additional charged hypermultiplets arise from decomposing the gauge group. Embedding a $G$ instanton into a simple factor $K$ of our gauge group leaves the commutant $G$ unbroken, where $G\times H\subset K$ is a maximal compact subgroup. Now we can decompose the adjoint representation of $K$ into tensor product representations
\begin{equation}
    \textup{adj}(K)=\sum_i\left(R^G_i,R^H_i\right)\;.
\end{equation}
The index theorem computes the net number of chiral spinors in the representation $R^H_i$ of the unbroken group $H$
\begin{equation}
    \label{eq:IndexFormula}
    N_h^{R^H_i}=\frac{1}{2}\textstyle{\int}_{K3}c_2(V)\textup{index}(R^G_i)-\textup{dim}(R^G_i)\;.
\end{equation}
Under favorable circumstances, one can use the resulting charged multiplets to Higgs the non-abelian part of the gauge group completely, corresponding to a dual with small $h^{11}$. By the usual rules of $N=2$ gauge theory, the number of neutral massless hypermultiplets after Higgsing is given by $N_h^\textup{Higgs}=\sum_iN_h^{R^H_i}\cdot \textup{dim}(R^{H_i})-\textup{dim}(H)$.

In the present case, the instanton numbers $(10,10,4)$ lead to
\begin{equation}
    N_h^\textup{grav}+N_h^\textup{bundle}=20+17+17+5=59\;.
\end{equation}
Since $E_7\times SU(2)\subset E_8$ is a maximal compact subgroup, we decompose
\begin{equation}
    \mathbf{248}\to (\mathbf{3},\mathbf{1}) \oplus (\mathbf{2},\mathbf{56}) \oplus (\mathbf{1},\mathbf{133})\;.
\end{equation}
This leads to
\begin{equation}
        N_h^{\mathbf{56}}=\frac{1}{2}\cdot 10\cdot 1-2=3\;.
\end{equation}
The spectrum allows for a complete Higgsing of the $E_7$ commutants, leading to an additonal
\begin{equation}
    N_h^\textup{Higgs}=2(3\cdot 56-133)=70\;.
\end{equation}
The $SU(2)$ factor is broken completely by the $SU(2)$ instanton background and gives no additional hypermultiplets. The final low energy spectrum contains
\begin{equation}
    N_h=N_h^\textup{grav}+N_h^\textup{bundle}+N_h^\textup{Higgs}=59+70=129
\end{equation}
neutral hypermultiplets. In addition, the $T^2$ contributes three vector multiplets. The spectrum $(N_v,N_h)=(3,129)$ is expected from a Calabi-Yau threefold with Hodge numbers $(h^{11},h^{21})=(N_v-1,N_h-1)=(2,128)$, which matches $\mathbb{P}^4_{11226}[12]$.

Let us now consider a heterotic vacuum on $T^2$ that has a gauge enhancement at the radius $T=1/\sqrt{2}$. Starting in 9D, from table 11 in~\cite{Font:2020rsk} we learn that one possible gauge group enhancement at $R_9=\tfrac{1}{\sqrt{2}}$ in 9D is
\begin{equation}
    \label{eq:LevelTwoHeteroticGaugeGroups}
        G=\left(E_7\times E_7\times SU(4)\right)/\mathbb{Z}_2\;,\qquad \vec{A}=\tfrac12\left(\vec{0}_6,-1,1;\vec{0}_6,-1,1\right)\;,
    \end{equation}
where $\vec{A}$ is the Wilson line background. Compactifying further to 8D, the gauge enhancement is expected to be preserved on the codimension one slice $2T=U$. For a rectangular torus with vanishing $B$-field, $T=iR_8R_9$ and $U=iR_8/R_9$, and the condition $T=2U$ indeed reduces to $R_9=\tfrac{1}{\sqrt{2}}$. The subgroup $SL(2,\mathbb{Z})_T\times SL(2,\mathbb{Z})_U$ of the heterotic T-duality group is broken on this slice to the level two Hecke congruence subgroup $\Gamma_0(2)_T$. Compared to our desired $\Gamma_0(2)_T^+$, this lacks the Fricke involution. We should not be too discouraged by this, as an enhancement of the naive T-duality group by Atkin-Lehner involutions has been observed in other contexts\cite{Persson:2015jka,Paquette:2016xoo,Paquette:2017xui,Persson:2017lkn}.

We proceed by looking for a suitable instanton embeddings into this gauge group, which should allow for a complete Higgsing. Let us first ignore the $\mathbb{Z}_2$ quotient and choose $SU(2)$ instanton numbers $(k_1,k_2,k_3)$ for this model. Under $E_7\to SO(12)\times SU(2)$ and $SU(4)\to SU(2)\times SU(2)\times U(1)$ we have
\begin{equation}
    \begin{gathered}
        \mathbf{133}\to (\mathbf{66},\mathbf{1})\oplus (\mathbf{32},\mathbf{2})\oplus(\mathbf{1},\mathbf{3})\;,\\
        \mathbf{15}\to(\mathbf{3},\mathbf{1})_0\oplus(\mathbf{2},\bar{\mathbf{2}})_{+2}\oplus(\bar{\mathbf{2}},\mathbf{2})_{-2}\oplus(\mathbf{1},\mathbf{3})_0\oplus(\mathbf{1},\mathbf{1})_0\;,
    \end{gathered}
\end{equation}
The K3 and bundle moduli add up to $59$ neutral hypermultiplets. From the index formula~\eqref{eq:IndexFormula} we obtain $k_1-4$ and $k_2-4$ half-hypermultiplets in the $\mathbf{12}$ of the respective $SO(32)$ factors. Assuming this spectrum allows for a complete Higgsing, the contribution from this sector is an additional $16(k_1+k_2)-260$ neutral hypermultiplets. Analogously, the Higgsing of the $SU(4)$-factor gives rise to $2k_3-10$ neutral hypermultiplets. Altogether, we obtain the spectrum
\begin{equation}
    N_V=3\;,\qquad N_H=171-14k_3\;.
\end{equation}
For $(k_1,k_2,k_3)=(9,9,6)$ we obtain the spectrum $(N_V,N_H)=(3,87)$, which is precisely the spectrum of type IIA string theory on a Calabi-Yau threefold with Hodge numbers $(h^{11},h^{21})=(2,86)$ such as $\mathbb{P}^4_{11222}[8]$.

While the matching of the spectrum is remarkable, some comments are in place. Normally, an $SU(N)$ factor requires instanton number $k=2N$ for complete Higgsing, but we have only $N+2$. We might hope that this problem is an artifact of ignoring the global structure of the gauge group, which includes a $\mathbb{Z}_2$ quotient. The global structure of $G$ is obviously important when considering $G$ bundles on K3. As a toy example, consider the group $SU(4)$ on its own. A complete Higgsing would require $k=8$ instantons. If the global structure of the gauge group were actually $SU(4)/\mathbb{Z}_2=SO(6)$, it would suffice to have $k=6$ instantons for complete breaking.

Similar considerations apply to the manifold $\mathbb{P}^5_{1^2 2^4}[4,6]$ from section~\ref{subsec:112222}. The corresponding gauge group and Wilson line are
\begin{equation}
        G=\left(E_6\times E_6\times SU(6)\right)/\mathbb{Z}_3\;,\qquad  \vec{A}=\tfrac13\left(\vec{0}_5,-1,-1,2;\vec{0}_5,-1,-1,2\right)\;,
\end{equation}
and the instanton embedding is $(8,8,8)$.

\bibliographystyle{JHEP}
\bibliography{ModularCurvesRDC}  

\providecommand{\href}[2]{#2}\begingroup\raggedright\begin{thebibliography}{100}

\bibitem{Vafa:2005ui}
C.~Vafa, \emph{{The String landscape and the swampland}},
  \href{https://arxiv.org/abs/hep-th/0509212}{{\ttfamily hep-th/0509212}}.

\bibitem{Ooguri:2006in}
H.~Ooguri and C.~Vafa, \emph{{On the Geometry of the String Landscape and the
  Swampland}},
  \href{https://doi.org/10.1016/j.nuclphysb.2006.10.033}{\emph{Nucl. Phys. B}
  {\bfseries 766} (2007) 21}
  [\href{https://arxiv.org/abs/hep-th/0605264}{{\ttfamily hep-th/0605264}}].

\bibitem{Brennan:2017rbf}
T.~D. Brennan, F.~Carta and C.~Vafa, \emph{{The String Landscape, the
  Swampland, and the Missing Corner}},
  \href{https://doi.org/10.22323/1.305.0015}{\emph{PoS} {\bfseries TASI2017}
  (2017) 015} [\href{https://arxiv.org/abs/1711.00864}{{\ttfamily
  1711.00864}}].

\bibitem{Palti:2019pca}
E.~Palti, \emph{{The Swampland: Introduction and Review}},
  \href{https://doi.org/10.1002/prop.201900037}{\emph{Fortsch. Phys.}
  {\bfseries 67} (2019) 1900037}
  [\href{https://arxiv.org/abs/1903.06239}{{\ttfamily 1903.06239}}].

\bibitem{vanBeest:2021lhn}
M.~van Beest, J.~Calder\'on-Infante, D.~Mirfendereski and I.~Valenzuela,
  \emph{{Lectures on the Swampland Program in String Compactifications}},
  \href{https://arxiv.org/abs/2102.01111}{{\ttfamily 2102.01111}}.

\bibitem{Grana:2021zvf}
M.~Gra\~na and A.~Herr\'aez, \emph{{The Swampland Conjectures: A bridge from
  Quantum Gravity to Particle Physics}},
  \href{https://arxiv.org/abs/2107.00087}{{\ttfamily 2107.00087}}.

\bibitem{Banks:1988yz}
T.~Banks and L.~J. Dixon, \emph{{Constraints on String Vacua with Space-Time
  Supersymmetry}},
  \href{https://doi.org/10.1016/0550-3213(88)90523-8}{\emph{Nucl. Phys. B}
  {\bfseries 307} (1988) 93}.

\bibitem{Kallosh:1995hi}
R.~Kallosh, A.~D. Linde, D.~A. Linde and L.~Susskind, \emph{{Gravity and Global
  Symmetries}}, \href{https://doi.org/10.1103/PhysRevD.52.912}{\emph{Phys. Rev.
  D} {\bfseries 52} (1995) 912}
  [\href{https://arxiv.org/abs/hep-th/9502069}{{\ttfamily hep-th/9502069}}].

\bibitem{Banks:2010zn}
T.~Banks and N.~Seiberg, \emph{{Symmetries and Strings in Field Theory and
  Gravity}}, \href{https://doi.org/10.1103/PhysRevD.83.084019}{\emph{Phys. Rev.
  D} {\bfseries 83} (2011) 084019}
  [\href{https://arxiv.org/abs/1011.5120}{{\ttfamily 1011.5120}}].

\bibitem{Harlow:2018tng}
D.~Harlow and H.~Ooguri, \emph{{Symmetries in quantum field theory and quantum
  gravity}}, \href{https://doi.org/10.1007/s00220-021-04040-y}{\emph{Commun.
  Math. Phys.} {\bfseries 383} (2021) 1669}
  [\href{https://arxiv.org/abs/1810.05338}{{\ttfamily 1810.05338}}].

\bibitem{Harlow:2020bee}
D.~Harlow and E.~Shaghoulian, \emph{{Global symmetry, Euclidean gravity, and
  the black hole information problem}},
  \href{https://doi.org/10.1007/JHEP04(2021)175}{\emph{JHEP} {\bfseries 04}
  (2021) 175} [\href{https://arxiv.org/abs/2010.10539}{{\ttfamily
  2010.10539}}].

\bibitem{Chen:2020ojn}
Y.~Chen and H.~W. Lin, \emph{{Signatures of global symmetry violation in
  relative entropies and replica wormholes}},
  \href{https://doi.org/10.1007/JHEP03(2021)040}{\emph{JHEP} {\bfseries 03}
  (2021) 040} [\href{https://arxiv.org/abs/2011.06005}{{\ttfamily
  2011.06005}}].

\bibitem{Heidenreich:2020pkc}
B.~Heidenreich, J.~McNamara, M.~Montero, M.~Reece, T.~Rudelius and
  I.~Valenzuela, \emph{{Chern-Weil Global Symmetries and How Quantum Gravity
  Avoids Them}},  \href{https://arxiv.org/abs/2012.00009}{{\ttfamily
  2012.00009}}.

\bibitem{Rudelius:2020orz}
T.~Rudelius and S.-H. Shao, \emph{{Topological Operators and Completeness of
  Spectrum in Discrete Gauge Theories}},
  \href{https://doi.org/10.1007/JHEP12(2020)172}{\emph{JHEP} {\bfseries 12}
  (2020) 172} [\href{https://arxiv.org/abs/2006.10052}{{\ttfamily
  2006.10052}}].

\bibitem{Heidenreich:2021tna}
B.~Heidenreich, J.~McNamara, M.~Montero, M.~Reece, T.~Rudelius and
  I.~Valenzuela, \emph{{Non-Invertible Global Symmetries and Completeness of
  the Spectrum}},  \href{https://arxiv.org/abs/2104.07036}{{\ttfamily
  2104.07036}}.

\bibitem{Polchinski:2003bq}
J.~Polchinski, \emph{{Monopoles, Duality, and String Theory}},
  \href{https://doi.org/10.1142/S0217751X0401866X}{\emph{Int. J. Mod. Phys. A}
  {\bfseries 19S1} (2004) 145}
  [\href{https://arxiv.org/abs/hep-th/0304042}{{\ttfamily hep-th/0304042}}].

\bibitem{McNamara:2019rup}
J.~McNamara and C.~Vafa, \emph{{Cobordism Classes and the Swampland}},
  \href{https://arxiv.org/abs/1909.10355}{{\ttfamily 1909.10355}}.

\bibitem{Cvetic:2020kuw}
M.~Cveti\v{c}, M.~Dierigl, L.~Lin and H.~Y. Zhang, \emph{{String Universality
  and Non-Simply-Connected Gauge Groups in 8d}},
  \href{https://doi.org/10.1103/PhysRevLett.125.211602}{\emph{Phys. Rev. Lett.}
  {\bfseries 125} (2020) 211602}
  [\href{https://arxiv.org/abs/2008.10605}{{\ttfamily 2008.10605}}].

\bibitem{Montero:2020icj}
M.~Montero and C.~Vafa, \emph{{Cobordism Conjecture, Anomalies, and the String
  Lamppost Principle}},
  \href{https://doi.org/10.1007/JHEP01(2021)063}{\emph{JHEP} {\bfseries 01}
  (2021) 063} [\href{https://arxiv.org/abs/2008.11729}{{\ttfamily
  2008.11729}}].

\bibitem{Dierigl:2020lai}
M.~Dierigl and J.~J. Heckman, \emph{{Swampland Cobordism Conjecture and
  Non-Abelian Duality Groups}},
  \href{https://doi.org/10.1103/PhysRevD.103.066006}{\emph{Phys. Rev. D}
  {\bfseries 103} (2021) 066006}
  [\href{https://arxiv.org/abs/2012.00013}{{\ttfamily 2012.00013}}].

\bibitem{Hamada:2021bbz}
Y.~Hamada and C.~Vafa, \emph{{8d Supergravity, Reconstruction of Internal
  Geometry and the Swampland}},
  \href{https://doi.org/10.1007/JHEP06(2021)178}{\emph{JHEP} {\bfseries 06}
  (2021) 178} [\href{https://arxiv.org/abs/2104.05724}{{\ttfamily
  2104.05724}}].

\bibitem{Arkani-Hamed:2006emk}
N.~Arkani-Hamed, L.~Motl, A.~Nicolis and C.~Vafa, \emph{{The String Landscape,
  Black Holes and Gravity as the Weakest Force}},
  \href{https://doi.org/10.1088/1126-6708/2007/06/060}{\emph{JHEP} {\bfseries
  06} (2007) 060} [\href{https://arxiv.org/abs/hep-th/0601001}{{\ttfamily
  hep-th/0601001}}].

\bibitem{Ibanez:2017kvh}
L.~E. Ibáñez, V.~Martín-Lozano and I.~Valenzuela, \emph{{Constraining
  Neutrino Masses, the Cosmological Constant and BSM Physics from the Weak
  Gravity Conjecture}},
  \href{https://doi.org/10.1007/JHEP11(2017)066}{\emph{JHEP} {\bfseries 11}
  (2017) 066} [\href{https://arxiv.org/abs/1706.05392}{{\ttfamily
  1706.05392}}].

\bibitem{Ibanez:2017oqr}
L.~E. Ibáñez, V.~Martín-Lozano and I.~Valenzuela, \emph{{Constraining the EW
  Hierarchy from the Weak Gravity Conjecture}},
  \href{https://arxiv.org/abs/1707.05811}{{\ttfamily 1707.05811}}.

\bibitem{Gonzalo:2018tpb}
E.~Gonzalo, A.~Herr\'aez and L.~E. Ib\'a\~nez, \emph{{AdS-phobia, the WGC, the
  Standard Model and Supersymmetry}},
  \href{https://doi.org/10.1007/JHEP06(2018)051}{\emph{JHEP} {\bfseries 06}
  (2018) 051} [\href{https://arxiv.org/abs/1803.08455}{{\ttfamily
  1803.08455}}].

\bibitem{Hamada:2017yji}
Y.~Hamada and G.~Shiu, \emph{{Weak Gravity Conjecture, Multiple Point Principle
  and the Standard Model Landscape}},
  \href{https://doi.org/10.1007/JHEP11(2017)043}{\emph{JHEP} {\bfseries 11}
  (2017) 043} [\href{https://arxiv.org/abs/1707.06326}{{\ttfamily
  1707.06326}}].

\bibitem{Reece:2018zvv}
M.~Reece, \emph{{Photon Masses in the Landscape and the Swampland}},
  \href{https://doi.org/10.1007/JHEP07(2019)181}{\emph{JHEP} {\bfseries 07}
  (2019) 181} [\href{https://arxiv.org/abs/1808.09966}{{\ttfamily
  1808.09966}}].

\bibitem{Montero:2019ekk}
M.~Montero, T.~Van~Riet and G.~Venken, \emph{{Festina Lente: EFT Constraints
  from Charged Black Hole Evaporation in de Sitter}},
  \href{https://doi.org/10.1007/JHEP01(2020)039}{\emph{JHEP} {\bfseries 01}
  (2020) 039} [\href{https://arxiv.org/abs/1910.01648}{{\ttfamily
  1910.01648}}].

\bibitem{Montero:2021otb}
M.~Montero, C.~Vafa, T.~Van~Riet and G.~Venken, \emph{{The FL bound and its
  phenomenological implications}},
  \href{https://arxiv.org/abs/2106.07650}{{\ttfamily 2106.07650}}.

\bibitem{Heidenreich:2015nta}
B.~Heidenreich, M.~Reece and T.~Rudelius, \emph{{Sharpening the Weak Gravity
  Conjecture with Dimensional Reduction}},
  \href{https://doi.org/10.1007/JHEP02(2016)140}{\emph{JHEP} {\bfseries 02}
  (2016) 140} [\href{https://arxiv.org/abs/1509.06374}{{\ttfamily
  1509.06374}}].

\bibitem{Heidenreich:2016aqi}
B.~Heidenreich, M.~Reece and T.~Rudelius, \emph{{Evidence for a sublattice weak
  gravity conjecture}},
  \href{https://doi.org/10.1007/JHEP08(2017)025}{\emph{JHEP} {\bfseries 08}
  (2017) 025} [\href{https://arxiv.org/abs/1606.08437}{{\ttfamily
  1606.08437}}].

\bibitem{Andriolo:2018lvp}
S.~Andriolo, D.~Junghans, T.~Noumi and G.~Shiu, \emph{{A Tower Weak Gravity
  Conjecture from Infrared Consistency}},
  \href{https://doi.org/10.1002/prop.201800020}{\emph{Fortsch. Phys.}
  {\bfseries 66} (2018) 1800020}
  [\href{https://arxiv.org/abs/1802.04287}{{\ttfamily 1802.04287}}].

\bibitem{Lee:2018urn}
S.-J. Lee, W.~Lerche and T.~Weigand, \emph{{Tensionless Strings and the Weak
  Gravity Conjecture}},
  \href{https://doi.org/10.1007/JHEP10(2018)164}{\emph{JHEP} {\bfseries 10}
  (2018) 164} [\href{https://arxiv.org/abs/1808.05958}{{\ttfamily
  1808.05958}}].

\bibitem{Lee:2018spm}
S.-J. Lee, W.~Lerche and T.~Weigand, \emph{{A Stringy Test of the Scalar Weak
  Gravity Conjecture}},
  \href{https://doi.org/10.1016/j.nuclphysb.2018.11.001}{\emph{Nucl. Phys. B}
  {\bfseries 938} (2019) 321}
  [\href{https://arxiv.org/abs/1810.05169}{{\ttfamily 1810.05169}}].

\bibitem{Lee:2019tst}
S.-J. Lee, W.~Lerche and T.~Weigand, \emph{{Modular Fluxes, Elliptic Genera,
  and Weak Gravity Conjectures in Four Dimensions}},
  \href{https://doi.org/10.1007/JHEP08(2019)104}{\emph{JHEP} {\bfseries 08}
  (2019) 104} [\href{https://arxiv.org/abs/1901.08065}{{\ttfamily
  1901.08065}}].

\bibitem{Lee:2019xtm}
S.-J. Lee, W.~Lerche and T.~Weigand, \emph{{Emergent Strings, Duality and Weak
  Coupling Limits for Two-Form Fields}},
  \href{https://arxiv.org/abs/1904.06344}{{\ttfamily 1904.06344}}.

\bibitem{Lee:2019wij}
S.-J. Lee, W.~Lerche and T.~Weigand, \emph{{Emergent Strings from Infinite
  Distance Limits}},  \href{https://arxiv.org/abs/1910.01135}{{\ttfamily
  1910.01135}}.

\bibitem{Klaewer:2020lfg}
D.~Kl{\"a}wer, S.-J. Lee, T.~Weigand and M.~Wiesner, \emph{{Quantum Corrections
  in 4d N=1 Infinite Distance Limits and the Weak Gravity Conjecture}},
  \href{https://arxiv.org/abs/2011.00024}{{\ttfamily 2011.00024}}.

\bibitem{Klaewer:2016kiy}
D.~Kl{\"a}wer and E.~Palti, \emph{{Super-Planckian Spatial Field Variations and
  Quantum Gravity}}, \href{https://doi.org/10.1007/JHEP01(2017)088}{\emph{JHEP}
  {\bfseries 01} (2017) 088}
  [\href{https://arxiv.org/abs/1610.00010}{{\ttfamily 1610.00010}}].

\bibitem{Lanza:2020qmt}
S.~Lanza, F.~Marchesano, L.~Martucci and I.~Valenzuela, \emph{{Swampland
  Conjectures for Strings and Membranes}},
  \href{https://doi.org/10.1007/JHEP02(2021)006}{\emph{JHEP} {\bfseries 02}
  (2021) 006} [\href{https://arxiv.org/abs/2006.15154}{{\ttfamily
  2006.15154}}].

\bibitem{Lanza:2021qsu}
S.~Lanza, F.~Marchesano, L.~Martucci and I.~Valenzuela, \emph{{The EFT stringy
  viewpoint on large distances}},
  \href{https://arxiv.org/abs/2104.05726}{{\ttfamily 2104.05726}}.

\bibitem{Grimm:2018ohb}
T.~W. Grimm, E.~Palti and I.~Valenzuela, \emph{{Infinite Distances in Field
  Space and Massless Towers of States}},
  \href{https://doi.org/10.1007/JHEP08(2018)143}{\emph{JHEP} {\bfseries 08}
  (2018) 143} [\href{https://arxiv.org/abs/1802.08264}{{\ttfamily
  1802.08264}}].

\bibitem{Blumenhagen:2018nts}
R.~Blumenhagen, D.~Kl\"awer, L.~Schlechter and F.~Wolf, \emph{{The Refined
  Swampland Distance Conjecture in Calabi-Yau Moduli Spaces}},
  \href{https://doi.org/10.1007/JHEP06(2018)052}{\emph{JHEP} {\bfseries 06}
  (2018) 052} [\href{https://arxiv.org/abs/1803.04989}{{\ttfamily
  1803.04989}}].

\bibitem{Erkinger:2019umg}
D.~Erkinger and J.~Knapp, \emph{{Refined swampland distance conjecture and
  exotic hybrid Calabi-Yaus}},
  \href{https://doi.org/10.1007/JHEP07(2019)029}{\emph{JHEP} {\bfseries 07}
  (2019) 029} [\href{https://arxiv.org/abs/1905.05225}{{\ttfamily
  1905.05225}}].

\bibitem{Joshi:2019nzi}
A.~Joshi and A.~Klemm, \emph{{Swampland Distance Conjecture for One-Parameter
  Calabi-Yau Threefolds}},
  \href{https://doi.org/10.1007/JHEP08(2019)086}{\emph{JHEP} {\bfseries 08}
  (2019) 086} [\href{https://arxiv.org/abs/1903.00596}{{\ttfamily
  1903.00596}}].

\bibitem{Bedroya:2019snp}
A.~Bedroya and C.~Vafa, \emph{{Trans-Planckian Censorship and the Swampland}},
  \href{https://doi.org/10.1007/JHEP09(2020)123}{\emph{JHEP} {\bfseries 09}
  (2020) 123} [\href{https://arxiv.org/abs/1909.11063}{{\ttfamily
  1909.11063}}].

\bibitem{EnriquezRojo:2020hzi}
M.~Enr\'\i{}quez~Rojo and E.~Plauschinn, \emph{{Swampland conjectures for type
  IIB orientifolds with closed-string U(1)s}},
  \href{https://doi.org/10.1007/JHEP07(2020)026}{\emph{JHEP} {\bfseries 07}
  (2020) 026} [\href{https://arxiv.org/abs/2002.04050}{{\ttfamily
  2002.04050}}].

\bibitem{Andriot:2020lea}
D.~Andriot, N.~Cribiori and D.~Erkinger, \emph{{The web of swampland
  conjectures and the TCC bound}},
  \href{https://doi.org/10.1007/JHEP07(2020)162}{\emph{JHEP} {\bfseries 07}
  (2020) 162} [\href{https://arxiv.org/abs/2004.00030}{{\ttfamily
  2004.00030}}].

\bibitem{Baume:2020dqd}
F.~Baume and J.~Calder\'on~Infante, \emph{{Tackling the SDC in AdS with CFTs}},
  \href{https://doi.org/10.1007/JHEP08(2021)057}{\emph{JHEP} {\bfseries 08}
  (2021) 057} [\href{https://arxiv.org/abs/2011.03583}{{\ttfamily
  2011.03583}}].

\bibitem{Perlmutter:2020buo}
E.~Perlmutter, L.~Rastelli, C.~Vafa and I.~Valenzuela, \emph{{A CFT Distance
  Conjecture}},  \href{https://arxiv.org/abs/2011.10040}{{\ttfamily
  2011.10040}}.

\bibitem{Calderon-Infante:2020dhm}
J.~Calder\'on-Infante, A.~M. Uranga and I.~Valenzuela, \emph{{The Convex Hull
  Swampland Distance Conjecture and Bounds on Non-geodesics}},
  \href{https://doi.org/10.1007/JHEP03(2021)299}{\emph{JHEP} {\bfseries 03}
  (2021) 299} [\href{https://arxiv.org/abs/2012.00034}{{\ttfamily
  2012.00034}}].

\bibitem{Demirtas:2020dbm}
M.~Demirtas, L.~McAllister and A.~Rios-Tascon, \emph{{Bounding the
  Kreuzer-Skarke Landscape}},
  \href{https://arxiv.org/abs/2008.01730}{{\ttfamily 2008.01730}}.

\bibitem{Baume:2016psm}
F.~Baume and E.~Palti, \emph{{Backreacted Axion Field Ranges in String
  Theory}}, \href{https://doi.org/10.1007/JHEP08(2016)043}{\emph{JHEP}
  {\bfseries 08} (2016) 043}
  [\href{https://arxiv.org/abs/1602.06517}{{\ttfamily 1602.06517}}].

\bibitem{Blumenhagen:2017cxt}
R.~Blumenhagen, I.~Valenzuela and F.~Wolf, \emph{{The Swampland Conjecture and
  F-term Axion Monodromy Inflation}},
  \href{https://doi.org/10.1007/JHEP07(2017)145}{\emph{JHEP} {\bfseries 07}
  (2017) 145} [\href{https://arxiv.org/abs/1703.05776}{{\ttfamily
  1703.05776}}].

\bibitem{Brodie:2021ain}
C.~R. Brodie, A.~Constantin, A.~Lukas and F.~Ruehle, \emph{{Swampland
  Conjectures and Infinite Flop Chains}},
  \href{https://arxiv.org/abs/2104.03325}{{\ttfamily 2104.03325}}.

\bibitem{Klawer:2019czy}
D.~Kläwer, \emph{{\href{https://doi.org/10.5282/edoc.24827}{Swampland
  Conjectures as Generic Predictions of Quantum Gravity}}}, Ph.D. thesis,
  Munich U., 2019.

\bibitem{Hebecker:2017lxm}
A.~Hebecker, P.~Henkenjohann and L.~T. Witkowski, \emph{{Flat Monodromies and a
  Moduli Space Size Conjecture}},
  \href{https://doi.org/10.1007/JHEP12(2017)033}{\emph{JHEP} {\bfseries 12}
  (2017) 033} [\href{https://arxiv.org/abs/1708.06761}{{\ttfamily
  1708.06761}}].

\bibitem{Conway:1979qga}
J.~H. Conway and S.~P. Norton, \emph{{Monstrous Moonshine}},
  \href{https://doi.org/10.1112/blms/11.3.308}{\emph{Bull. London Math. Soc.}
  {\bfseries 11} (1979) 308}.

\bibitem{Knapp:2021vkm}
J.~Knapp, E.~Scheidegger and T.~Schimannek, \emph{{On genus one fibered
  Calabi-Yau threefolds with 5-sections}},
  \href{https://arxiv.org/abs/2107.05647}{{\ttfamily 2107.05647}}.

\bibitem{Kimura:2016crs}
Y.~Kimura, \emph{{Discrete Gauge Groups in F-theory Models on Genus-One Fibered
  Calabi-Yau 4-folds without Section}},
  \href{https://doi.org/10.1007/JHEP04(2017)168}{\emph{JHEP} {\bfseries 04}
  (2017) 168} [\href{https://arxiv.org/abs/1608.07219}{{\ttfamily
  1608.07219}}].

\bibitem{Kimura:2019bzv}
Y.~Kimura, \emph{{Discrete gauge groups in certain F-theory models in six
  dimensions}}, \href{https://doi.org/10.1007/JHEP07(2019)027}{\emph{JHEP}
  {\bfseries 07} (2019) 027}
  [\href{https://arxiv.org/abs/1905.03775}{{\ttfamily 1905.03775}}].

\bibitem{Klemm:1995tj}
A.~Klemm, W.~Lerche and P.~Mayr, \emph{{K3 Fibrations and Heterotic-Type II
  String Duality}},
  \href{https://doi.org/10.1016/0370-2693(95)00937-G}{\emph{Phys. Lett. B}
  {\bfseries 357} (1995) 313}
  [\href{https://arxiv.org/abs/hep-th/9506112}{{\ttfamily hep-th/9506112}}].

\bibitem{Candelas:1993dm}
P.~Candelas, X.~De~La~Ossa, A.~Font, S.~H. Katz and D.~R. Morrison,
  \emph{{Mirror Symmetry for Two Parameter Models - 1}},
  \href{https://doi.org/10.1016/0550-3213(94)90322-0}{\emph{Nucl. Phys. B}
  {\bfseries 416} (1994) 481}
  [\href{https://arxiv.org/abs/hep-th/9308083}{{\ttfamily hep-th/9308083}}].

\bibitem{Kachru:1995wm}
S.~Kachru and C.~Vafa, \emph{{Exact Tesults for N=2 Compactifications of
  Heterotic Strings}},
  \href{https://doi.org/10.1016/0550-3213(95)00307-E}{\emph{Nucl. Phys. B}
  {\bfseries 450} (1995) 69}
  [\href{https://arxiv.org/abs/hep-th/9505105}{{\ttfamily hep-th/9505105}}].

\bibitem{Kachru:1995fv}
S.~Kachru, A.~Klemm, W.~Lerche, P.~Mayr and C.~Vafa, \emph{{Non-perturbative
  Results on the Point Particle Limit of N=2 Heterotic String
  Compactifications}},
  \href{https://doi.org/10.1016/0550-3213(95)00574-9}{\emph{Nucl. Phys. B}
  {\bfseries 459} (1996) 537}
  [\href{https://arxiv.org/abs/hep-th/9508155}{{\ttfamily hep-th/9508155}}].

\bibitem{Hosono:1994ax}
S.~Hosono, A.~Klemm, S.~Theisen and S.-T. Yau, \emph{{Mirror Symmetry, Mirror
  Map and Applications to Complete Intersection Calabi-Yau Spaces}},
  \href{https://doi.org/10.1016/0550-3213(94)00440-P}{\emph{Nucl. Phys. B}
  {\bfseries 433} (1995) 501}
  [\href{https://arxiv.org/abs/hep-th/9406055}{{\ttfamily hep-th/9406055}}].

\bibitem{Hori:2003ic}
K.~Hori, S.~Katz, A.~Klemm, R.~Pandharipande, R.~Thomas, C.~Vafa et~al.,
  \emph{{Mirror symmetry}}, vol.~1 of \emph{Clay mathematics monographs}. AMS,
  Providence, USA, 2003.

\bibitem{BayerTravesa:2007}
P.~Bayer and A.~Travesa, \emph{Uniformization of triangle modular curves},
  \href{https://doi.org/10.5565/PUBLMAT_PJTN05_03}{\emph{Publicacions
  matem{\`a}tiques} {\bfseries Extra} (2007) 43}.

\bibitem{Schimmrigk:1989tz}
R.~Schimmrigk, \emph{{Heterotic {RG} Flow Fixed Points With Nondiagonal Affine
  Invariants}}, \href{https://doi.org/10.1016/0370-2693(89)91162-3}{\emph{Phys.
  Lett. B} {\bfseries 229} (1989) 227}.

\bibitem{Lian:1995js}
B.~H. Lian and S.-T. Yau, \emph{{Mirror Maps, Modular Relations and
  Hypergeometric Series 1}},
  \href{https://arxiv.org/abs/hep-th/9507151}{{\ttfamily hep-th/9507151}}.

\bibitem{Harnad:1998hh}
J.~Harnad and J.~McKay, \emph{{Modular Solutions to Equations of Generalized
  Halphen Type}}, \href{https://doi.org/10.1098/rspa.2000.0517}{\emph{Proc.
  Roy. Soc. Lond. A} {\bfseries 456} (2000) 261}
  [\href{https://arxiv.org/abs/solv-int/9804006}{{\ttfamily
  solv-int/9804006}}].

\bibitem{Anderson:2017aux}
L.~B. Anderson, X.~Gao, J.~Gray and S.-J. Lee, \emph{{Fibrations in CICY
  Threefolds}}, \href{https://doi.org/10.1007/JHEP10(2017)077}{\emph{JHEP}
  {\bfseries 10} (2017) 077}
  [\href{https://arxiv.org/abs/1708.07907}{{\ttfamily 1708.07907}}].

\bibitem{Klemm:2004km}
A.~Klemm, M.~Kreuzer, E.~Riegler and E.~Scheidegger, \emph{{Topological string
  amplitudes, complete intersection Calabi-Yau spaces and threshold
  corrections}},
  \href{https://doi.org/10.1088/1126-6708/2005/05/023}{\emph{JHEP} {\bfseries
  05} (2005) 023} [\href{https://arxiv.org/abs/hep-th/0410018}{{\ttfamily
  hep-th/0410018}}].

\bibitem{Huybrechts:2016uxh}
D.~Huybrechts, \emph{{Lectures on K3 Surfaces}}. Cambridge University Press, 9,
  2016.

\bibitem{Mukai1}
S.~Mukai, \emph{{Curves, K3 Surfaces and Fano 3-folds of Genus {$\leq 10$}}},
  in \emph{\href{https://doi.org/10.1016/B978-0-12-348031-6.50026-7}{Algebraic
  Geometry and Commutative Algebra}}, pp.~357--377, Academic Press, (1988).

\bibitem{Mukai2}
S.~Mukai, \emph{{Biregular Classification of Fano 3-folds and Fano Manifolds of
  Coindex 3}}, \href{https://doi.org/10.1073/pnas.86.9.3000}{\emph{Proceedings
  of the National Academy of Sciences} {\bfseries 86} (1989) 3000}.

\bibitem{Mukai3}
S.~Mukai, \emph{{Curves and K3 Surfaces of Genus Eleven}},  in \emph{Moduli of
  Vector Bundles}, M.~Maruyama, ed., vol.~179 of \emph{Lecture Notes in Pure
  and Applied Mathematics}, pp.~189--197, Marcel Dekker, 1996.

\bibitem{Mukai4}
S.~Mukai, \emph{{Polarized K3 Surfaces of Genus 18 and 20}},  in
  \emph{\href{https://doi.org/10.1017/CBO9780511662652.019}{Complex Projective
  Geometry (Trieste, 1989/Bergen, 1989)}}, vol.~179, pp.~264--276, Cambridge
  University Press, (1992).

\bibitem{Mukai5}
S.~Mukai, \emph{{Polarized K3 Surfaces of Genus Thirteen}},  in \emph{Moduli
  Spaces and Arithmetic Geometry}, vol.~45 of
  \emph{\href{https://doi.org/10.2969/aspm/04510315}{Advanced Studies in Pure
  Mathematics}}, pp.~315--326, Mathematical Society of Japan, 2006.

\bibitem{Mukai6}
S.~Mukai, \emph{{K3 Surfaces of Genus Sixteen}},  in
  \emph{\href{https://doi.org/10.2969/aspm/07010379}{Minimal Models and
  Extremal Rays (Kyoto, 2011)}}, pp.~379--396, Mathematical Society of Japan,
  (2016).

\bibitem{Oguiso1}
K.~Oguiso, \emph{{On Algebraic Fiber Space Structures on a Calabi-Yau 3-fold}},
  \href{https://doi.org/10.1142/S0129167X93000248}{\emph{International Journal
  of Mathematics} {\bfseries 04} (1993) 439}.

\bibitem{Oguiso2}
K.~Oguiso, \emph{{On the Finiteness of Fiber-Space Structures on a Calabi-Yau
  3-fold}}, \href{https://doi.org/10.1023/A:1017959510524}{\emph{Journal of
  Mathematical Sciences} {\bfseries 106} (2001) 3320}.

\bibitem{Kanazawa2014}
A.~Kanazawa and P.~M.~H. Wilson, \emph{{Trilinear forms and Chern classes of
  Calabi-Yau threefolds}},
  {\emph{\href{https://projecteuclid.org/journals/osaka-journal-of-mathematics/volume-51/issue-1/Trilinear-forms-and-Chern-classes-of-Calabi--Yau-threefolds/ojm/1396966232.full}{Osaka
  Journal of Mathematics}} {\bfseries 51} (2014) 203}
  [\href{https://arxiv.org/abs/1201.3266}{{\ttfamily 1201.3266}}].

\bibitem{Wall1966}
C.~T.~C. Wall, \emph{{Classification problems in differential topology. V}},
  \href{https://doi.org/10.1007/BF01389738}{\emph{Inventiones mathematicae}
  {\bfseries 1} (1966) 355}.

\bibitem{Gross1993AFT}
M.~Gross, \emph{{A Finiteness Theorem for Elliptic Calabi-Yau Threefolds}},
  \href{https://doi.org/10.1215/S0012-7094-94-07414-0}{\emph{Duke Mathematical
  Journal} {\bfseries 74} (1993) 271}
  [\href{https://arxiv.org/abs/alg-geom/9305002}{{\ttfamily
  alg-geom/9305002}}].

\bibitem{Wilson2017}
P.~M.~H. Wilson, \emph{{Boundedness questions for Calabi-Yau threefolds}},
  \href{https://doi.org/10.1090/jag/781}{\emph{Journal of Algebraic Geometry}
  (2017) } [\href{https://arxiv.org/abs/1706.01268}{{\ttfamily 1706.01268}}].

\bibitem{Hajouji:2019vxs}
N.~Hajouji and P.-K. Oehlmann, \emph{{Modular Curves and Mordell-Weil Torsion
  in F-theory}}, \href{https://doi.org/10.1007/JHEP04(2020)103}{\emph{JHEP}
  {\bfseries 04} (2020) 103}
  [\href{https://arxiv.org/abs/1910.04095}{{\ttfamily 1910.04095}}].

\bibitem{eisenbud_harris_2016}
D.~Eisenbud and J.~Harris, \emph{3264 and All That: A Second Course in
  Algebraic Geometry}. Cambridge University Press, 2016,
  \href{https://doi.org/10.1017/CBO9781139062046}{10.1017/CBO9781139062046}.

\bibitem{mullet2006toric}
J.~P. Mullet, \emph{{On toric Calabi-Yau hypersurfaces fibered by weighted K3
  hypersurfaces}},
  \href{https://doi.org/10.4310/CAG.2009.V17.N1.A5}{\emph{Communications in
  Analysis and Geometry} {\bfseries 17} (2006) 107}
  [\href{https://arxiv.org/abs/math/0611338}{{\ttfamily math/0611338}}].

\bibitem{mullet_thesis}
J.~P. Mullet, \emph{{\href{http://hdl.handle.net/2142/86861}{Fibered Calabi-Yau
  Varieties and Toric Varieties}}}, Ph.D. thesis, University of Illinois at
  Urbana-Champaign, 2006.

\bibitem{fulton2016intersection}
W.~Fulton, \emph{{\href{https://doi.org/10.1007/978-1-4612-1700-8}{Intersection
  theory}}}. Princeton University Press, 2016.

\bibitem{Schubert2Source}
D.~R. Grayson, M.~E. Stillman, S.~A. Str{\o}mme, D.~Eisenbud and C.~Crissman,
  ``{Schubert2: characteristic classes for varieties without equations.
  Version~0.7}.'' A \emph{Macaulay2} package available at
  \url{http://www.math.uiuc.edu/Macaulay2/}.

\bibitem{M2}
D.~R. Grayson and M.~E. Stillman, ``Macaulay2, a software system for research
  in algebraic geometry.'' Available at \url{https://math.uiuc.edu/Macaulay2/}.

\bibitem{debarre2020hyperkahler}
O.~Debarre, \emph{Hyperk\"ahler manifolds},
  \href{https://arxiv.org/abs/1810.02087}{{\ttfamily 1810.02087}}.

\bibitem{Batyrev:2008rp}
V.~Batyrev and M.~Kreuzer, \emph{{Constructing new Calabi-Yau 3-folds and their
  mirrors via conifold transitions}},
  \href{https://doi.org/10.4310/ATMP.2010.v14.n3.a3}{\emph{Adv. Theor. Math.
  Phys.} {\bfseries 14} (2010) 879}
  [\href{https://arxiv.org/abs/0802.3376}{{\ttfamily 0802.3376}}].

\bibitem{Blumenhagen:2019vgj}
R.~Blumenhagen, M.~Brinkmann and A.~Makridou, \emph{{Quantum Log-Corrections to
  Swampland Conjectures}},
  \href{https://doi.org/10.1007/JHEP02(2020)064}{\emph{JHEP} {\bfseries 02}
  (2020) 064} [\href{https://arxiv.org/abs/1910.10185}{{\ttfamily
  1910.10185}}].

\bibitem{Hori:2013gga}
K.~Hori and J.~Knapp, \emph{{Linear sigma models with strongly coupled phases -
  one parameter models}},
  \href{https://doi.org/10.1007/JHEP11(2013)070}{\emph{JHEP} {\bfseries 11}
  (2013) 070} [\href{https://arxiv.org/abs/1308.6265}{{\ttfamily 1308.6265}}].

\bibitem{Caldararu:2017usq}
A.~Caldararu, J.~Knapp and E.~Sharpe, \emph{{GLSM realizations of maps and
  intersections of Grassmannians and Pfaffians}},
  \href{https://doi.org/10.1007/JHEP04(2018)119}{\emph{JHEP} {\bfseries 04}
  (2018) 119} [\href{https://arxiv.org/abs/1711.00047}{{\ttfamily
  1711.00047}}].

\bibitem{Knapp:2019cih}
J.~Knapp and E.~Sharpe, \emph{{GLSMs, joins, and nonperturbatively-realized
  geometries}}, \href{https://doi.org/10.1007/JHEP12(2019)096}{\emph{JHEP}
  {\bfseries 12} (2019) 096}
  [\href{https://arxiv.org/abs/1907.04350}{{\ttfamily 1907.04350}}].

\bibitem{Alvarez-Garcia:2021mzv}
R.~\'Alvarez-Garc\'\i{}a and L.~Schlechter, \emph{{Analytic Periods via Twisted
  Symmetric Squares}},  \href{https://arxiv.org/abs/2110.02962}{{\ttfamily
  2110.02962}}.

\bibitem{fricke1922elliptischen}
R.~Fricke, \emph{Die elliptischen Funktionen und ihre Anwendungen}. BG Teubner,
  1922.

\bibitem{FrickeSpringer1}
R.~Fricke, \emph{{Die elliptischen Funktionen und ihre Anwendungen -
  \href{https://doi.org/10.1007/978-3-642-19557-0}{Erster Teil: Die
  funktionentheoretischen und analytischen Grundlagen}}}. Springer-Verlag
  Berlin Heidelberg, 2012.

\bibitem{FrickeSpringer2}
R.~Fricke, \emph{{Die elliptischen Funktionen und ihre Anwendungen -
  \href{https://doi.org/10.1007/978-3-642-19561-7}{Zweiter Teil: Die
  algebraischen Ausführungen}}}. Springer-Verlag Berlin Heidelberg, 2012.

\bibitem{FrickeSpringer3}
R.~Fricke, \emph{{Die elliptischen Funktionen und ihre Anwendungen -
  \href{https://doi.org/10.1007/978-3-642-20954-3}{Dritter Teil:
  Anwendungen}}}. Springer-Verlag Berlin Heidelberg, 2012.

\bibitem{Aspinwall:1996mn}
P.~S. Aspinwall, \emph{{K3 Surfaces and String Duality}},  in
  \emph{{Theoretical Advanced Study Institute in Elementary Particle Physics
  (TASI 96): Fields, Strings, and Duality}}, pp.~421--540, 11, 1996,
  \href{https://arxiv.org/abs/hep-th/9611137}{{\ttfamily hep-th/9611137}}.

\bibitem{Braun:2016sks}
A.~P. Braun and T.~Watari, \emph{{Heterotic-Type IIA Duality and Degenerations
  of K3 Surfaces}}, \href{https://doi.org/10.1007/JHEP08(2016)034}{\emph{JHEP}
  {\bfseries 08} (2016) 034}
  [\href{https://arxiv.org/abs/1604.06437}{{\ttfamily 1604.06437}}].

\bibitem{Enoki:2019deb}
Y.~Enoki and T.~Watari, \emph{{Modular Forms as Classification Invariants of 4D
  $\mathcal{N}=2$ Heterotic-IIA Dual Vacua}},
  \href{https://doi.org/10.1007/JHEP06(2020)021}{\emph{JHEP} {\bfseries 06}
  (2020) 021} [\href{https://arxiv.org/abs/1911.09934}{{\ttfamily
  1911.09934}}].

\bibitem{Dolgachev:1996xw}
I.~V. Dolgachev, \emph{{Mirror symmetry for lattice polarized K3 surfaces}},
  \href{https://doi.org/10.1007/BF02362332}{\emph{Journal of Mathematical
  Sciences} {\bfseries 81} (1995) 2599}
  [\href{https://arxiv.org/abs/alg-geom/9502005}{{\ttfamily
  alg-geom/9502005}}].

\bibitem{Dolgachev:2005abc}
I.~V. Dolgachev and S.~Kondō,
  \emph{{\href{https://doi.org/10.1007/978-3-7643-8284-1_3}{Moduli Spaces of K3
  Surfaces and Complex Ball Quotients}}},  in \emph{Arithmetic and Geometry
  around Hypergeometric Functions}, pp.~43--100, Birkhäuser Basel, (2007),
  \href{https://arxiv.org/abs/math/0511051}{{\ttfamily math/0511051}}.

\bibitem{Doran_2019}
C.~F. Doran, A.~Harder, A.~Y. Novoseltsev and A.~Thompson, \emph{Calabi--yau
  threefolds fibred by high rank lattice polarized k3 surfaces},
  \href{https://doi.org/10.1007/s00209-019-02279-9}{\emph{Mathematische
  Zeitschrift} {\bfseries 294} (2019) 783}
  [\href{https://arxiv.org/abs/1701.03279}{{\ttfamily 1701.03279}}].

\bibitem{Narain:1985jj}
K.~S. Narain, \emph{{New Heterotic String Theories in Uncompactified Dimensions
  \ensuremath{<} 10}},
  \href{https://doi.org/10.1016/0370-2693(86)90682-9}{\emph{Phys. Lett. B}
  {\bfseries 169} (1986) 41}.

\bibitem{Narain:1986am}
K.~S. Narain, M.~H. Sarmadi and E.~Witten, \emph{{A Note on Toroidal
  Compactification of Heterotic String Theory}},
  \href{https://doi.org/10.1016/0550-3213(87)90001-0}{\emph{Nucl. Phys. B}
  {\bfseries 279} (1987) 369}.

\bibitem{Polchinski:1995df}
J.~Polchinski and E.~Witten, \emph{{Evidence for Heterotic - Type I String
  Duality}}, \href{https://doi.org/10.1016/0550-3213(95)00614-1}{\emph{Nucl.
  Phys. B} {\bfseries 460} (1996) 525}
  [\href{https://arxiv.org/abs/hep-th/9510169}{{\ttfamily hep-th/9510169}}].

\bibitem{Fraiman:2018ebo}
B.~Fraiman, M.~Gra\~na and C.~A. N\'u\~nez, \emph{{A new twist on heterotic
  string compactifications}},
  \href{https://doi.org/10.1007/JHEP09(2018)078}{\emph{JHEP} {\bfseries 09}
  (2018) 078} [\href{https://arxiv.org/abs/1805.11128}{{\ttfamily
  1805.11128}}].

\bibitem{Font:2020rsk}
A.~Font, B.~Fraiman, M.~Gra\~na, C.~A. N\'u\~nez and H.~P. De~Freitas,
  \emph{{Exploring the landscape of heterotic strings on $T^d$}},
  \href{https://doi.org/10.1007/JHEP10(2020)194}{\emph{JHEP} {\bfseries 10}
  (2020) 194} [\href{https://arxiv.org/abs/2007.10358}{{\ttfamily
  2007.10358}}].

\bibitem{Font:2021uyw}
A.~Font, B.~Fraiman, M.~Gra\~na, C.~A. N\'u\~nez and H.~Parra De~Freitas,
  \emph{{Exploring the landscape of CHL strings on $T^d$}},
  \href{https://arxiv.org/abs/2104.07131}{{\ttfamily 2104.07131}}.

\bibitem{Persson:2015jka}
D.~Persson and R.~Volpato, \emph{{Fricke S-duality in CHL models}},
  \href{https://doi.org/10.1007/JHEP12(2015)156}{\emph{JHEP} {\bfseries 12}
  (2015) 156} [\href{https://arxiv.org/abs/1504.07260}{{\ttfamily
  1504.07260}}].

\bibitem{Paquette:2016xoo}
N.~M. Paquette, D.~Persson and R.~Volpato, \emph{{Monstrous BPS-Algebras and
  the Superstring Origin of Moonshine}},
  \href{https://doi.org/10.4310/CNTP.2016.v10.n3.a2}{\emph{Commun. Num. Theor.
  Phys.} {\bfseries 10} (2016) 433}
  [\href{https://arxiv.org/abs/1601.05412}{{\ttfamily 1601.05412}}].

\bibitem{Paquette:2017xui}
N.~M. Paquette, D.~Persson and R.~Volpato, \emph{{BPS Algebras, Genus Zero, and
  the Heterotic Monster}},
  \href{https://doi.org/10.1088/1751-8121/aa8443}{\emph{J. Phys. A} {\bfseries
  50} (2017) 414001} [\href{https://arxiv.org/abs/1701.05169}{{\ttfamily
  1701.05169}}].

\bibitem{Persson:2017lkn}
D.~Persson and R.~Volpato, \emph{{Dualities in CHL-Models}},
  \href{https://doi.org/10.1088/1751-8121/aab489}{\emph{J. Phys. A} {\bfseries
  51} (2018) 164002} [\href{https://arxiv.org/abs/1704.00501}{{\ttfamily
  1704.00501}}].

\end{thebibliography}\endgroup
\end{document}